\documentclass{emulateapj}
\usepackage{apjfonts}
\usepackage{natbib}

\def\chandra{{\it Chandra\/}}
\def\xray{\hbox{X-ray}}

\def\asca{{\it ASCA\/}}

\def\flux{erg~cm$^{-2}$~s$^{-1}$}

\def\lsim{\mathrel{\rlap{\lower4pt\hbox{\hskip1pt$\sim$}}
    \raise1pt\hbox{$<$}}}                % less than or approx. symbol
\def\gsim{\mathrel{\rlap{\lower4pt\hbox{\hskip1pt$\sim$}}
    \raise1pt\hbox{$>$}}}                % greater than or approx. symbol

\def\cdfs{\hbox{CDF-S}}

\begin{document}
\slugcomment{Data and images available at http://www.astro.psu.edu/users/niel/cdfs/cdfs-chandra.html}

\title{The {\em Chandra} Deep Field-South Survey: 4 Ms Source Catalogs}

\author{
Y.~Q.~Xue,\altaffilmark{1,2}
B.~Luo,\altaffilmark{1,2}
W.~N.~Brandt,\altaffilmark{1,2}
F.~E.~Bauer,\altaffilmark{3,4}
B.~D.~Lehmer,\altaffilmark{5,6}
P.~S.~Broos,\altaffilmark{1}
D.~P.~Schneider,\altaffilmark{1}
D.~M.~Alexander,\altaffilmark{7}
M.~Brusa,\altaffilmark{8,9}
A.~Comastri,\altaffilmark{10}
A.~C.~Fabian,\altaffilmark{11}
R.~Gilli,\altaffilmark{10}
G.~Hasinger,\altaffilmark{12}
A.~E.~Hornschemeier,\altaffilmark{13}
A.~Koekemoer,\altaffilmark{14}
T.~Liu,\altaffilmark{15,16}
V.~Mainieri,\altaffilmark{17}
M.~Paolillo,\altaffilmark{18}
D.~A.~Rafferty,\altaffilmark{19}
P.~Rosati,\altaffilmark{17}
O.~Shemmer,\altaffilmark{20}
J.~D.~Silverman,\altaffilmark{21}
I.~Smail,\altaffilmark{22}
P.~Tozzi,\altaffilmark{15} and
C.~Vignali\altaffilmark{23}
}
\altaffiltext{1}{Department of Astronomy and Astrophysics, Pennsylvania State University, University Park, PA 16802, USA; xuey@astro.psu.edu}
\altaffiltext{2}{Institute for Gravitation and the Cosmos, Pennsylvania State University, University Park, PA 16802, USA}
\altaffiltext{3}{Pontificia Universidad Cat\'{o}lica de Chile, Departamento de Astronom\'{\i}a y Astrof\'{\i}sica, Casilla 306, Santiago 22, Chile}
\altaffiltext{4}{Space Science Institute, 4750 Walnut Street, Suite 205, Boulder, CO 80301, USA}
\altaffiltext{5}{The Johns Hopkins University, Homewood Campus, Baltimore, MD 21218, USA}
\altaffiltext{6}{NASA Goddard Space Flight Centre, Code 662, Greenbelt, MD 20771, USA}
\altaffiltext{7}{Department of Physics, Durham University, Durham, DH1 3LE, UK}
\altaffiltext{8}{Max-Planck-Institut f\"ur Extraterrestrische Physik, Giessenbachstrasse, D-85748 Garching, Germany}
\altaffiltext{9}{Department of Astronomy, University of Maryland, Baltimore County, 1000 Hilltop Circle, Baltimore, MD 21250, USA}
\altaffiltext{10}{INAF---Osservatorio Astronomico di Bologna, Via Ranzani 1, Bologna, Italy}
\altaffiltext{11}{Institute of Astronomy, Madingley Road, Cambridge, CB3 0HA, UK}
\altaffiltext{12}{Max-Planck-Institut f\"ur Plasmaphysik, Boltzmannstrasse 2, D-85748 Garching, Germany}
\altaffiltext{13}{Laboratory for X-ray Astrophysics, NASA Goddard Space Flight Center, Code 662, Greenbelt, MD 20771, USA}
\altaffiltext{14}{Space Telescope Science Institute, 3700 San Martin Drive, Baltimore, MD 21218, USA}
\altaffiltext{15}{INAF---Osservatorio Astronomico di Trieste, Via Tiepolo 11, I-34131 Trieste, Italy}
\altaffiltext{16}{CAS Key Laboratory for Research in Galaxies and Cosmology, Department of Astronomy, University of Science and Technology of China, Hefei, Anhui 230026, P. R. China}
\altaffiltext{17}{European Southern Observatory, Karl-Schwarzschild-Strasse 2, Garching, D-85748, Germany}
\altaffiltext{18}{Dipartimento di Scienze Fisiche, Universit\`a Federico II di Napoli, Via Cinthia, 80126 Napoli, Italy}
\altaffiltext{19}{Leiden Observatory, Leiden University, Oort Gebouw, P.O. Box 9513 RA, Leiden, The Netherlands}
\altaffiltext{20}{Department of Physics, University of North Texas, Denton, TX 76203, USA}
\altaffiltext{21}{Institute for the Physics and Mathematics of the Universe (IPMU), University of Tokyo, Kashiwanoha 5-1-5, Kashiwa-shi, Chiba 277-8568, Japan}
\altaffiltext{22}{Institute of Computational Cosmology, Durham University, Durham, DH1 3LE, UK}
\altaffiltext{23}{Universit\'a di Bologna, Via Ranzani 1, Bologna, Italy}

\begin{abstract}

We present source catalogs for the 4~Ms \chandra\ Deep Field-South (\hbox{CDF-S}),
which is the deepest \chandra\ survey to date
and covers an area of 464.5 arcmin$^{2}$.
We provide a main \chandra\ source catalog, which contains 740 \hbox{X-ray}  
sources that are detected with {\sc wavdetect} at a false-positive probability
threshold of $10^{-5}$ in at least one of three \hbox{X-ray} bands
(\hbox{0.5--8~keV}, full band; \hbox{0.5--2~keV}, soft band; and \hbox{2--8~keV}, hard band)
and also satisfy a 
binomial-probability source-selection criterion of \mbox{$P<0.004$}
(i.e., the probability of sources not being real is less than 0.004);
this approach is designed to maximize the number of reliable sources detected.
A total of 300 main-catalog sources are new compared to the previous 2~Ms
\hbox{CDF-S} main-catalog sources.
We determine \hbox{X-ray} source positions using centroid
and matched-filter techniques and obtain a median positional uncertainty of
$\approx0.42\arcsec$.
We also provide a supplementary catalog, which consists of 36 sources that
are detected with {\sc wavdetect} at a false-positive probability
threshold of $10^{-5}$, satisfy the condition of $0.004<P<0.1$, 
and have an optical counterpart with $R<24$.
Multiwavelength identifications, basic optical/infrared/radio photometry, and  
spectroscopic/photometric redshifts are provided for the \hbox{X-ray} sources
in the main and supplementary catalogs.
716 ($\approx 97\%$) of the 740 main-catalog sources have multiwavelength counterparts,
with 673 ($\approx 94\%$ of 716) having either spectroscopic or photometric redshifts.
The 740 main-catalog sources span broad ranges of full-band flux  
%($3.9\times 10^{-17}$--$1.1\times 10^{-13}$ \flux)
and \hbox{0.5--8 keV} luminosity; %($1.9\times 10^{39}$--$1.9\times 10^{45}$ erg~s$^{-1}$);
the 300 new main-catalog sources span similar ranges
although they tend to be systematically lower.
Basic analyses of the \hbox{X-ray} and multiwavelength properties of the sources
indicate that $>75\%$ of the main-catalog sources are AGNs;
of the 300 new main-catalog sources, about 35\% are likely normal and starburst galaxies, 
reflecting the rise of normal
and starburst galaxies at the very faint flux levels uniquely accessible to
the 4~Ms \hbox{CDF-S}.
Near the center of the 4~Ms \hbox{CDF-S} (i.e., within an off-axis angle of~$3\arcmin$),
the observed AGN and galaxy source densities have reached
$9800_{-1100}^{+1300}$~deg$^{-2}$ and 
$6900_{-900}^{+1100}$~deg$^{-2}$, respectively.
Simulations show that our main catalog is highly reliable and is reasonably complete.
The mean backgrounds (corrected for vignetting and exposure-time variations) are 0.063 and 0.178
count~Ms$^{-1}$~pixel$^{-1}$ (for a pixel size of $0.492\arcsec$) for the soft and hard bands, 
respectively; the majority of the pixels have zero background counts.
The 4~Ms \mbox{CDF-S} reaches
on-axis flux limits of
$\approx 3.2\times 10^{-17}$, $9.1\times 10^{-18}$, and $5.5\times 10^{-17}$ \flux\
for the full, soft, and hard bands, respectively.
An increase in the \hbox{CDF-S} exposure time
by a factor of \hbox{$\approx 2$--2.5} would provide further
significant gains and probe key unexplored discovery space.

\end{abstract}

\section{INTRODUCTION}\label{sec:intro}

Deep \hbox{X-ray} surveys indicate that the cosmic \hbox{X-ray} background (CXRB)
is largely due to accretion onto supermassive black holes integrated over
cosmic time. One of the greatest legacies of the {\it Chandra X-ray Observatory} 
(\chandra) is the characterization of the CXRB sources thanks to its extraordinary
sensitivity. The \chandra\ Deep Field-North 
and \chandra\ Deep Field-South (\hbox{CDF-N} and \hbox{CDF-S}, jointly CDFs)
are the two deepest \chandra\ surveys (see Brandt \& Hasinger 2005 and Brandt \& Alexander 2010
for reviews of deep extragalactic \hbox{X-ray} surveys), each covering $\approx 450$~arcmin$^2$
areas with tremendous multiwavelength observational investments.
Most of the CDF sources are active galactic nuclei (AGNs), often obscured,
at \hbox{$z\approx$~0.1--5.2}. The CDFs have found the
highest density of reliably identified AGNs on the sky, 
with an AGN source density approaching ten thousand sources per deg$^2$
(e.g., Bauer et~al. 2004). At faint fluxes, the CDFs are also detecting large numbers 
of starburst and normal galaxies at \hbox{$z\approx 0.1$--2}
as well as a few individual off-nuclear \hbox{X-ray} binaries 
at \hbox{$z\approx 0.05$--0.3}.

Deeper \xray\ observations not only further improve the
photon statistics that are required to understand better
the already detected sources via \xray\ spectral and variability constraints,
but also probe further down the \xray\ luminosity versus redshift plane
to characterize better the properties and evolution of typical AGNs and galaxies.
The recent extension of 
the \hbox{CDF-S} survey from 2~Ms (Luo et~al. 2008; hereafter L08) 
to 4~Ms of exposure, via a large Director's Discretionary Time project, has now 
provided our most sensitive \hbox{0.5--8~keV} view of the distant 
universe. These data, complemented by the recent $\approx 3.3$~Ms
XMM-{\it Newton} observations in the \hbox{CDF-S} (Comastri et~al. 2011), will enable detailed studies of AGN evolution, 
physics, and ecology as well as the \mbox{X-ray} properties of normal 
and starburst galaxies, groups and clusters of galaxies, large-scale 
structures, and Galactic stars.

In this paper, we present \chandra\ source catalogs and data
products derived from the full 4~Ms \hbox{CDF-S} data set
as well as details of the observations, data reduction, and technical 
analysis. We have made a number of methodological improvements in 
catalog production relative to past CDF catalogs. 
The structure of this paper is the following: 
in \S~\ref{sec:obs} we describe the observations and data reduction;
in \S~\ref{sec:cats} we detail the production of images, exposure maps,
and the candidate-list catalog;
in \S~\ref{main} and \S~\ref{sec:supp2} we present the main and supplementary source catalogs
as well as description of the adopted methodology, respectively;
in \S~\ref{sec:comp} we perform simulations to assess the completeness and
reliability of the main source catalog;
in \S~\ref{sec:bkg} we estimate
the background and sensitivity across the \mbox{CDF-S} and investigate
the prospects for longer \hbox{CDF-S} exposures;
and in \S~\ref{sec:summary} we summarize the results of this work.

Throughout this paper, we adopt a Galactic column density 
of \hbox{$N_{\rm H}=8.8\times 10^{19}$~cm$^{-2}$} 
(e.g., Stark et~al. 1992) along the line of sight to the
\hbox{CDF-S}. We use J2000.0 coordinates and a cosmology of 
$H_0=70.4$ km~s$^{-1}$~Mpc$^{-1}$, 
$\Omega_{\rm M}=0.272$, and 
$\Omega_\Lambda=0.728$ (e.g., Komatsu et~al. 2011).

\section{OBSERVATIONS AND DATA REDUCTION}\label{sec:obs}

\subsection{Observations and Observing Conditions}

Table~\ref{tbl-obs} summarizes basic information for the 31 \hbox{CDF-S} observations that were taken between 2010 March 18 and 2010 July 22, which comprise the second 2~Ms exposure. 
The first 2~Ms exposure consisted of 23 observations 
(see Table~1 of L08 for basic information) that 
were performed between 1999 October 15 and 2007 November 4; the corresponding
source catalogs were presented in L08.

All 54 \hbox{CDF-S} observations made use of the Advanced CCD Imaging Spectrometer imaging array 
(\hbox{ACIS-I}; Garmire et~al. 2003).
The \hbox{ACIS-I} is comprised of four $1024\times1024$ pixel CCDs (CCDs \hbox{I0--I3};
each has a pixel size of $0.492\arcsec$) and
is optimized for imaging wide fields (with a field of view of 
$16.9\arcmin \times 16.9\arcmin = 285.6$ arcmin$^2$).
The focal-plane temperature was $-110\degr$C during the first two 
observations (1431-0 and 1431-1; Giacconi et~al. 2002; L08) and $-120\degr$C during the others.
The 10 early \hbox{CDF-S} observations between 1999 November 23
and 2000 December 23 were taken in Faint mode (Giacconi et~al. 2002; L08);
all the later \hbox{CDF-S} observations as well as the earliest one (observation 1431-0)
were taken in Very Faint mode 
in order to improve the screening of background events and thus increase
the sensitivity of ACIS in detecting faint X-ray sources (Vikhlinin 2001).

We inspected the background light curves for all 54 \hbox{CDF-S} observations
using the \chandra\ Imaging and Plotting System (ChIPS)\footnote{See 
http://cxc.harvard.edu/ciao3.4/download/doc/chips\_manual/ for the ChIPS reference manual.} as well as
EVENT BROWSER in the Tools for ACIS Real-time Analysis ({\sc tara}; Broos et~al. 2000)
software package.\footnote{{\sc tara} is available at http://www.astro.psu.edu/xray/docs/TARA/.}
We find no significant flaring for all observations (the background is stable
within $\approx$20\% of typical quiescent \chandra\ values)
except observation 1431-0, during which a mild flare with a factor of $\approx3$ 
increase for $\approx5$~ks occurred.
We filtered the data on good-time intervals, removed the one mild flare,
and obtained a total exposure time of 3.872~Ms for the 54 \hbox{CDF-S} observations.

The entire \hbox{CDF-S} covers an area of 464.5 arcmin${^2}$; this is 
considerably larger than the \hbox{ACIS-I} field of view because the 
aim points and roll angles vary between observations.
The average aim point, weighted using the 54 individual exposure times, is $\alpha_{\rm J2000.0}=03^{\rm h}32^{\rm m}28.06^{\rm s}$, $\delta_{\rm J2000.0}=-27\degr48\arcmin26.4\arcsec$.

%\clearpage
%\LongTables
%\begin{landscape}
\begin{deluxetable*}{lcccccl}
%\tabletypesize{\small}
\tabletypesize{\footnotesize}
\tablecaption{Journal of New {\it Chandra} Deep Field-South Observations}
\tablehead{
\colhead{}                                 &
\colhead{Obs. Start}                                 &
\colhead{Exposure}                             &
\multicolumn{2}{c}{Aim Point$^{\rm b}$}                 &
\colhead{Roll Angle$^{\rm c}$}                                 &
%\colhead{Obs.}                                &
\colhead{Pipeline}                             \\
\cline{4-5}
\colhead{Obs. ID}                                 &
\colhead{(UT)}                         &
\colhead{Time$^{\rm a}$ (ks)}               &
\colhead{$\alpha$ (J2000.0)}                &
\colhead{$\delta$ (J2000.0)}                &
\colhead{(deg)}         &
%\colhead{Mode$^{\rm d}$}                             &
\colhead{Version$^{\rm d}$}                             
}
\tablewidth{0pt}
\startdata
\multicolumn{7}{c}{[The 23 observations made during the first 2~Ms exposure are listed in Table~1 of L08]} \\
12043\dotfill\ldots\ldots & 2010 Mar 18, 01:39 & 129.6 & 03 32 28.78 & $-$27 48 52.1 & 252.2 & 8.2.1 \\
12123\dotfill\ldots\ldots & 2010 Mar 21, 08:08 &  \phantom{0}24.8 & 03 32 28.78 & $-$27 48 52.1 & 252.2 & 8.2.1 \\
12044\dotfill\ldots\ldots & 2010 Mar 23, 11:31 &  \phantom{0}99.5 & 03 32 28.55 & $-$27 48 51.9 & 246.2 & 8.2.1 \\
12128\dotfill\ldots\ldots & 2010 Mar 27, 13:08 &  \phantom{0}22.8 & 03 32 28.55 & $-$27 48 51.9 & 246.2 & 8.2.1 \\
12045\dotfill\ldots\ldots & 2010 Mar 28, 16:38 &  \phantom{0}99.7 & 03 32 28.32 & $-$27 48 51.4 & 240.2 & 8.2.1 \\
12129\dotfill\ldots\ldots & 2010 Apr 03, 15:21 &  \phantom{0}77.1 & 03 32 28.33 & $-$27 48 51.4 & 240.2 & 8.2.1 \\
12135\dotfill\ldots\ldots & 2010 Apr 06, 09:36 &  \phantom{0}62.5 & 03 32 28.01 & $-$27 48 50.2 & 231.7 & 8.2.1 \\
12046\dotfill\ldots\ldots & 2010 Apr 08, 08:17 &  \phantom{0}78.0 & 03 32 28.01 & $-$27 48 50.2 & 231.7 & 8.2.1 \\
12047\dotfill\ldots\ldots & 2010 Apr 12, 13:21 &  \phantom{0}10.1 & 03 32 27.80 & $-$27 48 48.9 & 225.2 & 8.2.1 \\
12137\dotfill\ldots\ldots & 2010 Apr 16, 08:53 &  \phantom{0}92.8 & 03 32 27.59 & $-$27 48 47.2 & 219.2 & 8.2.1 \\
12138\dotfill\ldots\ldots & 2010 Apr 18, 12:40 &  \phantom{0}38.5 & 03 32 27.59 & $-$27 48 47.3 & 219.2 & 8.2.1 \\
12055\dotfill\ldots\ldots & 2010 May 15, 17:15 &  \phantom{0}80.7 & 03 32 26.72 & $-$27 48 32.3 & 181.4 & 8.2.1 \\
12213\dotfill\ldots\ldots & 2010 May 17, 14:22 &  \phantom{0}61.3 & 03 32 26.69 & $-$27 48 31.1 & 178.9 & 8.2.1 \\
12048\dotfill\ldots\ldots & 2010 May 23, 07:09 & 138.1 & 03 32 26.64 & $-$27 48 27.6 & 171.9 & 8.2.1 \\
12049\dotfill\ldots\ldots & 2010 May 28, 18:58 &  \phantom{0}86.9 & 03 32 26.61 & $-$27 48 24.4 & 165.5 & 8.2.1 \\
12050\dotfill\ldots\ldots & 2010 Jun 03, 06:47 &  \phantom{0}29.7 & 03 32 26.61 & $-$27 48 21.7 & 160.2 & 8.2.1 \\
12222\dotfill\ldots\ldots & 2010 Jun 05, 02:47 &  \phantom{0}30.6 & 03 32 26.61 & $-$27 48 21.7 & 160.2 & 8.2.1 \\
12219\dotfill\ldots\ldots & 2010 Jun 06, 16:30 &  \phantom{0}33.7 & 03 32 26.61 & $-$27 48 21.7 & 160.2 & 8.2.1 \\
12051\dotfill\ldots\ldots & 2010 Jun 10, 11:30 &  \phantom{0}57.3 & 03 32 26.63 & $-$27 48 19.2 & 155.2 & 8.2.1 \\
12218\dotfill\ldots\ldots & 2010 Jun 11, 10:18 &  \phantom{0}88.0 & 03 32 26.63 & $-$27 48 19.2 & 155.2 & 8.2.1 \\
12223\dotfill\ldots\ldots & 2010 Jun 13, 00:57 & 100.7 & 03 32 26.63 & $-$27 48 19.2 & 155.2 & 8.2.1 \\
12052\dotfill\ldots\ldots & 2010 Jun 15, 16:02 & 110.4 & 03 32 26.70 & $-$27 48 14.5 & 145.7 & 8.2.1 \\
12220\dotfill\ldots\ldots & 2010 Jun 18, 12:55 &  \phantom{0}48.1 & 03 32 26.70 & $-$27 48 14.5 & 145.7 & 8.2.1 \\
12053\dotfill\ldots\ldots & 2010 Jul 05, 03:12 &  \phantom{0}68.1 & 03 32 27.02 & $-$27 48 06.0 & 127.0 & 8.3 \\
12054\dotfill\ldots\ldots & 2010 Jul 09, 11:35 &  \phantom{0}61.0 & 03 32 27.02 & $-$27 48 06.1 & 127.0 & 8.3 \\
12230\dotfill\ldots\ldots & 2010 Jul 11, 03:52 &  \phantom{0}33.8 & 03 32 27.02 & $-$27 48 06.0 & 127.0 & 8.3 \\
12231\dotfill\ldots\ldots & 2010 Jul 12, 03:22 &  \phantom{0}24.7 & 03 32 27.16 & $-$27 48 03.6 & 121.2 & 8.3 \\
12227\dotfill\ldots\ldots & 2010 Jul 14, 21:04 &  \phantom{0}54.3 & 03 32 27.16 & $-$27 48 03.7 & 121.2 & 8.3 \\
12233\dotfill\ldots\ldots & 2010 Jul 16, 10:25 &  \phantom{0}35.6 & 03 32 27.16 & $-$27 48 03.7 & 121.2 & 8.3 \\
12232\dotfill\ldots\ldots & 2010 Jul 18, 19:53 &  \phantom{0}32.9 & 03 32 27.16 & $-$27 48 03.7 & 121.2 & 8.3 \\
12234\dotfill\ldots\ldots & 2010 Jul 22, 19:58 &  \phantom{0}49.1 & 03 32 27.19 & $-$27 48 03.3 & 120.2 & 8.3 \\
\enddata

\tablecomments{
The 4~Ms \hbox{CDF-S} consists of 54 observations, 
with the first 2~Ms exposure composed of 23 observations
(listed in Table~1 of L08; not listed here to avoid repetition) and
the second 2~Ms exposure composed of 31 observations (listed in this table;
these 31 observations were all taken with the Very Faint mode). 
Right ascension has units of hours, minutes, and seconds, and declination has units of 
degrees, arcminutes, and arcseconds.
}
\par \tablenotetext{a}{Each of the 54 observations was continuous. We filtered the data
on good-time intervals and removed one mild flare in observation 1431-0 (during 
the first 2~Ms exposure).
The summed exposure time for the 54 observations is 3.872~Ms.}

\tablenotetext{b}{The aim points of the individual observations are the nominal ones taken from the \chandra\ archive.
The average aim point, weighted by the 54 exposure times, is
$\alpha_{\rm J2000.0}=03^{\rm h}32^{\rm m}28.06^{\rm s}$, 
$\delta_{\rm J2000.0}=-27\degr48\arcmin26.4\arcsec$.}

\tablenotetext{c}{Roll angle, describing the orientation of the \chandra\ 
instruments
on the sky, ranges from 0$^{\circ}$ to 360$^{\circ}$ and increases to the west
of north (opposite to the sense of traditional position angle).}

\tablenotetext{d}{The version of the CXC pipeline software used for the basic 
processing of the data.}
\label{tbl-obs}
\end{deluxetable*}
%\clearpage
%\end{landscape}

\subsection{Data Reduction}

Table~\ref{tbl-obs} lists the versions of the \chandra\ \hbox{X-ray} Center (CXC) pipeline software used to process the basic archive data products for the 31 new observations (see Table~1 of L08 for the information for the first 23 observations).
We closely followed L08 in reducing and analyzing the data and refer 
readers to L08 for details.
Briefly, we utilized \chandra\ Interactive Analysis of Observations ({\sc ciao}; we used {\sc ciao} 4.2 and {\sc caldb} 4.3.0) tools 
and custom software, including the {\sc tara} package (version released on 2010 February 26), as appropriate.

We reprocessed each level~1 observation with the {\sc ciao} tool {\sc acis\_process\_events} 
to correct for the radiation damage sustained by the CCDs during the first few months of
\chandra\ operations using a Charge Transfer Inefficiency (CTI)
correction procedure (Townsley et~al. 2000, 2002)\footnote{The CXC CTI correction procedure
is only available for $-120\degr$C data and is thus not applied to 
observations 1431-0 and 1431-1.}, 
to remove the standard pixel randomization which causes point spread function (PSF) blurring,
and to apply a modified bad-pixel file.
We made use of a customized stripped-down bad-pixel file rather than the standard CXC
bad-pixel file because the latter excludes \hbox{$\approx6$--7\%} of the \hbox{ACIS-I} pixels 
on which a large fraction of events are valid for 
source searching as well as photometry and spectral analysis
(see \S~2.2 of L08 for details).
Our bad-pixel screening removed $\approx 1.3$\% of all events.
When cleaning background events, 
we set {\sc check\_vf\_pha=yes} in {\sc acis\_process\_events} for observations taken in Very Faint mode 
to utilize a $5\times 5$ pixel event island to search for potential cosmic-ray background events,
which typically removes \mbox{$\approx 20$--30\%} of the events of individual observations.

We used the {\sc ciao} tool {\sc acis\_detect\_afterglow} to remove cosmic-ray afterglows,
which is more stringent than the {\sc ciao} tool {\sc acis\_run\_hotpix} that often
fails to flag a substantial number of obvious cosmic-ray afterglows.
Even {\sc acis\_detect\_afterglow} fails
to reject all afterglows.
Working in CCD coordinates, we therefore utilized custom software to clean the data further by 
removing many additional faint afterglows with 3 or more total counts occurring within 20~s (or
equivalently 6 consecutive frames) on a pixel.\footnote{As shown later 
in Table~\ref{tbl-bkg}, the full-band (i.e., \hbox{0.5--8 keV}) mean background rate 
of the 4~Ms \hbox{CDF-S} is 0.252 count~Ms$^{-1}$~pixel$^{-1}$, 
which translates into a count rate of $5.04\times 10^{-6}$ counts per 20 s per pixel.
Given such a low background count rate, the probability of 3 or more counts
(that are not associated with cosmic-ray afterglows) occurring within 20 s
on a pixel by chance is negligible ($2.54\times 10^{-11}$).}
We removed a total of 176 additional faint afterglows across the full 4~Ms dataset
which, upon inspection,
were isolated and not associated with apparent legitimate \hbox{X-ray} sources.

As stated above, one significant deviation of our data reduction from the CXC reduction
of the \hbox{CDF-S} data set\footnote{The CXC \hbox{CDF-S} data products are available at
http://cxc.harvard.edu/cda/Contrib/CDFS.html.} is implementation of a customized stripped-down bad-pixel file, which
retains an appreciable number of valid events 
(accounting for \hbox{$\approx 5$\%} of all events) that would 
have been discarded using the standard CXC bad-pixel file.
As will be described in \S~\ref{sec:img} and \S~\ref{sec:list}, 
when creating the \cdfs\ data products (e.g., merged \xray\ images) and 
source catalogs (e.g., \xray\ source positions),
we registered individual \xray\ observations to a common optical/radio
astrometric frame (see \S~\ref{sec:img}) and refined the absolute astrometry
of the merged \xray\ images and source positions using high-quality radio data
(see \S~\ref{sec:list}), thereby producing sharp merged \xray\ images
and accurate \xray\ source positions (with $<0.2\arcsec$ astrometric shifts);   
in contrast, the CXC did not utilize multiwavelength data to register and refine
\xray\ astrometry.

\section{Images, exposure maps, and candidate-list catalog}\label{sec:cats}

While following the general procedure described in \S~3 of L08 in the production of our source catalogs,
we extensively made use of the ACIS Extract (AE; version released on 2010 February 26; Broos et~al. 2010)\footnote{See
http://www.astro.psu.edu/xray/docs/TARA/ae\_users\_guide.html for details on ACIS Extract.}
point-source analysis software that appropriately computes source properties when
multiple observations with different roll angles and/or aim points are being combined
(such as those analyzed here).
Significant improvements from the methodology of L08 include, e.g., 
(1) utilization of AE polygonal source-count extraction regions that
approximate the shape of the PSF and 
take into account the multi-observation nature of the data, and
(2) utilization of a two-stage approach to source detection, which
filters candidate sources according to binomial no-source probabilities 
(i.e., probabilities of sources not being real considering their local backgrounds)
calculated by AE.

We first generated a candidate-list catalog of sources detected by {\sc wavdetect} (Freeman et~al. 2002) on the combined images (see \S~\ref{sec:list})
at a false-positive probability threshold of $10^{-5}$.
We then pruned the candidate-list catalog to obtain a more conservative main catalog
by removing low-significance source candidates, according to the AE-computed binomial no-source probabilities.
As detailed later in \S~\ref{sec:list} and \S~\ref{main}, this approach not only produces source catalogs that are of similar quality to
those produced by running {\sc wavdetect} at the more typical false-positive probability threshold
of $10^{-6}$ or $10^{-7}$ used in previous CDF studies (e.g., Alexander et~al. 2003, hereafter A03; Lehmer et~al. 2005, hereafter L05; L08),
but also allows for flexibility in including additional legitimate sources that fall
below the $10^{-6}$ or $10^{-7}$ threshold.
This procedure has previously been employed in similar forms in a number of studies
(e.g., Getman et~al. 2005; Nandra et~al. 2005; Laird et~al. 2009; Lehmer et~al. 2009).

\subsection{Image and Exposure Map Creation}\label{sec:img}

To construct the combined event file we initially
ran {\sc wavdetect} at a false-positive probability threshold of
$10^{-6}$ on the individual cleaned \hbox{0.5--8}~keV
image of each observation to generate initial source lists
and used AE to determine centroid positions of each detected source.
We then registered the observations to a common astrometric frame
by matching \mbox{X-ray} centroid positions to optical sources detected in
deep $R$-band images taken with the Wide Field Imager (WFI) mounted on the 2.2-m  
Max Planck Gesellshaft/European Southern Observatory (ESO) telescope at La Silla (see \S~2 of 
Giavalisco et~al. 2004). 
We have manually shifted all the WFI $R$-band source positions by $0.175\arcsec$ 
in right ascension and $-0.284\arcsec$ in declination (also see Luo et~al. 2010; hereafter L10) 
to remove the systematic offsets
between the optical positions and the radio positions of sources in
the Very Large Array (VLA) 1.4~GHz radio catalog presented in 
Miller et~al. (2008).\footnote{Throughout this paper, we used the 
$5\sigma$ VLA \mbox{1.4 GHz} radio catalog (N. A. Miller 2010, private communication)
that has a limiting flux density of \mbox{$\approx 40$ $\mu$Jy}.\label{miller}}
We did not directly match \mbox{X-ray} centroid positions to the VLA radio catalog
because, for some observations,
there are too few common sources between the \mbox{X-ray} and radio source lists to ensure a robust astrometric solution, 
owing to the relatively low radio-source density and relatively small 
numbers of \chandra\ sources detected in {\it individual} observations.
However, as detailed in \S~\ref{sec:list}, we are able to lock the
absolute astrometry of the {\it combined} \hbox{X-ray} images to the VLA radio catalog
because of the larger number of \hbox{X-ray} sources detected.
We performed \xray/$R$-band matching and astrometric re-projection using the {\sc ciao} tools {\sc reproject\_aspect} and
{\sc wcs\_update} with a 3$\arcsec$ matching radius and a
residual rejection limit\footnote{This is a parameter used in {\sc wcs\_update}
to remove source pairs based on pair positional offsets.} of $0.6\arcsec$.
Typically, \hbox{60--150} \xray/$R$-band matches were used in each observation for the astrometric solutions. 
When using {\sc wcs\_update}, linear translations range from $0.032\arcsec$ to $0.525\arcsec$, 
rotations range from $-0.048\degr$ to $0.035\degr$, and scale changes range from 1.00004 to 1.00145.
Individual registrations are accurate to $\approx 0.3\arcsec$.
We then reprojected all the observations to the frame of observation 2406, 
which is one of the observations that requires the smallest translation to be aligned with the optical astrometric frame; 
however, we note that 
it does not matter which observation is used as the reference frame for reprojection
once each observation is analyzed consistently.

\begin{figure*}
\centerline{
\includegraphics[scale=0.75]{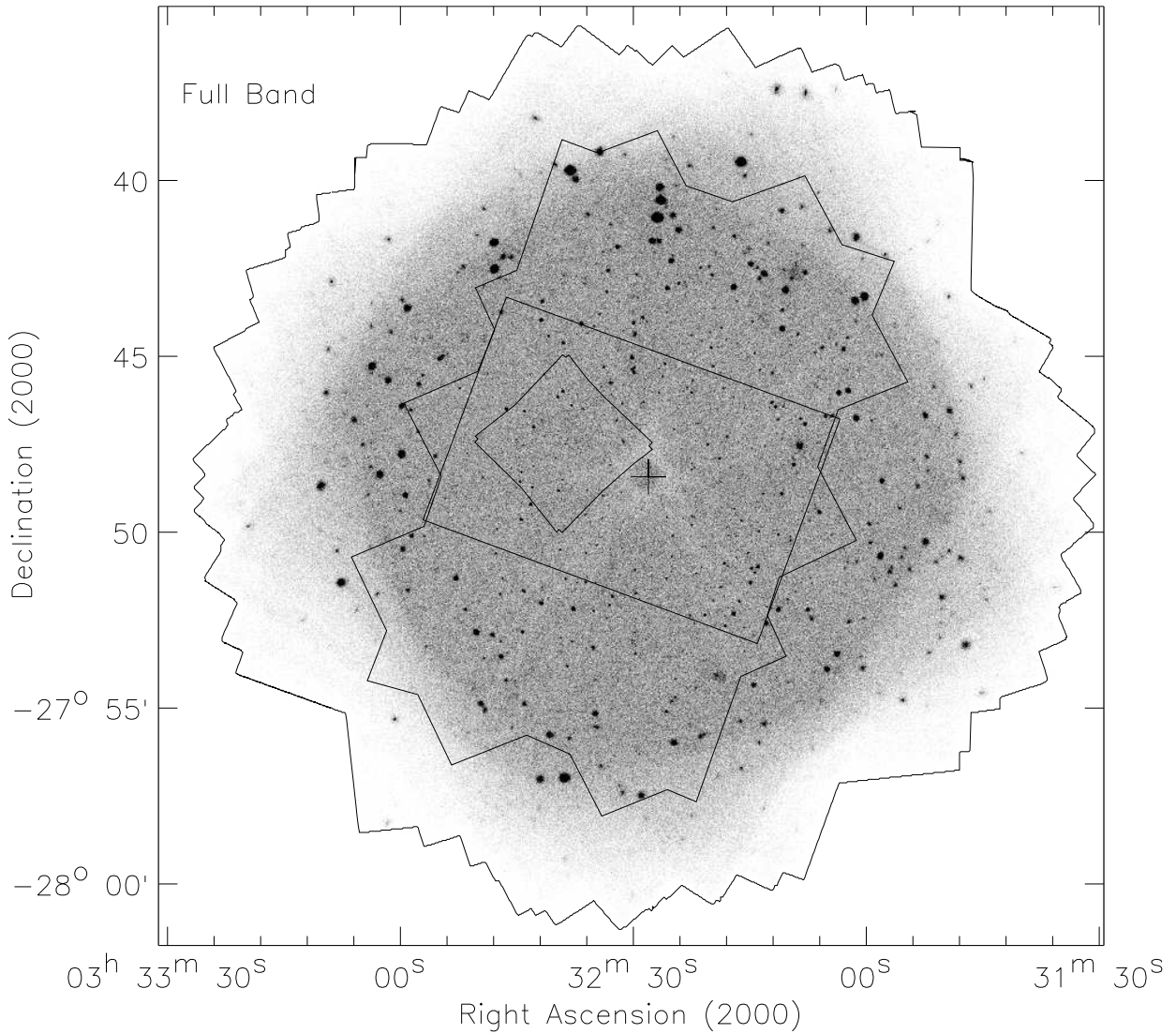}
}
\figcaption{Full-band (0.5--8.0~keV) raw image of the 4~Ms \hbox{CDF-S} displayed with linear gray scales.
The segmented boundary surrounding the image shows the coverage of the entire \hbox{CDF-S}.
The large polygon, the rectangle, and the central small polygon indicate
the regions for the GOODS-S (Giavalisco et~al. 2004), the planned CANDELS GOODS-S (5-orbit {\it HST}/WFC3; see \S~\ref{sec:summary} for more details about CANDELS), and
the {\it Hubble} Ultra Deep Field (UDF; Beckwith et~al. 2006), respectively. 
The central plus sign indicates the average aim point, weighted by exposure time (see
Table~\ref{tbl-obs}).
The pale ring-like area near the field center is caused by 
the ACIS-I CCD gaps in which the effective exposures are lower
than in the nearby non-gap areas (see Fig.~\ref{fbemap}).
The apparent scarcity of sources near the field center is mainly due
to the small size of the on-axis PSF (see Figs.~\ref{clrimg} and \ref{pos} for clarification).
\label{fbimg}}
\end{figure*}

\begin{figure}
\centerline{
\includegraphics[scale=0.40]{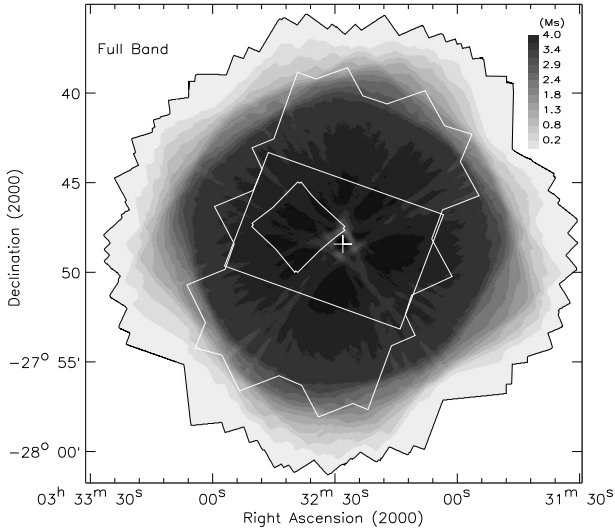}%{figs/fb-emap.eps}
}
\figcaption{
Full-band (0.5--8.0~keV) effective-exposure map of
the 4~Ms \hbox{CDF-S} displayed with linear gray scales that are indicated by the inset scale bar (effective exposure times are in units of seconds).
The darkest areas represent the highest effective exposure times, with a maximum of 3.811~Ms.
The distributions of the \hbox{ACIS-I} CCD gaps can be clearly identified (indicated by the radial trails).
The regions and the plus sign are the same as those in Fig.~\ref{fbimg}.
\label{fbemap}}
\end{figure}

\begin{figure}
\centerline{
\includegraphics[scale=0.5]{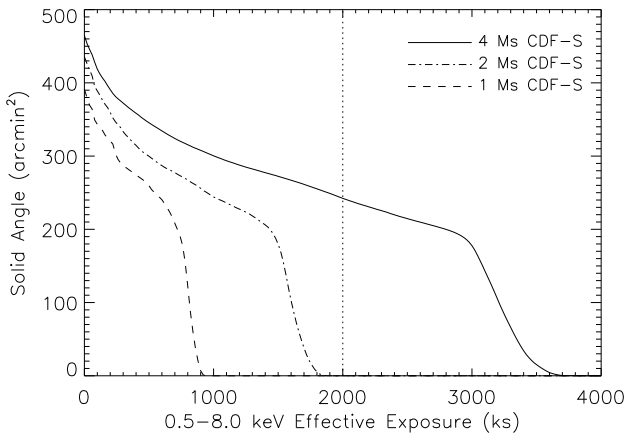}
}
\figcaption{
Plot of survey solid angle as a function of minimum full-band effective exposure for the 4~Ms \hbox{CDF-S}
({\it solid curve}).
The maximum exposure is 3.811~Ms. The vertical dotted
line indicates an effective exposure of 2~Ms. 
Approximately 242.3 arcmin$^2$
($\approx 52\%$) of the \hbox{CDF-S} survey area has $>2$~Ms effective exposure. 
For comparison, %the $\approx2$~Ms \hbox{CDF-N} (A03; {\it dashed curve}) and 
the 1~Ms \hbox{CDF-S} result ({\it dashed curve}) and the 2~Ms \hbox{CDF-S} result 
({\it dash-dot curve}), both of which are obtained using the procedures in this paper,
are also shown in the plot.
\label{emapcum}}
\end{figure}

We utilized the {\sc ciao} tool {\sc dmmerge} to produce a merged event file by combining the individual event files. 
We constructed images from this merged event file using the standard
\asca\ grade set (\asca\ grades 0, 2, 3, 4, 6) for three standard bands:
\hbox{0.5--8.0~keV} (full band; FB), \hbox{0.5--2.0~keV} (soft band; SB), and \hbox{2--8~keV} (hard band; HB).\footnote{We compared \xray\ source catalogs made with the two
upper energy cuts of \hbox{7 keV} and \hbox{8 keV}
(i.e., the set of energy bands of \hbox{0.5--7.0},
\hbox{0.5--2.0}, and \hbox{2--7~keV} versus the set of energy bands adopted here;
see \S~\ref{sec:list} and \S~\ref{sec:srcselect} for the details of catalog production).
We found no clear statistical difference between catalogs;
the \xray\ sources that are unique in each catalog
are faint (i.e., close to or right on source-detection limits) and account
for only $\approx 3\%$ of all detected sources.
We thus adopted the traditional standard 
bands (i.e., using the upper energy cut of \hbox{8 keV}) to maintain continuity with 
past catalogs (e.g., A03; L05; L08).}
Figure~\ref{fbimg} shows the raw full-band image.
We generated effective-exposure maps for the three standard bands following the basic procedure
outlined in \S3.2 of Hornschemeier et~al. (2001) 
and normalized them to the effective exposures of a pixel located at the average aim point.
This procedure takes into account the effects of vignetting, gaps between the CCDs,
bad-column filtering, bad-pixel filtering, and the spatial and time
dependent degradation in quantum efficiency
due to contamination on the ACIS optical-blocking filters;
thus, the derived effective exposures are typically smaller than the nominal exposures (i.e., durations of observations).
When creating the effective-exposure maps, we assumed a photon index of
$\Gamma=1.4$, the slope of the cosmic \hbox{2--10~keV} \mbox{X-ray} background
(e.g., Marshall et~al. 1980; Gendreau et~al. 1995; Hasinger et~al. 1998; Hickox \& Markevitch 2006).
Figure~\ref{fbemap} shows the full-band effective-exposure map, and
Figure~\ref{emapcum} displays the survey solid angle as a
function of the minimum full-band effective exposure. 
According to Fig.~\ref{emapcum}, about 52\% and 38\% of the
\hbox{CDF-S} field has a full-band effective exposure
greater than 2~Ms and 3~Ms, respectively; the maximum effective exposure is 3.811~Ms,
which is slightly smaller than the 3.872~Ms total exposure since the
locations of the aim points of individual observations vary.
For a given full-band effective exposure,
the survey solid angle is up to a factor of $\approx 1.5$ times larger than that of the
%$\approx2$~Ms \hbox{CDF-N} (A03; Fig.~\ref{emapcum}, {\it dashed curve})
2~Ms \hbox{CDF-S} (L08; Fig.~\ref{emapcum}, {\it dash-dot curve}) 
at the low end of effective exposure (\mbox{$<1.5$ Ms}),
and it is much larger 
than that of %the $\approx2$~Ms \hbox{CDF-N} and 
the 2~Ms \hbox{CDF-S} above \mbox{1.5 Ms} effective exposure.
Thus in addition to the fact that the 4~Ms \hbox{CDF-S} can detect
new sources that have lower fluxes than the 2~Ms sources, it can also detect
new sources that have a similar flux distribution to the 2~Ms sources
over as much as 50\% more area.

\begin{figure*}
\centerline{
\includegraphics[scale=1.10]{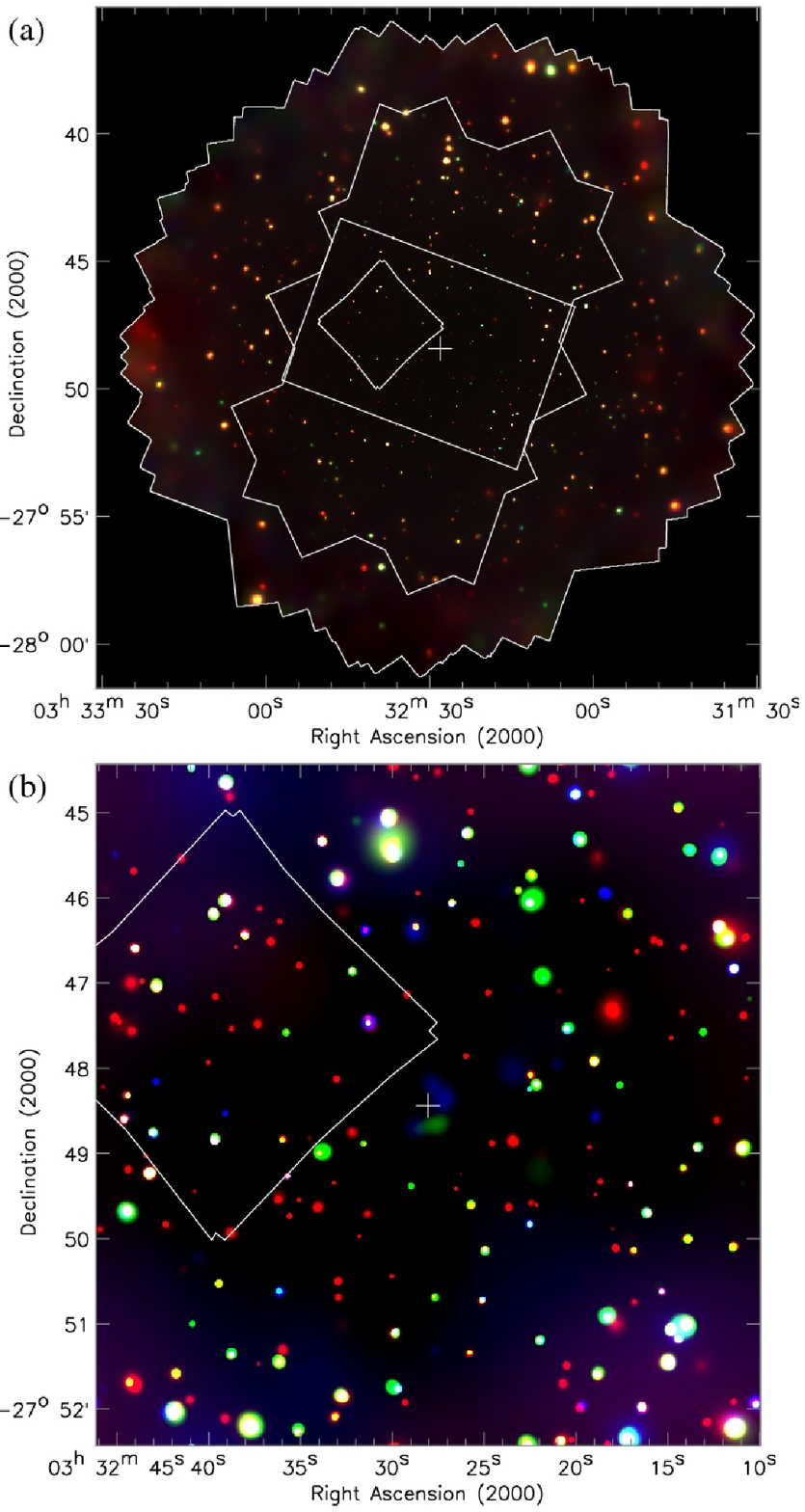}
}
\figcaption{
(a) \chandra\ ``false-color'' image of the 4~Ms \hbox{CDF-S}, which
is a color composite of the  exposure-corrected and adaptively smoothed images
in the 0.5--2.0 keV ({\it red}), 2--4 keV ({\it green}), and 4--8
keV ({\it blue}) bands. (b) An expanded view of the \chandra\ ``false-color'' image of the
central $8\arcmin \times 8\arcmin$ region (note that a slightly different
contrast ratio from that for the full image is used here in order to render the faint sources
more clearly).
The apparent smaller size and lower brightness of sources near the
field center is due to the smaller size of the on-axis PSF.
The regions and the plus sign are the same as those in Fig.~\ref{fbimg}.
\label{clrimg}}
\end{figure*}

We followed \S~3.3 of Baganoff et~al. (2003) to construct exposure-corrected smoothed images.
We first produced the raw images and effective-exposure maps in the \hbox{0.5--2.0~keV}, 
\hbox{2--4~keV}, and \hbox{4--8~keV} bands, using the aforementioned procedures.
We then adaptively smoothed the raw images and effective-exposure maps using the {\sc ciao} tool {\sc csmooth} (Ebeling, White, \& Rangarajan 2006).
Finally, we divided the smoothed images by their corresponding smoothed effective-exposure maps and combined the exposure-corrected smoothed images together to produce
a full-band color composite, as shown in Fig.~\ref{clrimg}
(note that this color composite is not background-subtracted); an expanded view of the central $8\arcmin\times 8\arcmin$ region is also shown in Fig.~\ref{clrimg}.
Note that we ran {\sc wavdetect} only on the raw images for source searching,
although many detected \hbox{X-ray} sources appear more clearly in the adaptively smoothed images.

\subsection{Candidate-List Catalog Production}\label{sec:list}

We ran {\sc wavdetect} on each combined raw image in the three standard bands\footnote{We note that
{\sc wavdetect} was run on the {\it combined} raw images
where the average aim point (given in \S~\ref{sec:obs}) is a good approximation of
the image center for the purpose of computing PSFs.
Given that we used multiple wavelet scales, the Mexican-Hat wavelet patterns (adopted
by {\sc wavdetect}) provide reasonable first-order approximations of the 
multi-observation PSFs.}
to perform source searching
and to construct a candidate-list catalog,
using a ``$\sqrt{2}$~sequence'' of wavelet scales (i.e.,\ 1, $\sqrt{2}$, 2,
$2\sqrt{2}$, 4, $4\sqrt{2}$, 8, $8\sqrt{2}$, and 16 pixels) and a false-positive probability threshold of $10^{-5}$.
We expect the use of a false-positive probability threshold of $10^{-5}$ to introduce
a non-negligible number of spurious sources that have \mbox{$\lsim 2$--3} source counts.
However, as pointed out by Alexander et~al. (2001), using a more stringent
source-detection threshold (e.g., $10^{-6}$, $10^{-7}$, or $10^{-8}$)
can lose an appreciable number of real sources.
In \S~\ref{main}, we create a more conservative main catalog by determining the
detection significances of each candidate-list source in the three standard
bands and discarding sources with significances below an adopted threshold value.

Our candidate-list catalog consists of 892 \mbox{X-ray} source candidates;
each candidate was detected in at least one of the three standard bands with
{\sc wavdetect} at a false-positive probability threshold of $10^{-5}$.
We adopted, in order of priority, full-band, soft-band, or hard-band source positions for candidate sources.
We performed cross-band matching using a $2.5\arcsec$ matching radius for 
sources within $6\arcmin$ of the average aim point (i.e., off-axis angle $\theta<6\arcmin$)
and a $4.0\arcsec$ matching radius for sources located at larger off-axis angles (i.e., $\theta \ge 6\arcmin$). 
The choice of these matching radii was made based on inspection of histograms
that show the number of matches as a function of angular separation
(e.g., see \S2 of Boller et~al. 1998).
With these matching radii, the mismatch probability is $\approx 1$\% over the entire field.
We removed a few duplicate sources due to false matches near the edge of the field through visual inspection.\footnote{For a few sources that
lie near the edge of the field,
the offset between the \xray\ positions determined from different bands by {\sc wavdetect} is $>4\arcsec$;
such a source will be counted twice (i.e., treated as two sources) 
according to our matching approach
(i.e., a $4.0\arcsec$ matching radius at $\theta \ge 6\arcmin$).
We removed the duplicated sources in these few cases.}

We improved the above {\sc wavdetect} source positions utilizing the centroid and matched-filter positions computed by AE.
The matched-filter positions are obtained by correlating the full-band image
in the neighborhood around each source with the source's combined PSF.
The combined PSF is generated by combining the individual PSFs of a
source for each relevant observation, weighted by the number of detected counts. 
This technique takes into account the fact that, due to the complex PSF at large off-axis
angles, the \mbox{X-ray} source position is not always located at the peak of the \mbox{X-ray} emission. 
The {\sc wavdetect}, centroid, and matched-filter positions
have comparable accuracy on-axis, while the matched-filter positions
have better accuracy off-axis.
Thus, we adopted centroid positions for sources with $\theta <8\arcmin$
and matched-filter positions for sources with $\theta \ge 8\arcmin$.

We refined the absolute astrometry of the raw \hbox{X-ray} images
by matching the candidate-list sources to the $5\sigma$ VLA 1.4~GHz
radio-catalog sources (see \S~\ref{sec:img}).
There are 359 radio sources across the \hbox{CDF-S} field with positions 
accurate to $\lsim 0.1\arcsec$.
We performed cross-matching between the 892 candidate-list catalog \mbox{X-ray} sources and the
359 radio sources in the field using a $2\arcsec$ matching radius and found 141 matches.
We estimated the expected false matches by
manually shifting the \hbox{X-ray} source positions in right
ascension and declination by \hbox{$\pm$(5--60$\arcsec)$} in steps of $5\arcsec$
(i.e., in unique directions) and recorrelating with the radio sources. 
The average number of false matches is $\approx 2.3$ ($\approx 1.7\%$)
and the median offset of these false matches is $1.41\arcsec$. 
Of the 141 matches, we identified five extended radio sources
upon inspecting the radio image.
We excluded two of these five extended radio sources
for the astrometry refinement analysis because these two matches are spurious
with positional offsets greater than $1.5\arcsec$
(see \S~\ref{sec:dpos} for more details on these two extended radio sources);
the other three matches are robust with small positional offsets ($<0.7\arcsec$) and were
included for the subsequent analysis.
Using these 139 matches, we found small shift and plate-scale corrections when comparing the \mbox{X-ray}
and radio source positions and applied these corrections to all 
the combined \hbox{X-ray} images and source positions, which results in
small ($<0.2\arcsec$) astrometric shifts.

We utilized AE to perform photometry for the candidate-list catalog sources.
Compared to ``traditional'' circular-aperture photometry (e.g., L08),
the most important difference in the AE-computed photometry is the use of
polygonal source-extraction regions.\footnote{The polygonal 
source-extraction regions typically become more non-circular 
toward larger off-axis angles. In particular, the source-extraction regions for
crowded sources at large off-axis angles are reduced from \hbox{$\approx 90$\%}
to \hbox{$\approx 40$--75\%} encircled-energy fractions (EEFs) and thus 
represent the most dramatic examples of deviation from circular apertures (see, e.g., Fig.~6 of 
Broos et~al. 2010 for such an example).}
AE models the \chandra\ High Resolution Mirror Assembly (HRMA) using
the MARX\footnote{MARX is available at http://space.mit.edu/CXC/MARX/index.html.\label{marx}}
ray-tracing simulator (version 4.4.0) to obtain the PSF model.
It then constructs a polygonal extraction region that approximates the
$\approx 90$\% encircled-energy fraction (EEF) contour of a local PSF measured at 1.497 keV
(note that AE also constructs PSFs at energies of 0.277, 4.510, 6.400, and 8.600~keV).
When dealing with crowded sources having overlapping polygonal extraction regions,
AE utilized smaller extraction regions (corresponding to \mbox{$\approx$40--75\%} EEFs) 
that were chosen to be as large as possible without overlapping.
Less than 6\% of the 892 candidate-list sources are crowded by this definition.
For background extraction, we adopted the AE ``BETTER\_BACKGROUNDS'' algorithm.
This algorithm models the spatial distributions of flux for the source of interest and
its neighboring sources using unmasked data.
It then computes local background counts 
within background regions that subtract 
contributions from the source and its neighboring sources.
In our AE usage, the background-extraction region
is typically a factor of $\approx 16$
larger than the source-extraction region
and contains at least 100 background counts.
As discussed in \S~7.15 of the AE manual,
AE also imposes an explicit requirement that the uncertainty in the estimate of net counts 
be dominated by the uncertainty in the extracted source counts in order to ensure 
photometric accuracy; this requirement leads to enlargement of 
background regions/counts when necessary.
As a result, the median number of full-band background counts extracted for the main-catalog sources
(see \S~\ref{sec:maincat}) is 780, with an interquartile range of \hbox{278--2621}.
This algorithm produces accurate background extractions, 
which are particularly critical for crowded sources.
For sources that are not crowded, this algorithm produces essentially the same
background-extraction results as the traditional AE ``EXTRACT\_BACKGROUNDS''
algorithm; the latter algorithm computes local background counts 
by masking all the sources and then searching around each source 
for the smallest circular region that contains a desired number of background counts.
AE analyzes individual observations independently (including, e.g., the use of MARX for PSF modeling and source and background extractions) and merges the data to produce photometry for each source.\footnote{For this work, we did not use the optional AE ``MERGE\_FOR\_PHOTOMETRY'' algorithm, as discussed in Broos et~al. (2010), that allows AE to discard some extractions during a merge of AE products from individual observations.}
The resulting combined PSFs at 1.497 keV have typical FWHMs of
0.68\arcsec, 1.07\arcsec, 1.76\arcsec, 2.79\arcsec, and 3.61\arcsec\
at off-axis angles of 1\arcmin, 3\arcmin, 5\arcmin, 7\arcmin, and 9\arcmin,
respectively; these FWHM values represent typical angular resolutions of the 4~Ms images.

AE estimates an energy-dependent aperture correction for each source and applies the correction to the effective area calibration file used for spectral modeling.
For this work, we chose to apply aperture corrections to the background-subtracted photometry as follows.
For the soft (hard) band, we derived an effective PSF fraction for each source 
by weighting PSF measurements at 1.497 (4.510)~keV
by the exposures for the individual observations.
Given that the full band
is a combination of the soft and hard bands,
we derived the full-band effective PSF fraction 
based on the derived soft-band and hard-band effective PSF fractions:
(1) if a source was detected both in the soft and hard bands, we derived the full-band effective PSF fraction
by weighting the soft and hard-band effective PSF fractions with the soft and hard-band background-subtracted counts;
(2) if a source was detected in the soft or hard band (but not both), 
we set the full-band effective PSF
fraction to the soft or hard band effective PSF fraction, respectively; and
(3) if a source was detected in neither the soft band nor the hard band,
we took the average of the soft and hard band effective PSF fractions as the full-band
effective PSF fraction.
The median aperture corrections for the full, soft, and hard bands are 0.875,
0.898, and 0.826, respectively.
We then applied aperture corrections by dividing the background-subtracted source counts
by the derived effective PSF fractions.
Since our candidate-list catalog was constructed using {\sc wavdetect} with a liberal false-positive probability
threshold of $10^{-5}$, many candidate sources have \mbox{$\lsim 2$--3} (background-subtracted) source counts.
In the next section, we evaluate the reliability of candidate sources on a source-by-source basis
to produce a more robust main source catalog.

\section{Main Chandra Source Catalog}\label{main}

\subsection{Selection of Main-Catalog Sources}\label{sec:srcselect}

As discussed above, we expect our candidate-list catalog of 892 \mbox{X-ray} 
sources to include a significant number of false sources since we ran
{\sc wavdetect} at a liberal false-positive probability threshold of $10^{-5}$. 
If we conservatively treat the three standard-band images as independent, 
we can estimate the number of expected false sources in the candidate-list catalog 
for the case of a uniform background by multiplying the
{\sc wavdetect} threshold of $10^{-5}$ by the sum of pixels in the three bands
(i.e., \mbox{$\approx 2.07 \times 10^7$}).  
However, such a false-source estimate is conservative, since 
over the majority of the field, a single pixel will 
not be considered a source-detection cell. In particular, at large off-axis angles
{\sc wavdetect} suppresses fluctuations on scales smaller than the PSF. 
As quantified in \S~3.4.1 of A03, 
the number of false-sources is likely \hbox{$\approx$2--3}
times smaller than the above conservative estimate.
We refer readers to \S~\ref{sec:comprel} for relevant discussions.

To produce a more reliable main \chandra\ source catalog, 
we evaluated for each source the binomial probability $P$ that no source exists 
given the measurements of the source and local background.
As discussed in \S~5.10.3 of the AE manual
(also see Appendix~A2 of Weisskopf et~al. 2007 for further details),
the binomial no-source probability $P$ can be calculated using
the following equation:
\begin{equation}
P(X\ge S)=\sum_{X=S}^N \frac{N!}{X!(N-X)!} p^X (1-p)^{N-X}.
%P = binomial \bigg(S, S+B_{\rm ext}, \frac{B_{\rm src}}{B_{\rm ext}+B_{\rm src}}\bigg)
\label{equ:bi}
\end{equation}
\noindent In this equation, $S$
is the total number of counts in the source-extraction region without subtraction of the background
counts $B_{\rm src}$ in this region;
$N=S+B_{\rm ext}$, where $B_{\rm ext}$ is the total extracted background counts
within a background-extraction region that is 
typically a factor of $\approx 16$
larger than the source-extraction region in our AE usage (see \S~\ref{sec:list}); and
$p=1/(1+BACKSCAL)$ is the probability that a photon
lies in the source-extraction region (thus contributing to $S$), where
$BACKSCAL=B_{\rm ext}/B_{\rm src}$ with a typical value of $\approx 16$, as stated earlier.
$P$ is computed by AE in each of the three standard bands.
For a source to be included in our main catalog, we
required $P < 0.004$ in at least one of the three standard bands.
We identified multiwavelength counterparts for the \hbox{X-ray} sources
(see \S~\ref{sec:id})
and studied the identification rate as a function of the $P$ value, given
that \hbox{X-ray} sources without identifications in ultradeep multiwavelength data are more
likely to be false detections (see, e.g., L10).
The requirement of $P < 0.004$ was empirically chosen as a compromise
to keep the fraction of potential false sources small while
recovering the largest number of real sources.
Using this criterion of $P<0.004$, our main catalog contains a total of 740 sources.
We note that for a different choice of source-detection criterion of $P<0.01$, 
a total of 33 additional sources with $0.004\le P<0.01$ would be included; however, 
only $\approx 64\%$ (i.e., 21) of these 33 sources have multiwavelength counterparts,
as opposed to an identification rate of $\approx 97\%$ for the main catalog (see \S~\ref{sec:id}).
We refer readers to \S~\ref{sec:comprel} for a detailed discussion on the completeness and 
reliability of the main catalog based on simulations.

Our adopted cataloging procedure,
with the utilization of AE,
has a number of
advantages over a ``traditional'' {\sc wavdetect}-only approach:
(1) the more detailed treatment of complex source-extraction regions (i.e., using
polygonal regions, as opposed to elliptical apertures, to simulate the PSF) that is more suitable for the case of multiple observations 
with different aim points and roll angles, 
(2) the better source-position determination that maximizes the signal-to-noise ratio
and leads to more accurate count estimates, 
(3) the more careful background estimates that take into account the effects of
all the neighboring sources and CCD gaps, and
(4) the more immediately transparent mathematical criterion (i.e.,
the binomial probability) that is utilized for source detection.
We will demonstrate below that our adopted procedure
recovers almost all of the sources detected with {\sc wavdetect} at a
false-positive probability threshold of $10^{-6}$ and a significant number of
additional real sources detected at $10^{-5}$.

In order to give a more detailed {\sc wavdetect}-based perspective on source
significance, we also ran {\sc wavdetect} on the three standard-band images
at false-positive probability thresholds of $10^{-6}$, $10^{-7}$, and $10^{-8}$,
and found detections for 659 (73.9\%), 569
(63.8\%), and 502 (56.3\%) of the 892
candidate-list catalog sources, respectively.
Among the 152 candidate-list sources that failed the selection cut of $P<0.004$, 
and thus were not included in the main catalog, 
40 ($\approx 4.5$\% of the 892 candidate-list sources) had {\sc wavdetect} false-positive 
probability detection thresholds of $\le 10^{-6}$.
Meanwhile, our main catalog includes 121 sources that had minimum {\sc wavdetect} probabilities of $10^{-5}$.\footnote{The minimum {\sc wavdetect} probability 
represents the {\sc wavdetect} significance of a source, with lower values
indicating higher significances.
For example, if a source was detected with {\sc wavdetect} in at least one of the
three standard bands at a
false-positive probability threshold of $10^{-7}$ but was not 
detected in any of the three standard bands at a threshold of $10^{-8}$, then the minimum {\sc wavdetect} probability
of this source is $10^{-7}$.\label{siglev}}
Therefore, our adopted procedure, as opposed to a direct {\sc wavdetect}-based approach,
has a ``net gain'' of 81 sources.
We note that a larger net gain of sources could be achieved if we adopted 
a less conservative no-source probability cut (e.g., $P<0.01$) at the expense 
of introducing more spurious sources.

\begin{figure*}
\centerline{
\includegraphics[scale=0.9]{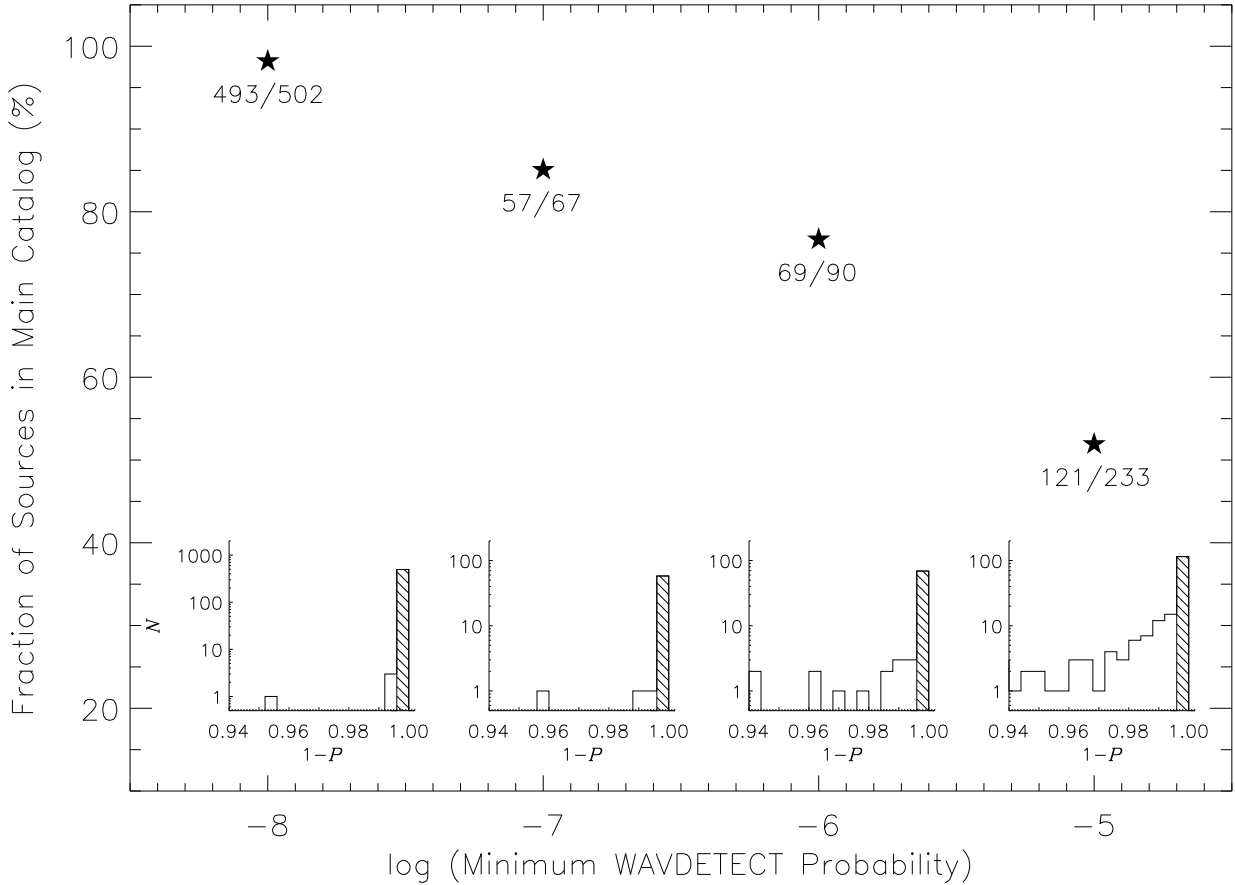}
}
\figcaption{The fraction of sources in the candidate-list catalog with an AE 
binomial no-source probability $P<0.004$, which were included in 
the main catalog, as a function of minimum {\sc wavdetect} probability$^{\ref{siglev}}$
(shown as five-pointed stars). 
The number of sources with $P<0.004$ versus the number of 
candidate-list catalog 
sources detected at each minimum {\sc wavdetect} probability are
annotated in the figure (note that, in this figure,
502+67+90+233=892 and 493+57+69+121=740).
The fraction of candidate-list catalog sources included in the
main catalog falls from 98.2\% to 51.9\% 
between minimum {\sc wavdetect} probabilities of $10^{-8}$ 
and $10^{-5}$.
Shown in the insets are the histograms of $1-P$ for the 
candidate-list catalog sources at each minimum {\sc wavdetect} 
probability, with shaded areas highlighting those included in
the main catalog (i.e., having $1-P>0.996$).\label{psig}}
\end{figure*}

Figure~\ref{psig} shows the fraction of candidate-list sources included in the main catalog 
and the $1-P$ distribution of candidate-list sources as a function of the minimum {\sc wavdetect} probability.
The fraction of candidate-list sources included in the main catalog is
98.2\%, 85.1\%, 76.7\%, and 51.9\%
for a minimum {\sc wavdetect} probability of $10^{-8}$, $10^{-7}$, $10^{-6}$,
and $10^{-5}$, respectively.
As shown later in \S~\ref{sec:id}, 
we find that 716 (96.8\%) of the 740 main-catalog sources have secure multiwavelength counterparts (with a false-matching probability of $\approx 2.1\%$),
where the identification rate is 98.1\% (90.1\%) for the 
619 (121) sources with a minimum {\sc wavdetect} probability of $\le 10^{-6}$
($10^{-5}$) in the main catalog.
Given the relatively small false-matching rate,
the above high identification rates indicate that the vast majority of the
main-catalog sources are real \mbox{X-ray} sources (see, e.g., L10).
Thus, our main-catalog selection provides an effective identification of
real \mbox{X-ray} sources including those falling below the traditional $10^{-6}$ 
{\sc wavdetect} searching threshold.

\subsection{X-ray Source Positional Uncertainty}\label{sec:dpos}

\begin{figure}
\centerline{
\includegraphics[scale=0.6]{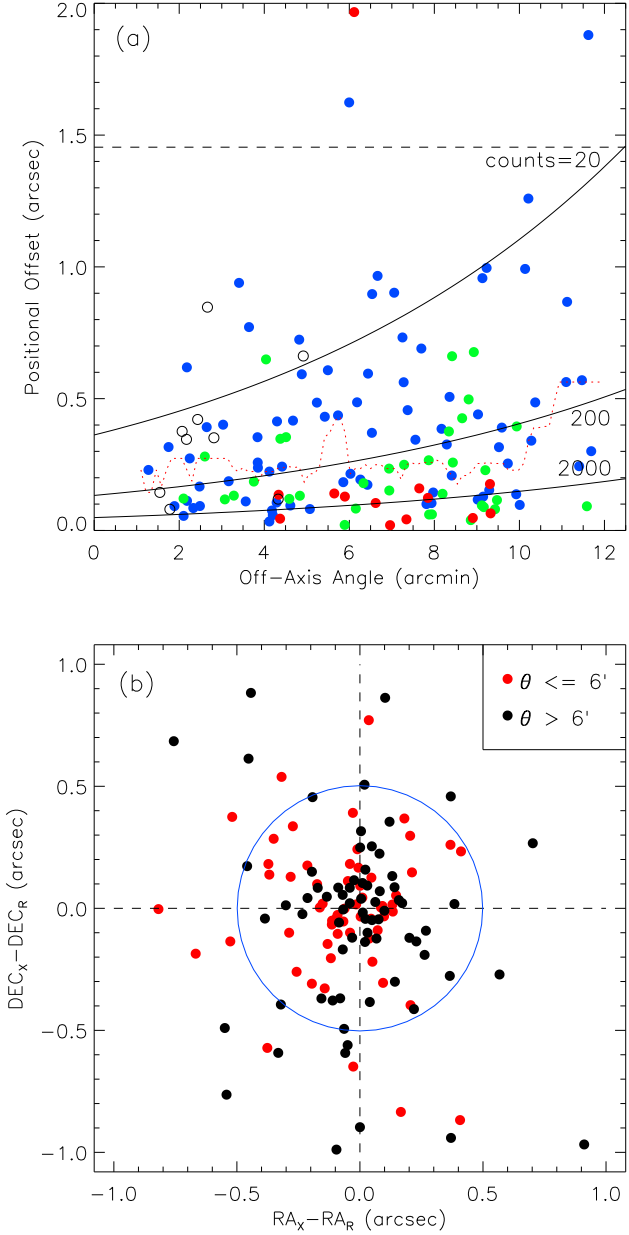}
}
\figcaption{
(a) Positional offset vs. off-axis angle for
the 135 main-catalog sources that have counterparts in the
$5\sigma$ VLA 1.4~GHz radio catalog using a matching radius of $2\arcsec$
(see \S~\ref{sec:dpos} for descriptions of the three sources with $>1.5\arcsec$ positional offsets).
Red filled, green filled, blue filled, and black open circles represent
\mbox{X-ray} sources with $\ge2000$, $\ge200$, $\ge20$, and $<20$
counts in the energy band where the source position was determined, respectively.
The red dotted curve shows the running median of positional offset in
bins of $2\arcmin$. 
The horizontal dashed line indicates the median offset ($1.45\arcsec$) 
of the expected false matches.
We used these data to derive the $\approx68\%$ confidence-level \mbox{X-ray} source
positional uncertainties, i.e., eq.~(\ref{equ:dpos}).
Three solid curves indicate the $\approx68\%$ confidence-level positional
uncertainties for sources with 20, 200 and 2000 counts.
(b) Positional residuals between the \xray\ and radio positions for the 132 
main-catalog sources that have radio counterparts within a radius of $1.5\arcsec$
[see Panel (a)]. Red and black filled circles indicate sources with an off-axis angle
of $\le 6\arcmin$ and $>6\arcmin$, respectively.
A large blue circle with a radius of $0.5\arcsec$ is drawn at the center
as a guide to the eyes.
[{\it see the electronic edition of the Supplement for a color version of this figure.}]
\label{dpos}}
\end{figure}

\begin{figure}
\centerline{
\includegraphics[scale=0.5]{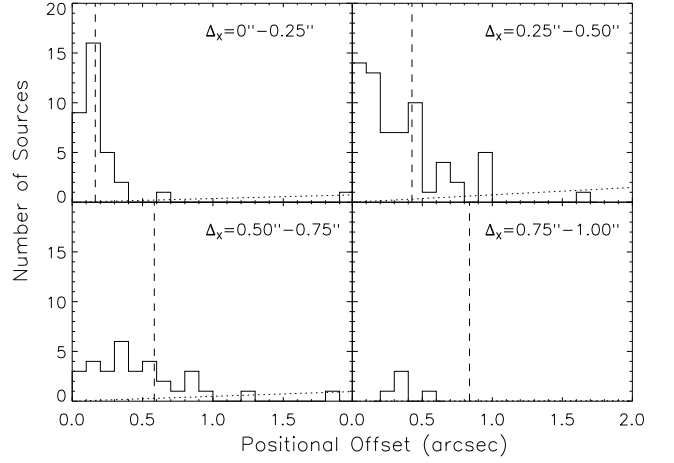}
}
\figcaption{
Histograms showing the distributions of positional offset for the 135 main-catalog
sources that have counterparts in the $5\sigma$ VLA 1.4~GHz radio catalog 
using a matching radius of $2\arcsec$.
These 135 sources were divided into four bins according to their positional 
uncertainties estimated using eq.~(\ref{equ:dpos}): 
$0\arcsec$--$0.25\arcsec$, $0.25\arcsec$--$0.50\arcsec$,
$0.50\arcsec$--$0.75\arcsec$, and $0.75\arcsec$--$1\arcsec$.
The vertical dashed line in each panel indicates the median \mbox{X-ray} positional uncertainty in each bin.
The dotted line shows the total expected number of random radio sources
as a function of the positional offset.
$\lsim 35\%$ of the radio counterparts lie beyond the median \mbox{X-ray} positional 
uncertainty in each bin.
\label{poshist}}
\end{figure}

As in \S~\ref{sec:list},
we cross matched the 740 main-catalog sources with the 359 radio sources in the field 
using a $2\arcsec$ matching radius and found 135 matches.\footnote{We note that 6 (i.e., $141-135=6$; also see \S~\ref{sec:list}) candidate-list \xray\ sources that
have a radio counterpart were not included in the main catalog;
these sources are likely real \xray\ sources that fail to satisfy our relatively
stringent source-selection criterion of $P<0.004$ (see \S~\ref{sec:srcselect}).}
We estimated on average $\approx 2.0$ ($\approx 1.5\%$) false matches and
a median offset of $1.45\arcsec$ for these false matches.
Figure~\ref{dpos}(a) shows the positional offset between
the \hbox{X-ray} sources and their radio counterparts as a function of off-axis angle.
The median positional offset is $0.24\arcsec$.
There are three sources in Fig.~\ref{dpos}(a) that have
positional offsets greater than $1.5\arcsec$:
(1) the one with the largest offset ($1.97\arcsec$) mistakenly matches to
one of the two lobes of a radio galaxy due to the fact that the radio core,
which is likely the real counterpart for this \mbox{X-ray} source,
was not detected in the radio catalog;
(2) the one with the second largest offset ($1.88\arcsec$)
is likely a false match because such an offset is much larger than
its expected positional uncertainty [see eq.~(\ref{equ:dpos}) below] considering its
off-axis angle ($11.6\arcmin$) and source-counts ($\approx 100$); and
(3) the one with the third largest offset ($1.62\arcsec$) has one radio source,
which is the core of a radio galaxy,
and a few optical sources within its $2\arcsec$ radius, with the radio counterpart
not matching to the likely real optical counterpart of this \mbox{X-ray} source 
(thus being a false match).
Excluding the above three sources, we then estimated \hbox{X-ray} positional
uncertainties using the remaining 132 \hbox{X-ray} 
detected radio sources.
Figure~\ref{dpos}(b) shows the positional residuals between the \xray\
and radio positions for these 132 sources; the ``scatter cloud'' of positional residual
appears circular, with no residual distortions.
As shown in Fig.~\ref{dpos}(a), there are clear off-axis angle and source-count dependencies for these 132 sources,
with the former due to the degradation of the \chandra\ PSF
at large off-axis angles
and the latter due to statistical limitations in finding the centroid of a faint 
\hbox{X-ray} source.
Implementing the parametrization provided by Kim et~al. (2007),\footnote{We note
that the Kim et~al. (2007) parametrization fits our data adequately 
(i.e., the AE-derived positions and photometry), although it was originally 
based on {\sc wavdetect}-derived positions and photometry.}
we derived an empirical relation for the positional uncertainty of our
\mbox{X-ray} sources by fitting to these 132 \mbox{X-ray} sources that
have radio counterparts within a radius of $1.5\arcsec$. The relation is
\begin{equation}
\log \Delta_{\rm X}=0.0484 \theta-0.4356\log C+0.1258,\label{equ:dpos}
\end{equation}
\noindent
where $\Delta_{\rm X}$ is the \mbox{X-ray} positional uncertainty in arcseconds, 
$\theta$ is the off-axis angle in arcminutes, and $C$ is the source counts in the energy
band where the source position was determined
(see the description of Columns~\hbox{8--16} of the main catalog in \S~\ref{sec:maincat} for details on photometry calculation).
We set an upper limit of 2000 on $C$ since the positional accuracy
does not improve significantly above that level.
As a guide to the derived relation,
we show positional uncertainties for $C=20$, 200, and 2000 in Fig.~\ref{dpos}.
The stated positional uncertainties are for the
$\approx 68\%$ confidence level, which are smaller than the {\sc wavdetect}
positional uncertainties, particularly at large off-axis angles, due to 
our adopted positional refinement.
In Figure~\ref{poshist}, we show the distributions of positional
offset in four bins of X-ray positional uncertainty, as well
as the expected false matches assuming a uniform spatial
distribution of radio sources.
For each histogram in Fig.~\ref{poshist}, as expected,
$\gsim 65\%$ of the positional offsets between the \xray\ sources and their radio counterparts
are less than the corresponding median \xray\ positional uncertainty.

Owing to the factor of $\approx 2$ increase in exposure/source counts from 2~Ms to 4~Ms,
the areas of source positional error regions are expected to be reduced
by $\approx 30\%$ on average (see \S~4.2 of L10).
We thus compared our positional uncertainties with the positional uncertainties
for the 440 main-catalog sources that
were previously detected in the L08 main catalog (see the description of Column~59
of the main catalog in \S~\ref{sec:maincat}).
We find a median ratio of 0.82 between our and the L08
positional uncertainties\footnote{In L08, the \xray\ positional uncertainties
are quoted at the $\approx 85$\% confidence level. 
For straightforward comparison,
we thus adopted the $\approx 68$\% confidence-level positional uncertainties
reported in Table~2 of L10 that were used in the L10 
likelihood-ratio matching procedure (see \S~\ref{sec:id} for more details).} (corresponding to a median ratio of 0.67 between 
areas of our and the L08 positional error regions); 
such an improvement is in agreement with the above expectation.
We also cross matched the 462 L08 main-catalog sources with the 359
radio sources in the field using a matching radius of $2\arcsec$, taking
into account the systematic positional offsets between the optical
catalogs and the VLA radio catalog (see \S~\ref{sec:img}).
The median positional offset is $0.40\arcsec$ between the L08
main-catalog sources and their radio counterparts for a total of 94 matches,
as opposed to $0.24\arcsec$ in our case.
This significant improvement is not only because of the improved photon statistics,
but also because we locked the astrometry of the combined \hbox{X-ray} images to the VLA radio 
sources rather than the WFI $R$-band sources that were adopted by L08.

\subsection{Multiwavelength Identifications}\label{sec:id}

We utilized the likelihood-ratio matching procedure presented in \S~2 of L10 
to identify the optical/near-infrared/infrared/radio (ONIR) counterparts 
for the main-catalog \xray\ sources.
Briefly, the likelihood-ratio 
technique (e.g., Sutherland \& Saunders 1992; Ciliegi et~al. 2003; Brusa et~al. 2005, 2007)
searches for probable counterparts taking into account 
the positional accuracy of both the 
ONIR and \chandra\ \xray\ sources and also the expected magnitude distribution
of the counterparts. 
Compared to a simple matching method that searches for the
nearest counterpart within a given radius, the likelihood-ratio method 
significantly reduces the false-match probability toward faint
ONIR magnitudes (see, e.g., \S~2.4 of L10).

We used seven ONIR catalogs for identification purposes
(see Table~1 of L10 for further details):
\begin{enumerate}
\item The ESO 2.2-m WFI $R$-band catalog (denoted as ``WFI'';
Giavalisco et~al. 2004),
with a $5\sigma$ limiting AB magnitude (Oke \& Gunn 1983) of $27.3$;
\item The \mbox{GOODS-S} {\it Hubble Space Telescope} ({\it HST}) version r2.0z
$z$-band catalog (denoted as ``\mbox{GOODS-S}''; Giavalisco et~al. 2004), with
a $5\sigma$ limiting AB magnitude of $28.2$;
\item The GEMS {\it HST} $z$-band catalog (denoted as ``GEMS''; Caldwell et~al. 2008),
with a $5\sigma$ limiting AB magnitude of $27.3$;
\item The GOODS-S MUSIC catalog (denoted as ``MUSIC''; Grazian et~al. 2006; 
we used the $K$-selected sources in the V2 catalog that was presented in Santini et~al. 2009)
based on the Retzlaff et~al. (2010) VLT/ISAAC data,
with a limiting $K$-band AB magnitude of $23.8$ (at 90\% completeness);
\item The MUSYC $K$-band catalog (denoted as ``MUSYC''; Taylor et~al. 2009),
with a $5\sigma$ limiting AB magnitude of $22.4$;
\item The SIMPLE Spitzer/IRAC 3.6 $\mu$m catalog (denoted as ``SIMPLE''; Damen et~al. 2011), with a
$5\sigma$ limiting AB magnitude of 23.8; and
\item The VLA 1.4 GHz radio catalog (denoted as ``VLA'';
Miller et~al. 2008), with a 5$\sigma$ limiting flux density of $\approx 40~\mu$Jy.
\end{enumerate}
As mentioned in \S~\ref{sec:img}, we find systematic positional offsets between the
optical/near-infrared catalogs and the radio catalog and have chosen to shift
all the optical and near-infrared/infrared source positions throughout this paper by
$0.175\arcsec$ in right ascension and $-0.284\arcsec$ in declination
to be consistent with the radio astrometry.

We found that 716 (96.8\%) of the 740 main-catalog 
sources have ONIR counterparts.
For an X-ray source having multiple counterparts from the
likelihood-ratio matching (108 such cases), we chose
a primary counterpart from, in
order of priority, the VLA, \hbox{GOODS-S}, GEMS, MUSIC, WFI, MUSYC, or
SIMPLE catalog. This order is chosen based on several related factors:
the positional accuracy,   
angular resolution (to minimize any blending effects),
false-match probability, and catalog depth. Manual adjustments were
made to a few sources based on visual inspection (e.g., 
we selected the optical position rather than the VLA radio
position if the radio counterpart is clearly extended; see \S~2.3 of L10
for more details).

We used the Monte Carlo approach described in Broos et~al. (2007, 2011)
to estimate the false-match probability for each ONIR catalog.
The main-catalog \xray\ sources are considered to consist of two populations:
an ``associated population'' for which true counterparts are expected
in an ONIR catalog, and an ``isolated population'' for which 
no counterparts are expected (e.g., the true counterparts may be too faint or 
blended with other sources and thus not included) in an ONIR catalog.
We estimated the false-match probability for the associated population
by producing a mock ONIR counterpart for each X-ray source and running
the likelihood-ratio matching procedure to find the counterpart recovery fraction.
The offset between the mock counterpart and the X-ray source is selected
randomly based on the positional uncertainties, and
the magnitude of the mock counterpart is drawn randomly from
the expected magnitude distribution of the counterparts (derived previously in the
likelihood-ratio matching procedure). 
The mock ONIR catalog is thus 
composed of the mock counterparts and the original ONIR catalog with
source positions shifted and potential counterparts removed.
To estimate the false-match probability for the isolated population, we
shifted the \xray\ source positions and recorrelated the shifted sources
with the ONIR sources using likelihood-ratio matching. 
The above simulations were performed 100 times for each
\xray\ source population, and the results were used to solve for the 
final false-match probability for each ONIR catalog
(see Broos et~al. 2011 for details). 
The false-match probability for the 
associated population is 
generally smaller than that for the isolated population, and the 
final false-match probability for each ONIR catalog is $<4\%$.
The expected mean false-match probability for the main-catalog sources
is \hbox{$\approx2.1\%$}, derived by weighting
the false-match probabilities of individual ONIR catalogs 
with the number of primary counterparts in each catalog.
We note that the high identification 
rate, combined with the small false-match rate, 
provides independent evidence that the vast majority of our X-ray detections are robust.

For the 24 main-catalog sources that do not have highly significant multiwavelength
counterparts, we visually inspected the \hbox{X-ray} images and found that
the majority of them have apparent or strong \hbox{X-ray} signatures.
Of these 24 sources,
19 were detected in the full band, with a median number of full-band counts of 49.8;
17 were detected in the soft band, with a median number of soft-band counts of 40.5;
9 were detected in the hard band, with a median number of hard-band counts of 58.7;
and 17 were detected in at least two of the three standard bands.
We also investigated the \chandra\ events for these 24 sources and
concluded that they were not compromised by short-lived cosmic-ray afterglows.
Of these 24 unidentified sources,
5 were previously detected in the L08 main catalog,
3 were previously detected in the L08 supplementary \mbox{CDF-S} plus E-CDF-S \chandra\
catalog,
and 16 were only detected in the 4~Ms observations.
As for the nature of these 24 unidentified sources,
we refer readers to \S~4.1 of L10 and references therein for detailed discussion of the possibilities.
For example, 5 of these 24 unidentified sources are probably related to
off-nuclear \hbox{X-ray} sources associated with nearby galaxies
(e.g., Hornschemeier et~al. 2004; Lehmer et~al. 2006;
note that, in this paper, we did not attempt a thorough identification of
off-nuclear \xray\ sources).

\subsection{Main-Catalog Details}\label{sec:maincat}

We summarize in Table~\ref{tab:cols} 
the columns (a total of 79) in the main \chandra\ \hbox{X-ray} source catalog;
the main catalog itself is presented in Table~\ref{tab:main}.
The details of the 79 columns are given below.

\begin{deluxetable*}{ll}
\tabletypesize{\footnotesize}
\tablecaption{Overview of Columns in the Main \chandra\ Source Catalog}
\tablehead{
\colhead{Column} & \colhead{Description}}
%
%\tablewidth{0pt}
\startdata
1 & Source sequence number (i.e., XID) \\
2, 3 & Right ascension and declination of the \mbox{X-ray} source \\
4 & Minimum value of $\log P$ among the three standard bands ($P$ is the AE-computed binomial no-source probability) \\
5 & Logarithm of the minimum {\sc wavdetect} false-positive probability detection threshold \\
6 & $\approx68\%$ confidence-level \mbox{X-ray} positional uncertainty \\
7 & Off-axis angle of the \hbox{X-ray} source \\ 
8--16 & Aperture-corrected net (i.e., background-subtracted) source counts and the corresponding errors for the three standard bands \\
17 & Flag of whether a source shows any evidence for spatial extent\\
18, 19 & Right ascension and declination of the optical/near-infrared/infrared/radio (ONIR) counterpart \\
20 & Offset between the \mbox{X-ray} source and ONIR counterpart \\
21 & AB magnitude of the ONIR counterpart \\
22 & Name of the ONIR catalog from which the primary counterpart has been taken \\
23--43 & Right ascension, declination, and AB magnitude of the counterpart in seven ONIR catalogs\\
44--46 & Spectroscopic redshift, redshift quality flag, and the reference for the redshift \\
47--57 & Photometric-redshift information taken from sources in the literature \\
58 & Preferred redshift adopted in this paper \\
59 & Corresponding 2~Ms \hbox{CDF-S} source number from the main and supplementary \chandra\ catalogs presented in L08 \\
60, 61 & Right ascension and declination of the corresponding L08 source \\ 
62 & Corresponding 250~ks \hbox{E-CDF-S} source number from the main and supplementary \chandra\ catalogs presented in L05\\
63, 64 & Right ascension and declination of the corresponding L05 \hbox{E-CDF-S} source \\
65--67 & Effective exposure times derived from the exposure maps for the three standard bands\\
68--70 & Band ratio and the corresponding errors \\
71--73 & Effective photon index with the corresponding errors \\
74--76 & Observed-frame fluxes for the three standard bands\\
77 & Absorption-corrected, rest-frame \hbox{0.5--8~keV} luminosity \\
78 & Estimate of likely source type\\
79 & Notes on the source
\enddata
\label{tab:cols}
\end{deluxetable*}

\begin{deluxetable*}{lllcccccccccc}
\tabletypesize{\scriptsize}
\tablewidth{0pt}
\tablecaption{Main {\it Chandra} Source Catalog}

\tablehead{
\colhead{} &
\multicolumn{2}{c}{X-ray Coordinates} &
\multicolumn{2}{c}{Detection Probability} &
\colhead{}                   &
\colhead{}                   &
\multicolumn{6}{c}{Counts}      \\
\\ \cline{2-3} \cline{4-5} \cline{8-13} \\
\colhead{No.}                    &
\colhead{$\alpha_{2000}$}       &
\colhead{$\delta_{2000}$}       &
\colhead{$\log P$} &
\colhead{{\sc wavdetect}} &
\colhead{Pos Err}       &
\colhead{Off-Axis}       &
\colhead{FB}          &
\colhead{FB Upp Err}          &
\colhead{FB Low Err}          &
\colhead{SB}          &
\colhead{SB Upp Err}          &
\colhead{SB Low Err}          \\
\colhead{(1)}         &
\colhead{(2)}         &
\colhead{(3)}         &
\colhead{(4)}         &
\colhead{(5)}         &
\colhead{(6)}         &
\colhead{(7)}         &
\colhead{(8)}         &
\colhead{(9)}        &
\colhead{(10)}        &
\colhead{(11)}        &
\colhead{(12)}        &
\colhead{(13)}
}

\startdata
1 \dotfill \ldots & 03 31 35.79 &$-$27 51 36.0 &  $-$99.0 &  $-$8 &  0.5 &  11.98 &   186.8 &  19.0 &  17.9 &   117.8 &  13.5 &  12.4 \\
2 \dotfill \ldots & 03 31 40.12 &$-$27 47 46.6 &  $-$30.9 &  $-$8 &  0.5 &  10.62 &   155.7 &  19.8 &  18.6 &   101.5 &  13.4 &  12.2 \\
3 \dotfill \ldots & 03 31 41.01 &$-$27 44 34.7 &  $-$15.6 &  $-$8 &  0.6 &  11.10 &    \phantom{0}96.5 &  15.7 &  14.5 &    \phantom{0}31.5 &   \phantom{0}8.5 &   \phantom{0}7.3 \\
4 \dotfill \ldots & 03 31 43.25 &$-$27 54 05.6 &  \phantom{0}$-$6.5 &  $-$5 &  0.8 &  11.41 &    \phantom{0}54.1 &  13.8 &  12.6 &    \phantom{0}19.9 &  $-$1.0 &  $-$1.0 \\
5 \dotfill \ldots & 03 31 43.42 &$-$27 51 03.8 &  \phantom{0}$-$5.9 &  $-$8 &  0.5 &  10.21 &   109.0 &  27.5 &  25.7 &    \phantom{0}38.1 &  14.7 &  12.9 \\
\enddata
\tablecomments{
Units of right
ascension are hours, minutes, and seconds, and units of declination are
degrees, arcminutes, and arcseconds.
Table~\ref{tab:main} is presented in its entirety in the electronic edition. 
A portion is shown here for guidance regarding its form and content. The full table 
contains 79 columns of 
information for the 740 \hbox{X-ray} sources.}
\label{tab:main}

\end{deluxetable*}

1. Column 1 gives the source sequence number (i.e., XID). We list sources in order of 
increasing right ascension.

2. Columns 2 and 3 give the right ascension and declination of the \mbox{X-ray}
source, respectively.
We determined source positions following the procedure detailed 
in \S~\ref{sec:list}. 
To avoid truncation error, we quote the positions to higher precision 
than in the International Astronomical Union (IAU) registered names that begin 
with the acronym ``CXO CDFS''.

3. Columns 4 and 5 give the minimum value of $\log P$
($P$ is the AE-computed binomial no-source probability) among the three standard bands,
and the logarithm of the minimum {\sc wavdetect} false-positive probability
detection threshold, respectively.
More negative values of $\log P$ (Column~4) and 
false-positive probability threshold (Column~5) 
indicate a more significant source detection.
We set $\log P=-99.0$ for sources with $P=0$.
For the main-catalog sources, the median value of $\log P$ is $-8.9$
(note that $P<0.004$, corresponding to $\log P<-2.4$, is the condition for a source to be included in the main catalog).
There are 493, 57, 69, and 121 sources with minimum {\sc wavdetect} probabilities$^{\ref{siglev}}$
of $10^{-8}$, $10^{-7}$, $10^{-6}$, and $10^{-5}$, respectively (see Fig.~\ref{psig}).

4. Column 6 gives the $\approx68\%$ confidence-level \mbox{X-ray}
positional uncertainty in arcseconds computed using eq.~(\ref{equ:dpos}), which is dependent on both off-axis angle
and aperture-corrected net source counts.
The $\approx68\%$ confidence-level \mbox{X-ray}
positional uncertainty was used in the likelihood-ratio matching procedure
(see \S~\ref{sec:id}).
The positional uncertainty for the main-catalog sources ranges from
$0.10\arcsec$ to $1.51\arcsec$, with a median value of $0.42\arcsec$.

5. Column 7 gives the off-axis angle of the \hbox{X-ray} source in arcminutes,
which is the angular separation between the \hbox{X-ray} source (coordinates given
in Columns 2 and 3) and the \mbox{CDF-S} average aim point (given in Table~\ref{tbl-obs}).
The off-axis angle for the main-catalog sources ranges from
$0.33\arcmin$ to $12.36\arcmin$, with a median value of $5.82\arcmin$.
The maximum off-axis angle of $12.36\arcmin$ is slightly larger than
a half of the diagonal size of the \mbox{ACIS-I} field of view ($11.95\arcmin$),
due to the fact that the \mbox{CDF-S} observations have varying aim points and roll angles,
as shown in Table~\ref{tbl-obs}.

6. Columns 8--16 give the aperture-corrected net (i.e., background-subtracted) source counts and
the corresponding $1\sigma$ upper and lower statistical errors (Gehrels 1986)
for the three standard bands, respectively.
The photometry was calculated by AE using the position given in Columns 2 and 3 
for all bands and following the procedure described in \S~\ref{sec:list},
and was not corrected for vignetting or exposure time variations.
To be consistent with our source detection criterion (i.e., $P<0.004$),
we considered a source to be ``detected'' for photometry purposes
in a given band only if the AE-computed binomial no-source probability for that band 
is less than 0.004.
For sources not detected in a given band,
we calculated upper limits and placed $-1.00$ in the corresponding error columns. 
When the total number of counts within the polygonal extraction region of
an undetected source was $\le 10$,
we computed the upper limit using the Bayesian method of 
Kraft et~al. (1991) for a 99\% confidence level; otherwise,
we computed the upper limit at the $3\sigma$ level for Poisson statistics (Gehrels~1986).

7. Column 17 gives a flag indicating whether a source shows any evidence 
for spatial extent in basic testing.
In \S~\ref{sec:list}, we ran {\sc wavdetect} using 9 wavelet scales up to 16 pixels,
which potentially allows detection of sources that are extended on such scales.
We utilized the following procedure to assess extent.
We first derived a set of cumulative EEFs
by extracting the PSF power within a series of circular apertures (centered at the source position) up to
a 90\% EEF radius from the merged PSF image.
We then derived another set of cumulative EEFs
by extracting source counts within a series of circular apertures (also centered at the source
position) up to the same 90\% EEF radius from the merged source image.
Finally, we used a Kolmogorov-Smirnov (K-S) test suitable for two distributions to compute 
the probability ($\rho_{\rm KS}$) that the two sets of
cumulative EEFs are consistent with each other. 
Of the 740 main-catalog sources,
7 have $\rho_{\rm KS}\le 0.01$ (i.e., the merged PSF and source images are
inconsistent with each other at or above a 99\% confidence level)
and have the value of this column set to 2;
24 have $0.01<\rho_{\rm KS}\le 0.05$ 
and have the value of this column set to 1;
all the remaining sources have the value of this column set to 0.
A total of 31 main-catalog sources are flagged as 1 or 2 that corresponds to a $\ge 95\%$
confidence level,
which is comparable to the expected number of 
false-positive determinations, i.e., $37=740\times(1-95\%)$.
These 31 sources are located across the entire \cdfs\ field
 and do not show the likely expected pattern of central clustering
(since the PSF is sharpest near the field center), which might also indicate that
many of these sources could be false positives.
Moreover, we did not find any significant signature of extension for
these 31 sources upon visual inspection.
For the sources that truly have slight extents or are point sources sitting 
on top of highly extended sources,
our AE-computed photometry should be reasonably accurate, as detailed in \S~\ref{sec:list}.  
We note that a few highly extended sources in the \cdfs\ (e.g., Giacconi et~al. 2002; L05)
cannot be identified here because these sources have larger extents than the maximum value
of our adopted wavelet scales (i.e., 16~pixels); 
a full study of such extended sources is beyond the scope of this paper 
and will be presented in A.~Finoguenov et al. (in preparation).

8. Columns 18 and 19 give the right ascension and declination
of the ONIR counterpart
(see \S~\ref{sec:id} for the details of multiwavelength identifications).
Sources without multiwavelength identifications have these right ascension 
and declination values set to \hbox{``0 00 00.00''} and \hbox{``$-$00 00 00.0''}.

9. Column 20 gives the measured offset between the \mbox{X-ray} source and
ONIR counterpart in arcseconds.
Sources without multiwavelength identifications have a value set to $-1.00$.

10. Column 21 gives the AB magnitude of the ONIR counterpart,
measured in the counterpart-detection band.\footnote{The AB
magnitudes for the radio counterparts were converted from the radio flux densities,
$m({\rm AB}) =-2.5 \log(f_\nu)-48.60$.}
Sources without counterparts have a value set to $-1.00$.

11. Column 22 gives the name of the ONIR catalog (i.e., VLA, 
GOODS-S, GEMS, MUSIC, WFI, MUSYC, or SIMPLE) from which the primary counterpart
has been taken.
Sources without counterparts have this column set to ``...''.

12. Columns 23--43 give the right ascension, declination, and AB magnitude of the
counterpart in the above seven ONIR catalogs
that are used for identifications
(i.e., WFI, \mbox{GOODS-S}, GEMS, MUSIC, MUSYC, SIMPLE, and VLA).
We cross matched the positions of primary ONIR counterparts (i.e., Columns 17 and 18)
with the seven ONIR catalogs using likelihood-ratio matching.
Sources without counterparts have corresponding right ascension 
and declination values set to \hbox{``0 00 00.00''} and \hbox{``$-$00 00 00.0''}
and AB magnitudes set to $-1.00$.
We find $\approx 75\%$, 61\%, 72\%, 55\%, 70\%, 88\%,  and 18\% of the main-catalog \hbox{X-ray}
sources have WFI, \mbox{GOODS-S}, GEMS, MUSIC, MUSYC, SIMPLE, and VLA 
counterparts,\footnote{Note that the \mbox{GOODS-S} and MUSIC catalogs
cover $\approx 39\%$ of the \hbox{CDF-S} while the other five catalogs
cover the entire \hbox{CDF-S} (see Table~1 of L10 for more details);
$\approx 70\%$ of the main-catalog sources are in the \mbox{GOODS-S}/MUSIC area 
[see Fig.~\ref{pos}(a)].} respectively, 
with a false-match probability of $<2\%$ for each ONIR catalog (see \S~\ref{sec:id} for details).

13. Columns 44--46 give the spectroscopic redshift ($z_{\rm spec}$), redshift quality flag, and the 
reference for the redshift.
Spectroscopic redshifts were collected from
Le F\`{e}vre et~al. (2004),
Szokoly et~al. (2004),
Zheng et~al. (2004),
Mignoli et~al. (2005),
Ravikumar et~al. (2007),
Vanzella et~al. (2008),
Popesso et~al. (2009),
Treister et~al. (2009)\footnote{We flagged the spectroscopic redshifts from Treister et~al. (2009) as ``Insecure'' since Treister et~al. (2009) did not provide redshift quality flags.},
Balestra et~al. (2010), and 
Silverman et~al. (2010)
with the reference numbers of 1--10 in Column~46, respectively.
We cross matched the positions of primary ONIR counterparts (i.e., Columns 18 and 19)
with the above catalogs of spectroscopic redshifts 
using a matching radius of $0.5\arcsec$.
Of the 716 main-catalog sources that have multiwavelength identifications,
419 (58.5\%) have spectroscopic redshift measurements.
343 (81.9\%) of these 419 spectroscopic redshifts are secure,
i.e., they are measured at $\gsim 95\%$ confidence levels with multiple secure spectral features (flagged as ``Secure'' in Column 45);
76 (18.1\%) of these 419 spectroscopic redshifts are insecure (flagged as ``Insecure'' in Column 45).
We estimated the false-match probability to be $\lsim 1\%$ in all cases.
Sources without spectroscopic redshifts
have these three columns set to
$-1.000$, ``None'', and $-1$, respectively.

\begin{figure*}
\centerline{
\includegraphics[scale=0.75]{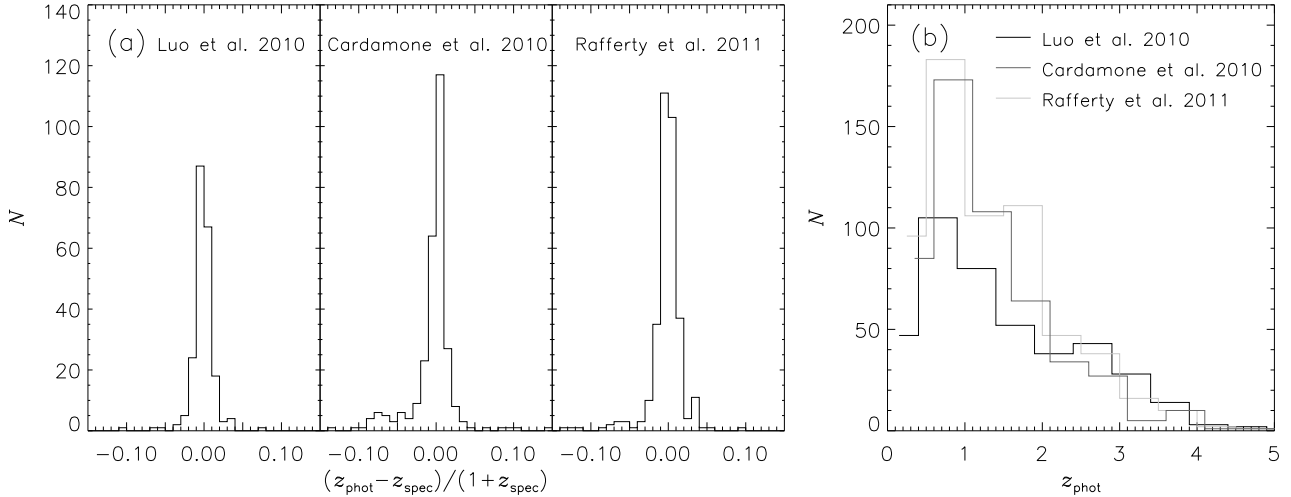}%{figs/zphot-comp.eps}
}
\figcaption{(a) 
Histogram of $(z_{\rm phot}-z_{\rm spec})/(1+z_{\rm spec})$
for L10 (218 sources), Cardamone et~al. (2010; 314 sources), and
Rafferty et~al. (2011; 339 sources).
(b) Histogram of $z_{\rm phot}$ for L10 (black histogram; 417 sources), 
Cardamone et~al. (2010; dark-gray histogram; 508 sources), and
Rafferty et~al. (2011; light-gray histogram; 611 sources). 
The histograms have been slightly shifted for clarity.
\label{zphot}}
\end{figure*}

14. Columns 47--57 give the photometric-redshift ($z_{\rm phot}$) information taken from
sources in the literature.
Columns 47--50 give the photometric redshift, the corresponding $1\sigma$
lower and upper bounds,\footnote{The photometric-redshift errors derived with
the Zurich Extragalactic Bayesian Redshift Analyzer
(ZEBRA; Feldmann et al. 2006) generally underestimate the real errors by
factors of $\approx 3$ and $\approx 6$ for the spectroscopic and
non-spectroscopic samples, respectively (see, e.g., \S~3.4 of L10). Therefore, 
multiplying the photometric-redshift errors presented here by these corresponding factors
(i.e., $\approx 3$ and $\approx 6$ for the spectroscopic and
non-spectroscopic samples, respectively) 
will roughly give realistic $1\sigma$ errors.\label{zerr}} and the alternative photometric redshift (set to
$-1.000$ if not available) from L10.
Columns 51--54 give the photometric redshift, the corresponding $1\sigma$
lower and upper bounds, and the corresponding quality flag $Q_z$ 
(smaller values of $Q_z$ indicate better quality; \hbox{$0<Q_z\lsim 1$--3}
indicates a reliable photometric-redshift estimate) from Cardamone et~al. (2010).
Columns 55--57 give the photometric redshift and the corresponding $1\sigma$
lower and upper bounds$^{\ref{zerr}}$ from Rafferty et~al. (2011).
We chose the above photometric-redshift catalogs because they utilized
extensive multiwavelength photometric data and produced accurate photometric redshifts.
L10 derived high-quality photometric redshifts for the 462 L08 main-catalog \hbox{X-ray}
sources with a treatment of photometry that included
utilizing likelihood-matching, manual source deblending,
and appropriate upper limits.
Cardamone et~al. (2010) employed new medium-band Subaru photometry and a PSF-matching technique to
create a uniform photometric catalog and derived photometric redshifts
for over 80,000 sources in the \hbox{E-CDF-S};
their photometric redshifts are of high quality, in particular for bright sources.
Rafferty et~al. (2011) derived photometric redshifts for over 100,000
sources in the \hbox{E-CDF-S}, using a compiled photometric catalog that
probes fainter magnitudes than the Cardamone et~al. (2010) catalog
by including sources in the \hbox{GOODS-S} MUSIC catalog 
(Grazian et~al. 2006; Santini et~al. 2009);
their photometric redshifts are accurate down to faint fluxes.
We cross matched the positions of primary ONIR counterparts (i.e., Columns 18 and 19)
with the above photometric-redshift catalogs 
using a matching radius of $0.5\arcsec$.
Of the 716 main-catalog sources that have multiwavelength identifications,
668 (93.3\%) have photometric-redshift estimates from at least one source
(this number excludes sources identified as stars, given in Column~78, that
have all these columns set to $-1.000$).
We estimate the false-match probability to be $\lsim 1\%$ in all cases.
Sources without photometric redshifts have all these columns set to $-1.000$.
We show in Fig.~\ref{zphot} 
the histograms of (a) $(z_{\rm phot}-z_{\rm spec})/(1+z_{\rm spec})$ and (b) $z_{\rm phot}$
for the above three sources of photometric redshift.
It seems clear that the photometric redshifts from each of these three sources
have high quality\footnote{In Fig.~\ref{zphot}(a),
the photometric redshifts from both L10 and Rafferty et~al. (2011) appear to have smaller 
outlier percentages than those from Cardamone et~al. (2010) because 
the spectral energy distribution templates were optimized using the spectroscopic-redshift
information before template fitting in both L10 and Rafferty et~al. (2011).
Blind-test results show that the actual outlier percentages from L10 and Rafferty et~al. (2011)
are comparable to those from Cardamone et~al. (2010) (see, e.g., \S~3.4 of L10 for the details
of blind tests).} in terms of accuracy and outlier percentage [see Fig.~\ref{zphot}(a)]
and cover a similar range of $z\approx 0$--5 [see Fig.~\ref{zphot}(b)].
We refer readers to the cited references for the respective details of
the photometric-redshift derivations, the advantages of the adopted methodologies,
and the caveats when using these photometric redshifts.

15. Column 58 gives the preferred redshift adopted in this paper.
We chose redshifts, in order of preference, as follows:
(1) secure spectroscopic redshifts;
(2) insecure spectroscopic redshifts that are in agreement with
at least one of the L10, Cardamone et~al. (2010), or Rafferty et~al. (2011)
photometric-redshift estimates [i.e., $|(z_{\rm spec}-z_{\rm phot})/(1+z_{\rm spec})|\le 0.15$,
where $z_{\rm spec}$/$z_{\rm phot}$ is the spectroscopic/photometric redshift];
(3) the L10 photometric redshifts;
(4) the Cardamone et~al. (2010) photometric redshifts; and
(5) the Rafferty et~al. (2011) photometric redshifts.
Of the 716 main-catalog sources that have multiwavelength identifications,
673 (94.0\%) have spectroscopic or photometric redshifts.

16. Column 59 gives the corresponding 2~Ms \hbox{CDF-S} source
number from the main and supplementary \chandra\ catalogs presented in L08.
We matched our \mbox{X-ray} source positions (i.e., Columns 2 and 3)
to L08 source positions (corrected for the systematic positional shifts
described in \S~\ref{sec:img}) using a $2.5\arcsec$ matching radius for sources
with off-axis angle $\theta <6\arcmin$ and a $4.0\arcsec$ matching radius
for sources with $\theta \ge 6\arcmin$.
The mismatch probability is $\approx 1\%$ using this approach.
For the 740 main-catalog sources, we find
\begin{enumerate}
\item[(a)] 440 have matches to the 462 L08 main-catalog sources (the value of Column~59 is 
that from col.~[1] of Table~2 in L08; see \S~\ref{sec:comp2ms} for more details);
\item[(b)] 41 have matches to the 86 L08 supplementary \mbox{CDF-S} plus E-CDF-S \chandra\
catalog sources (the value of Column~59 is that from col.~[1] of Table~5 in L08 with a
prefix of ``SP1\_'', e.g., SP1\_1);
\item[(c)] 22 have matches to the 30 L08 supplementary optically bright \chandra\ catalog sources
(the value of Column~59 is that from col.~[1] of Table~6 in L08 with a prefix of 
``SP2\_'', e.g., SP2\_1);
\item[(d)] 6 were outside of the 2~Ms \hbox{CDF-S} footprint of L08
(the value of Column~59 is set to $-1$);
the detection of these sources is simply due to the new sky coverage 
(rather than the improved sensitivity) of the
4~Ms \hbox{CDF-S}; and
\item[(e)] 231 have no match in any of the L08 main and supplementary \chandra\ catalogs;
these sources were inside the 2~Ms \hbox{CDF-S} footprint
but are only detected now due to the improved sensitivity of the 
4~Ms observations (the value of Column~59 is set to 0).
\end{enumerate}
In summary, of the 740 main-catalog sources, 503 were detected previously
in the 2~Ms \hbox{CDF-S} observations (the value of Column~59 is greater than 0)
and 237 were detected only in the 4~Ms observations
(the value of Column~59 is either $-1$ or 0).
Compared to the L08 main catalog, there are 300 (i.e., $740-440=300$) new main-catalog 
sources (see \S~\ref{sec:new} for more details of these 300 sources).

17. Columns 60 and 61 give the right ascension and declination of
the corresponding L08 source (corrected for the systematic positional shifts
described in \S~\ref{sec:img}) indicated in Column~59. 
Sources without an L08 match have right ascension and
declination values set to \hbox{``0 00 00.00''} and \hbox{``$-$00 00 00.0''}.

18. Column 62 gives the corresponding 250~ks \hbox{E-CDF-S} source
number from the main and supplementary \chandra\ catalogs presented in 
L05.
We adopted the same matching approach between \mbox{X-ray} catalogs as used for
Column~59, again with the \hbox{E-CDF-S} source positions corrected for the 
systematic positional shifts described in \S~\ref{sec:img}.
For the 740 main-catalog sources, we find
(1) 239 have matches in the \hbox{E-CDF-S} main \chandra\ catalog 
(the value of Column~62 is 
that from col.~[1] of Table~2 in L05);
(2) 5 have matches in the \hbox{E-CDF-S} supplementary optically bright \chandra\ catalog
(the value of Column~62 is that from col.~[1] of Table~6 in L05 with a prefix of 
``SP\_'', e.g., SP\_1);
and
(3) 496 have no match in either of the \hbox{E-CDF-S} main or supplementary 
\chandra\ catalogs (the value of Column~62 is set to 0).

19. Columns 63 and 64 give the right ascension and declination of
the corresponding L05 \hbox{E-CDF-S} source 
(corrected for the systematic positional shifts
described in \S~\ref{sec:img}) indicated in Column~62. 
Sources without an \hbox{E-CDF-S} match have right ascension and
declination values set to \hbox{``0 00 00.00''} and \hbox{``$-$00 00 00.0''}.

20. Columns 65--67 give the effective exposure times derived from the
exposure maps (detailed in \S~\ref{sec:img}) for the full, soft, and hard bands.
Dividing the counts in Columns~8--16 by the corresponding effective
exposure times will provide effective count rates 
that have been corrected for vignetting, quantum-efficiency
degradation, and exposure time variations.

21. Columns 68--70 give the band ratio and the corresponding upper and lower errors, 
respectively.
We defined the band ratio as the ratio of counts between the hard and soft bands,
correcting for differential vignetting between the hard and soft bands 
using the appropriate exposure maps.
We followed the numerical error-propagation method described in \S1.7.3 of Lyons (1991) to compute
band-ratio errors.
This method avoids the failure of the standard approximate variance 
formula when the number of counts is small and the error distribution 
is non-Gaussian (e.g., see \S2.4.5 of Eadie et~al. 1971).
We calculated upper limits for sources detected in the soft
band but not the hard band and lower limits for sources detected
in the hard band but not the soft band.
For these sources, we set the upper and lower
errors to the computed band ratio.
We set band ratios and corresponding errors to $-1.00$ for sources detected only 
in the full band.

22. Columns 71--73 give the effective photon index ($\Gamma$) with the corresponding
upper and lower errors, respectively, for a power-law model with the Galactic
column density given in \S~\ref{sec:intro}. 
We calculated the effective photon index based on the band ratio in Column~68,
using a conversion between the effective photon index and the band ratio.
We derived this conversion using the band ratios and photon indices
calculated by the AE-automated XSPEC-fitting procedure for relatively bright \mbox{X-ray} sources
(with full-band counts greater than 200; this ensures reliable XSPEC-fitting results).
This approach takes into account the
multi-epoch \chandra\ calibration information and thus has an advantage over 
methods using only single-epoch calibration information such as the
CXC's Portable, Interactive, Multi-Mission Simulator (PIMMS) method used by L08. 
We calculated upper limits for sources detected in the hard band but not the soft band
and lower limits for sources detected in the soft band but not the hard band.
For these sources, we set the upper and lower errors to the computed effective
photon index.
For low-count sources, we are unable to determine the effective photon
index reliably; we therefore assumed $\Gamma=1.4$, 
which is a representative value for faint
sources that should yield reasonable fluxes,
and set the corresponding upper and lower
errors to 0.00.
We defined sources with a low number of counts as being (1) detected in the
soft band with $<30$ counts and not detected in the hard band, (2) detected in
the hard band with $<15$ counts and not detected in the soft band, (3) detected
in both the soft and hard bands, but with $<15$ counts in each, or (4) detected
only in the full band.

23. Columns 74--76 give observed-frame fluxes in units of \flux\ in the full,
soft, and hard bands.
We computed fluxes using the counts in Columns~8--16, the
appropriate exposure maps (Columns~65--67), and the
effective power-law photon indices 
given in Column~71.
We did not correct fluxes for absorptions by Galactic material or material
intrinsic to the source.
Negative flux values indicate upper limits. 
We note that, due to the
Eddington bias, sources with low net counts (given in Columns~\hbox{8--16}) could
have true fluxes lower than those computed here
(see, e.g., Vikhlinin et~al. 1995; Georgakakis et~al. 2008).
We do not attempt to correct for the
Eddington bias, since we aim
to provide only observed fluxes here.
Determining more accurate fluxes for these sources would require (1) using
a number-count distribution prior to estimate the flux probabilities for
sources near the sensitivity limit and/or (2) directly fitting the \xray\
spectra for each observation; these analyses are beyond the scope of 
this paper.

24. Column 77 gives a basic estimate of the absorption-corrected, rest-frame 
\hbox{0.5--8~keV} luminosity
($L_{\rm 0.5-8\ keV}$) in units of \hbox{erg s$^{-1}$}.
We calculated $L_{\rm 0.5-8\ keV}$ using the procedure detailed in \S~3.4 of 
Xue et~al. (2010).
Briefly, this procedure
models the \hbox{X-ray} emission using a power-law 
with both intrinsic and Galactic absorption (i.e., $zpow\times wabs\times zwabs$ in XSPEC)
to find the intrinsic column density that reproduces the observed band ratio (given in
Column~68), assuming
a typical power-law photon index of $\Gamma_{\rm int}=1.8$ for intrinsic AGN spectra.
It then corrects for both Galactic and intrinsic absorption to obtain the 
absorption-corrected
flux ($f_{\rm 0.5-8\ keV,int}$; as opposed to the observed flux given in Column~74), and follows the equation $L_{\rm 0.5-8\ keV}=4\pi d^2_L f_{\rm 0.5-8\ keV,int} (1+z)^{\Gamma_{\rm int} -2}$ to derive $L_{\rm 0.5-8\ keV}$ 
(where $d_L$ is the luminosity distance and $z$ is the adopted redshift 
given in Column~58).
In this procedure, we set the observed band ratio to a value that
corresponds to $\Gamma=1.4$ for sources detected only in the full band;
for sources having upper or lower limits on the band ratio, we
adopted their upper or lower limits for this calculation.
Basic luminosity estimates derived in this manner are generally found to agree
with those from direct spectral fitting to within 
a factor of $\approx 30\%$\footnote{We caution that our basic $L_{\rm 0.5-8\ keV}$ estimates could be subject to
larger uncertainties for heavily obscured AGNs. This is not only due to  
the increasing difficulty in determining the intrinsic column density from the observed band ratio,
but also due to the fact that other components (e.g., reflection, and scattering) 
become stronger in such heavily obscured sources.};  
the direct spectral-fitting approach should produce more reliable estimates, but is beyond the scope of this paper.
Sources without redshift estimates have this column set to $-1.000$;
negative luminosity values other than $-1.000$ indicate upper limits.

25. Column 78 gives a basic estimate of likely source type.
We categorized the \hbox{X-ray} sources into three basic types: ``AGN'', ``Galaxy'', and ``Star''.
We utilized four criteria that are based on distinct AGN physical properties 
and one criterion that is based on optical spectroscopic information to identify AGN candidates, which must satisfy at least one of these five criteria.
We briefly describe these criteria below:
\begin{enumerate}
\item[(a)] A source with an intrinsic \hbox{X-ray} luminosity (given in Column~77) of
$L_{\rm 0.5-8\ keV}\ge 3\times 10^{42}$ \hbox{erg s$^{-1}$} will be identified as a 
luminous AGN.
\item[(b)] A source with an effective photon index (given in Column~71) of $\Gamma \le 1.0$
will be identified as an obscured AGN.
\item[(c)] A source with an \hbox{X-ray}-to-optical flux ratio of
$\log(f_{\rm X}/f_R)>-1$ (where $f_{\rm X}=f_{\rm 0.5-8\ keV}$, $f_{\rm 0.5-2\ keV}$, or $f_{\rm 2-8\ keV}$)
will be identified as an AGN.
\item[(d)] A source with excess (i.e., a factor of $\ge 3$) 
\xray\ emission over the level expected from pure
star formation will be identified as an AGN, i.e., with 
$L_{\rm 0.5-8\ keV}\gsim 3\times (8.9\times 10^{17}L_{\rm R})$,
where $L_{\rm R}$ is the rest-frame 1.4~GHz monochromatic luminosity in units of W~Hz$^{-1}$ and
$8.9\times 10^{17}L_{\rm R}$ is the expected \xray\ emission level that
originates from starburst galaxies (see Alexander et~al. 2005 for the details of this criterion).
\item[(e)] A source with optical spectroscopic AGN features such as broad emission lines and/or
high-excitation emission lines will be identified as an AGN; 
we cross matched the sources (using the ONIR counterpart positions given in Columns 18 and 19) with the spectroscopically identified AGNs 
in Szokoly et~al. (2004),
Mignoli et~al. (2005), and Silverman et~al. (2010), 
using a matching radius of $0.5\arcsec$.
\end{enumerate}
We note that the above five criteria are effective but not complete in identifying AGNs and
refer readers to, e.g., Bauer et~al. (2004), Alexander et~al. (2005), Lehmer et~al. (2008), and
Xue et~al. (2010) for discussions and caveats
(e.g., low-luminosity and/or highly obscured AGNs may still not be identified
through the criteria presented here).
We also identified likely stars by cross matching the sources
(using the ONIR counterpart positions given in Columns 18 and 19) 
with (1) the spectroscopically identified stars in Szokoly et~al. (2004),
Mignoli et~al. (2005), and Silverman et~al. (2010),
(2) the likely stars with stellarity indices greater than 0.7
in the GEMS {\it HST} catalog (Caldwell et~al. 2008),
and (3) the likely stars with best-fit stellar templates
in the MUSYC photometric-redshift catalog (Cardamone et~al. 2010), 
using a matching radius of $0.5\arcsec$.
We inspected each of the sources identified as stars in the {\it HST} images and
retrieved sources that appear to be galaxies (i.e., set our classification to galaxy).
The sources that were not identified as AGNs or stars are classified as ``galaxies''.
Of the 740 main-catalog sources, 
568 (76.8\%), 162 (21.9\%), and 10 (1.3\%)
are identified as AGNs, galaxies, and stars, respectively.
Of the 568 AGNs in the main catalog,
65.1\%, 40.3\%, 91.7\%, 14.8\%, and 1.1\% satisfy the
criteria (a), (b), (c), (d), and (e), respectively.

26. Column 79 gives notes on the sources.
We annotated sources at the field edge
that lie partially outside of the survey area with ``E'' (one source only)
and sources in close doubles or triples with ``C'' (a total of 35 sources;
these 35 sources have overlapping polygonal extraction regions that correspond
to \hbox{$\approx 40$--75\%} EEFs; see \S~\ref{sec:list}).
Sources not annotated have this column set to ``...''.

\subsection{Comparison with 2~Ms CDF-S Main-Catalog Sources}\label{sec:comp2ms}

We summarize in Table~\ref{tbldet} the source detections in the three standard bands.
In total 740 sources are detected, 
with 634, 650, and 403 detected in the full, soft, and hard band, respectively.
As stated earlier in \S~\ref{sec:maincat} (see the description of Column~59),
503 of the main-catalog sources were detected in the L08 main or supplementary 
catalogs, among which 440 were detected in the L08 main catalog.
For these 440 common sources, we find general agreement between the 
derived \hbox{X-ray} photometry presented here and in L08. 
For instance, the median ratio between our full-band count rates and the L08
full-band count rates for the 387 full-band detected sources (among
these 440 common sources) is 1.03,
with an interquartile range of 0.91--1.14.
The $\approx 3$\% increase in the full-band count rates is mainly 
caused by a few updates to the ancillary response file (ARF)
and contamination model in the {\sc caldb} data\footnote{For example, 
there was a recalibration of the \hbox{ACIS-I} ARF in 
{\sc caldb}~4.1.1 (released in January 2009),
yielding a flat $\approx 9\%$ reduction in the effective area below 2~keV and 
a \hbox{$\approx 0$--8\%} reduction between 2 and 5~keV (see
http://cxc.harvard.edu/ciao/why/caldb4.1.1\_hrma.html).} since the production of the L08 catalogs.
The detailed differences (e.g., scattering) in the derived \hbox{X-ray} photometry
are mainly due to source variability
and/or the above {\sc caldb} updates (e.g., sources with
different \xray\ spectral shapes are affected differently by these {\sc caldb} updates).
The approximately doubled exposure improves the
source positions and spectral constraints significantly.
Hence, the 4~Ms \hbox{CDF-S} catalogs presented here supersede those in L08.

\begin{deluxetable}{lccccc}
\tabletypesize{\small}
\tablewidth{0pt}
\tablecaption{Summary of {\it Chandra} Source Detections \label{tbldet}}

\tablehead{
\colhead{} &
\colhead{Number of} &
\multicolumn{4}{c}{Detected Counts Per Source} \\
\cline{3-6}   
\colhead{Band (keV)} &
\colhead{Sources} &
\colhead{Maximum} &
\colhead{Minimum} &
\colhead{Median} &
\colhead{Mean}
}

\startdata
Full (0.5--8.0)   & 634 & 35657.0 & 11.4 & 101.4 & 497.4 \\
Soft (0.5--2.0)  & 650 & 25470.7 & \phantom{0}6.0 & \phantom{0}45.1 & 293.8  \\
Hard (2--8)  & 403 & 10219.3 & 10.7 & \phantom{0}99.9 & 302.8 
\enddata
\end{deluxetable}

Twenty-two (i.e., $462-440=22$) of the 462 sources detected in the L08 main catalog are not included in our main catalog,
among which 3 are included in our supplementary catalog (see \S~\ref{sec:supp2}).
Thus, there are a total of 19 ``missing'' L08 main-catalog sources not included in the
4~Ms main or supplementary catalogs.
Among these 19 missing sources, there are two cases where 
a source was previously listed as being in a close pair but is now removed  
due to no apparent signature of a close pair in the 4~Ms images.
Of the remaining 17 sources,
12, 3, and 2 have a logarithm of the minimum {\sc wavdetect} false-positive
probability detection threshold of $-6$, $-7$, and $-8$ in the L08 main catalog, 
respectively.
Among these 17 sources,
9 have no multiwavelength counterparts and have no emission
clearly distinct from the background in the 4~Ms images,
which indicates that most of these 9 sources are likely false 
detections.\footnote{We note that L08 estimated the number of false detections
in their main catalog to be $\approx 18$, which is a conservative estimate;
the real number of false detections is likely \hbox{$\approx 2$--3} times smaller,
i.e., \hbox{$\approx 6$--9} (see \S~3.2 of L08).}
For the other 8 sources that have relatively faint multiwavelength counterparts,
they also have no apparent \hbox{X-ray} signatures in the 4~Ms images
although a few of them have full-band counts
of \hbox{$\gsim 20$--30} in the L08 catalog;
these 8 sources are likely real \hbox{X-ray} sources,
but they are not detected in the 4~Ms images probably due to 
source variability and/or background fluctuations,
as the second 2~Ms exposure was taken $\approx 2.5$ years
after the completion of the first 2~Ms exposure.
Indeed, all of these 8 sources were variable
(at $\ge 99.7\%$ confidence levels based on K-S tests)
and became fainter (i.e., having a factor of $\gsim 2$ smaller count rates) during the second
2~Ms of observations; consequently, the addition of background counts
diluted their signals from the first 2~Ms of observations.\footnote{Of 
these 8 sources, only 3 satisfy the $P<0.004$ criterion 
during the first 2~Ms exposure, while none satisfies the $P<0.004$ criterion
during the second 2~Ms exposure.}
We note that source variability is not uncommon among the CDF sources:
over short timescales (days-to-weeks) 
a $\approx 35\%$ median flux variability for the sources in the first
2~Ms data set has been observed;
over long timescales (years) source fluxes 
could vary by up to a factor of \hbox{$\approx$5--10} in a few extreme cases
(Paolillo et~al. 2004; M.~Paolillo et~al. 2011, in preparation).

We summarize in Table~\ref{tblundet}
the number of sources detected in one band but not another. 
There are 21, 101, and 5 sources detected only in the full, soft, and hard band, respectively,
as opposed to 31, 56, and 3 sources detected only in the full, soft, and hard band
in the L08 main catalog.

\begin{deluxetable}{lccc}
\tabletypesize{\small}
\tablewidth{0pt}
\tablecaption{Sources Detected in One Band but not Another \label{tblundet}}
\tablehead{
\colhead{Detection Band} &
\multicolumn{3}{c}{Nondetection Energy Band} \\
\cline{2-4}
\colhead{(keV)} &
\colhead{Full} &
\colhead{Soft} &
\colhead{Hard} 
}
\startdata
Full (0.5--8.0)  & \ldots & 85 & 236 \\
Soft (0.5--2.0)  & ~~~~101~~~~ & \ldots & 316 \\
Hard (2--8)   & ~~~~\phantom{0}5~~~~ & 69  & \ldots \\
\enddata
\tablecomments{For example, there were 85 sources detected in the full band
but not in the soft band.}
\end{deluxetable}

\subsection{Properties of Main-Catalog Sources}\label{sec:prop}

\begin{figure}
\centerline{\includegraphics[scale=0.5]{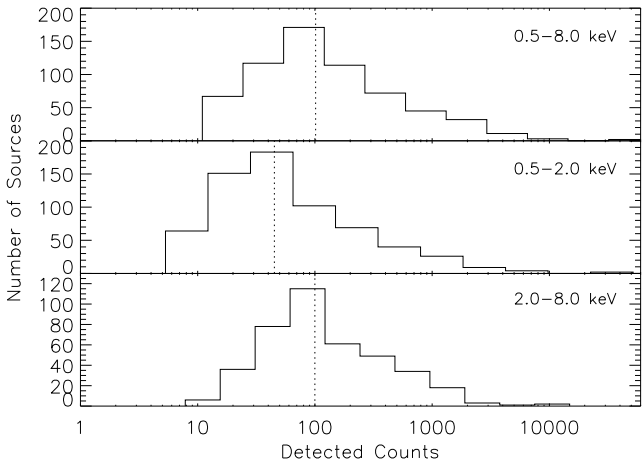}
} \figcaption{Histograms of detected source
counts for the main-catalog sources 
in the full ({\it top}), soft ({\it middle}), and hard ({\it bottom}) bands. 
Sources with upper limits have not been included in the plots.
The vertical dotted lines indicate median numbers of counts in each band
(see Table~\ref{tbldet}).
\label{cnthist}}
\end{figure}

In Figure~\ref{cnthist} we show the distributions of detected
counts in the three standard bands for the main-catalog sources. 
The median number of counts is $\approx 101$, 45, and 100
for the full, soft, and hard band, respectively.
There are 319 sources with $>100$ full-band counts, for which basic
spectral analyses are possible;
there are 202, 101, and 60 sources with $>200$, 500, 1000 full-band counts, respectively.

\begin{figure}
\centerline{\includegraphics[scale=0.5]{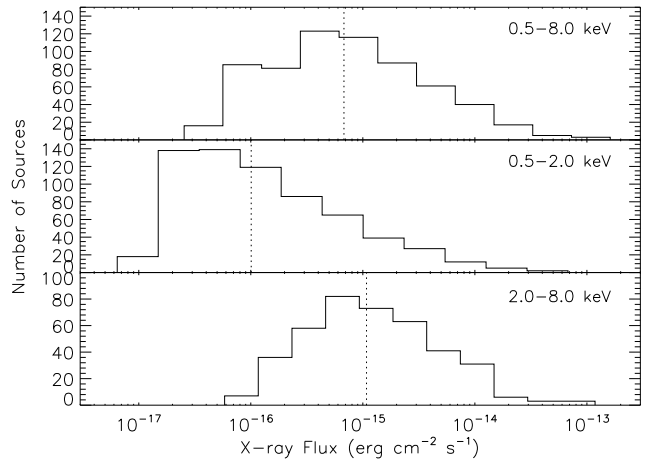}}
\figcaption{Histograms of \hbox{X-ray} fluxes for the main-catalog
sources in the full ({\it top}), soft ({\it middle}), and hard ({\it
bottom}) bands.
Sources with upper limits have not been included in the plots.
The vertical dotted lines indicate the median fluxes of
$6.8\times10^{-16}$, $1.0\times10^{-16}$ and $1.1\times10^{-15}$ \flux\
for the full, soft, and hard bands, respectively.
\label{fluxhist}}
\end{figure}

In Figure~\ref{fluxhist} we show the distributions of \hbox{X-ray} flux
in the three standard bands for the main-catalog sources. 
The \hbox{X-ray} fluxes span roughly four orders of
magnitude, with a median value of 
$6.8\times 10^{-16}$, $1.0\times 10^{-16}$, and $1.1\times 10^{-15}$ \flux\
for the full, soft, and hard band, respectively.

\begin{figure}
\centerline{\includegraphics[scale=0.52]{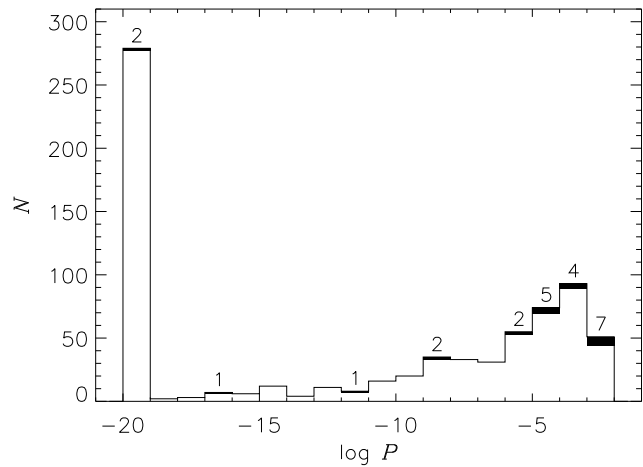}}
\figcaption{Histogram of the AE-computed binomial
no-source probability, $P$, for the
main-catalog sources.
For the purpose of illustration, we set the values of
$P<10^{-20}$ to $P=10^{-20}$ in this plot.
The shaded areas indicate sources that have no multiwavelength counterparts,
with the numbers of these unidentified sources listed above the corresponding shaded areas.
\label{pid}}
\end{figure}

We show in Figure~\ref{pid} the distribution of the AE-computed 
no-source probability $P$ (given in Column~4) for the main-catalog sources;
sources without multiwavelength counterparts (given in Columns~18 and 19) are highlighted by shaded areas.
It is clear that the majority of the main-catalog sources have
low no-source probabilities (i.e., with $\log P\le -6$). 
We find that 1.3\% of the $\log P\le -6$ sources 
have no multiwavelength counterparts, as opposed to the 6.6\% of $\log P> -6$
sources that lack multiwavelength counterparts.
Combined with the small false-match rate (see \S~\ref{sec:id}),
the above observations suggest that an \xray\ source having a secure multiwavelength counterpart is an effective indicator of it being real.

\begin{figure*}
\centerline{\includegraphics[scale=1.250]{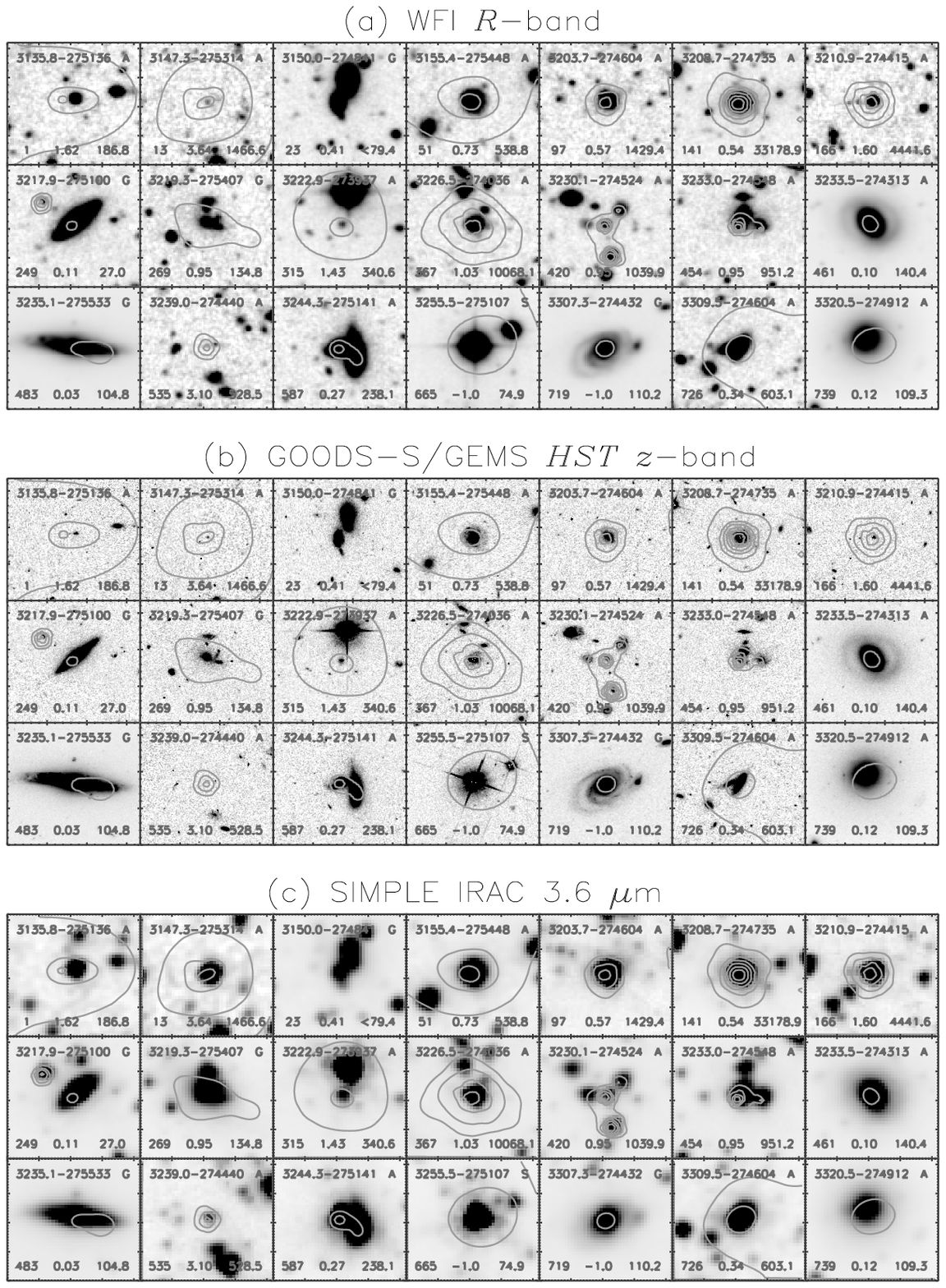}}
\figcaption{Typical postage-stamp images from (a) the WFI $R$-band, (b) the GOODS-S/GEMS {\it HST} $z$-band, and (c) the SIMPLE IRAC \hbox{3.6~$\mu$m} band for the main-catalog sources
with full-band adaptively smoothed \hbox{X-ray} contours overlaid.
The contours have a logarithmic scale and range from \hbox{$\approx$0.003\%--30\%}
of the maximum pixel value.
The labels at the top of each image give the
source name (for right ascension, the hours ``03'' have been omitted for succinctness) derived from the source coordinates and the source type
(``A'' denotes ``AGN''; ``G'' denotes ``Galaxy''; and ``S'' denotes ``Star'').
The numbers at the bottom of each image
indicate the source number, the adopted redshift,
and the full-band counts or upper limit (with a ``$<$'' sign).
There are several cases where no \hbox{X-ray} contours are
present, either because these sources were not detected in the full band or
their full-band counts are low resulting in their observable emission in the adaptively
smoothed images being suppressed by {\sc csmooth}.
Each image is $25\arcsec\times 25\arcsec$,
with the source of interest located at the center.
The cutouts of all the main-catalog sources are available in the electronic edition.
[{\it See the electronic edition of the Supplement for a complete version of this figure.}]
\label{ps}}
\end{figure*}

We show in Figure~\ref{ps} ``postage-stamp'' images from the WFI
$R$-band, the GOODS-S/GEMS {\it HST} $z$-band, and the SIMPLE IRAC \hbox{3.6~$\mu$m} band with adaptively smoothed full-band \xray\ contours overlaid for
the main-catalog sources. 
The size of \hbox{X-ray} sources in these images spans a wide range
largely due to PSF broadening with off-axis angle.

\subsection{Properties of the 300 New Main-Catalog Sources}\label{sec:new}

In this section we examine the properties of the 300 main-catalog sources 
that were not detected in the L08 main catalog (hereafter new sources), 
putting emphasis on the comparison
with the sources previously detected in the L08 main catalog (hereafter old sources;
a total of $740-300=440$ sources).

\begin{figure*}
\centerline{\includegraphics[scale=0.7]{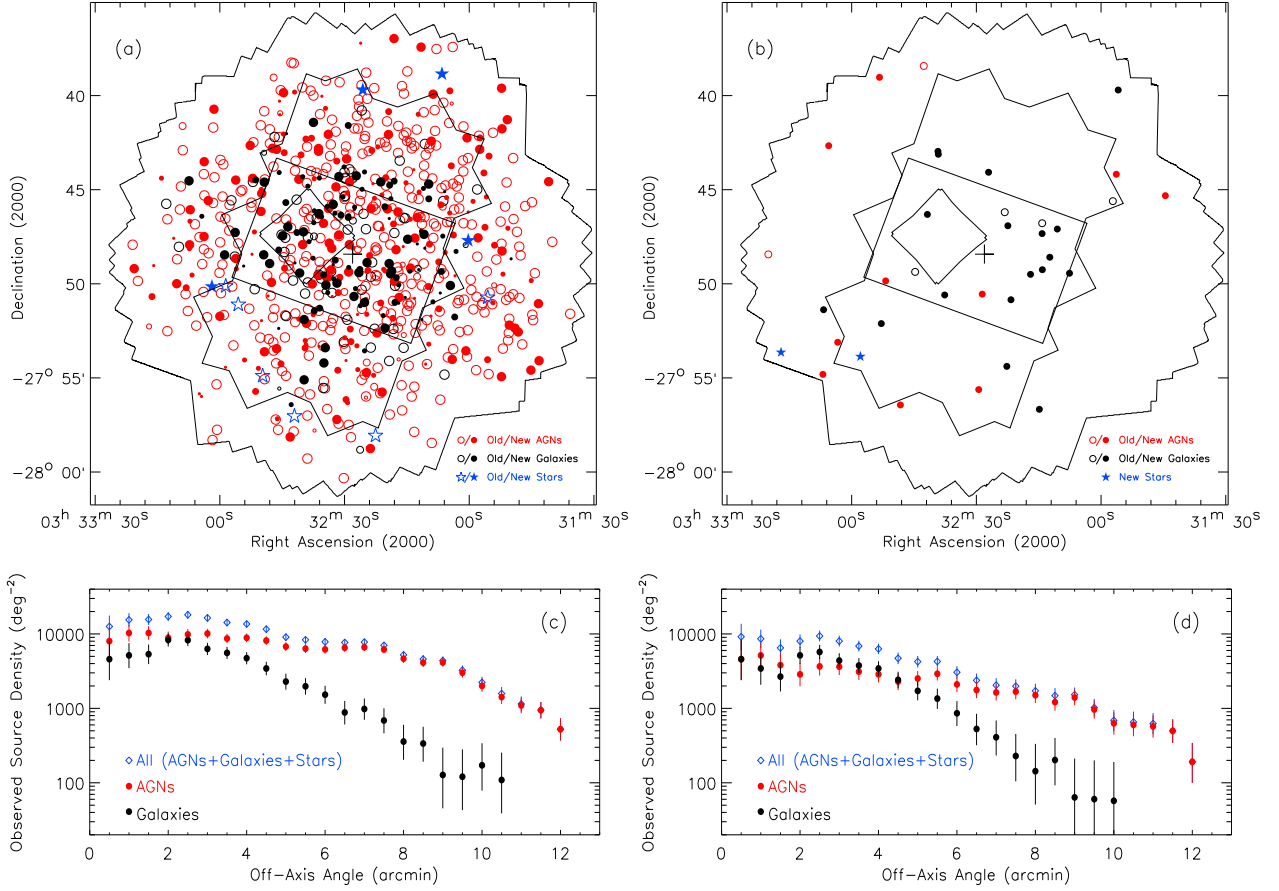}}
\figcaption{
(Top) Source spatial distributions for (a) the main catalog
and (b) the supplementary optically bright catalog.
Sources that are considered AGNs, galaxies, and stars (given in Column~78)
are colored red, black, and blue, respectively.
Open circles indicate AGNs/galaxies
that were previously detected in (a) the L08 main catalog or 
(b) the L08 main or supplementary optically bright catalog;
open stars in (a) indicate stars 
that were previously detected in the L08 main catalog;
filled circles and stars indicate new AGNs/galaxies and stars, respectively.
The regions and the plus sign are the same as those in Fig.~\ref{fbimg}.
In panel (a),
the sizes of the circles and stars indicate the AE binomial no-source probabilities,  
with larger sizes indicating lower no-source probabilities:
as the size becomes smaller, the AE binomial no-source probability $P$ moves from
$\log P\le -5$, $-5<\log P\le -4$, $-4<\log P\le -3$, to $\log P >-3$.
In panel (b), all sources have $\log P >-3$ and are plotted as circles/stars of the same size. 
(Bottom) Observed source density for different source types as a function of off-axis angle 
for (c) all the main-catalog sources and (d) the new main-catalog sources, 
as computed in bins of $\Delta\theta=1\arcmin$.
$1\sigma$ errors are calculated using Poisson statistics.
[{\it see the electronic edition of the Supplement for a color version of this figure.}]
\label{pos}}
\end{figure*}

Figure~\ref{pos}(a) shows the positions of the new sources (shown as
filled symbols) and the old sources (shown as open symbols),
with source types (given in Column~78) being color-coded (red for AGNs, black for galaxies, and 
blue for stars, respectively).
Different symbol sizes represent different AE binomial no-source probabilities
(see Column~4 of Table~\ref{tab:main}),
with larger sizes indicating lower no-source probabilities (i.e., higher source-detection significances).
In the \hbox{GOODS-S} region,
there are 512 main-catalog sources, with 221 being new;
in the \hbox{CANDELS} region,
there are 258 main-catalog sources, with 123 being new; and
in the \hbox{UDF} region,
there are 45 main-catalog sources, with 20 being new.
The source densities of both new and old sources decline toward large off-axis angles
as the sensitivity decreases (see \S~\ref{sec:smap});
such a trend appears more apparent among new sources than among old sources, e.g.,
22.0\% of new sources and while only 14.8\% of old sources have $\theta<3\arcmin$, and
62.0\% of new sources and while only 46.1\% of old sources have $\theta<6\arcmin$, respectively.
Figure~\ref{pos}(c) presents the observed 
source density as
a function of off-axis angle for all the main-catalog sources.
Overall, AGNs have larger observed source densities than galaxies.
However, since the slope of the observed galaxy number counts
at faint fluxes is steeper than that of
the observed AGN number counts (e.g., Bauer et~al. 2004), 
the galaxy source density approaches the AGN source density
toward smaller off-axis angles (i.e., toward lower flux levels).
This can also be seen in Fig.~\ref{pos}(d) that plots the observed source density 
versus off-axis angle for new sources; within $\theta=3\arcmin$, the new galaxies 
already outnumber the new AGNs (36 versus 30).
Near the center of the 4~Ms \hbox{CDF-S} (within $\theta=3\arcmin$), as shown in Fig.~\ref{pos}(c),
the overall observed AGN and galaxy source densities have reached
$9800_{-1100}^{+1300}$~deg$^{-2}$ and 
$6900_{-900}^{+1100}$~deg$^{-2}$, respectively.
We note that detailed analyses of the overall source densities for different
source types,
which consider effects such as the Eddington bias and incompleteness,
are beyond the scope of this work.

\begin{figure*}
\centerline{\includegraphics[scale=0.82]{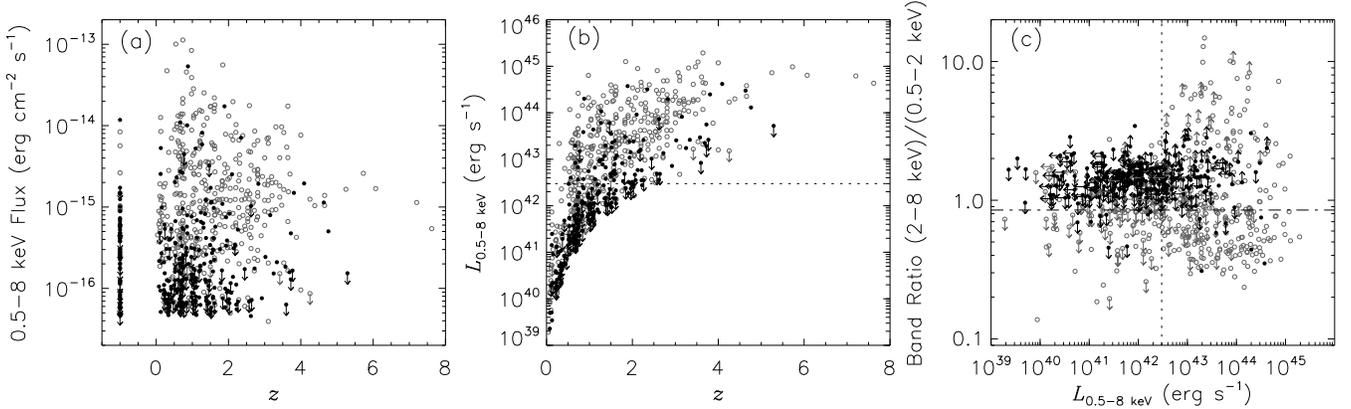}
}
\figcaption{Plots of (a) observed-frame full-band flux vs. redshift,
(b) absorption-corrected, rest-frame \hbox{0.5--8 keV} luminosity vs. redshift,
and (c) band ratio vs. absorption-corrected, rest-frame \hbox{0.5--8 keV} luminosity,
for the main-catalog sources.
Gray open circles indicate the main-catalog sources that were previously detected
in the L08 main catalog;
black filled circles indicate the new main-catalog sources that were not previously
detected in the L08 main catalog.
Arrows indicate limits. 
Several sources shown in Panels (a) and (b) have photometric redshifts
greater than $\approx 4.5$; these photometric redshifts are probably not
very reliable due to poor photometric coverage (see \S~3.3 of L10 for more discussion).
In Panel (b), sources without redshift estimates have not been included in the plot;
in Panel (c), sources without redshift estimates and sources with only
full-band detections have not been included in the plot.
The dotted lines in Panels (b, c) and the dashed-dot line in Panel (c)
indicate the threshold values of two AGN-identification criteria 
(i.e., $L_{\rm 0.5-8\ keV}\ge 3\times 10^{42}$ \hbox{erg s$^{-1}$} and $\Gamma \le 1.0$;
see the description of Column~78 for details).
\label{f-lx-z-br}}
\end{figure*}

We show in Figure~\ref{f-lx-z-br} plots of (a) observed-frame full-band flux
(given in Column~74) vs. redshift (given in Column~58),
(b) absorption-corrected, rest-frame \hbox{0.5--8 keV} luminosity (given in Column~77) vs. redshift,
and (c) band ratio (given in Column~68) vs. absorption-corrected, rest-frame \hbox{0.5--8 keV} luminosity,
for new sources (shown as filled circles) and old sources (shown as open circles), respectively.
Compared to old sources,
new sources typically have smaller 
full-band fluxes and \hbox{0.5--8 keV} luminosities 
[see their clustering
at the faint-flux end in Fig.~\ref{f-lx-z-br}(a) and 
at the low-luminosity end in Fig.~\ref{f-lx-z-br}(b)],
which is expected since the 4~Ms \cdfs\ has fainter flux limits than the 2~Ms \cdfs.
The existence of a small number of new sources at the high-flux/luminosity end
leads to the full range of flux/luminosity for new sources being similar to
that for old sources; 
these bright/luminous sources are typically located at 
relatively large off-axis angles.
As shown in Fig.~\ref{f-lx-z-br}(a),
there is no apparent correlation between full-band flux and
redshift for either new or old sources, and
the 4~Ms \hbox{CDF-S} is detecting an appreciable number of  
the faintest sources at least up to $z\approx 3$.
According to Fig.~\ref{f-lx-z-br}(b),
the \hbox{0.5--8 keV} luminosity spans
a very broad range (roughly six orders of magnitude) for both new and old sources;
13.6\% of the main-catalog sources are very luminous (with $L_{\rm 0.5-8\ keV}>10^{44}$ 
\hbox{erg s$^{-1}$}; most are old sources), among which there are a number of sources that
are highly obscured [see the upper right corner of Fig.~\ref{f-lx-z-br}(c)].  
As seen in Fig.~\ref{f-lx-z-br}(c),
new sources could potentially have a similar
range or distribution of band ratio to that of old sources, given that 82.7\% of new sources have 
either lower limits (19.0\%) or upper limits (81.0\%) on their band ratios
(see relevant discussions on this point later in this section).

\begin{figure}
\centerline{\includegraphics[scale=0.5]{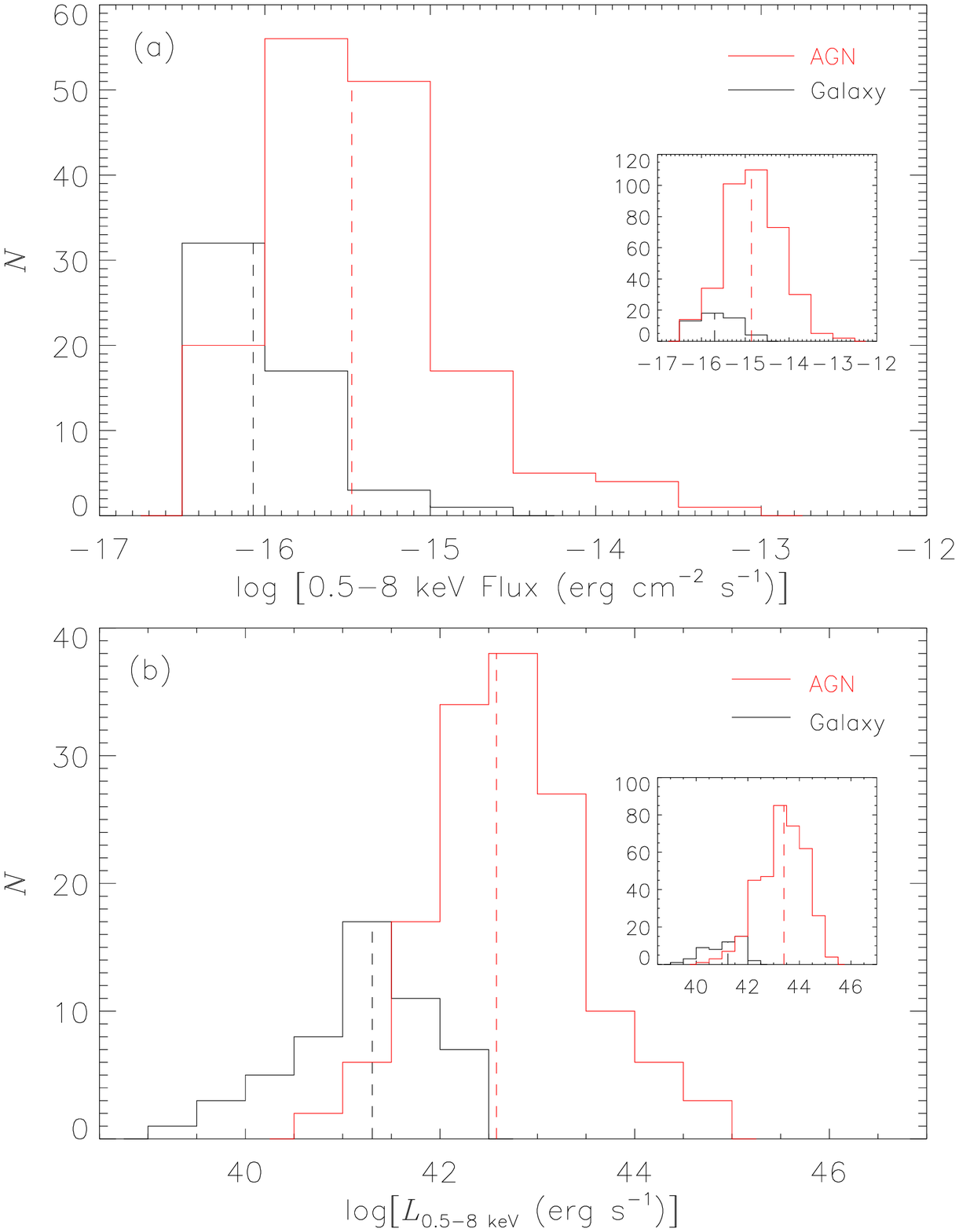}
}
\figcaption{Distributions of (a) observed-frame full-band flux and (b) absorption-corrected,
rest-frame \hbox{0.5--8 keV} luminosity for the new main-catalog sources.
The red and black histograms indicate AGNs and galaxies, respectively.
The vertical red and black dashed lines indicate the median values
for AGNs and galaxies, respectively.
Sources with upper limits on full-band fluxes have not been included in the plotting
for panel (a); sources without estimates of \xray\ luminosities 
(due to no available redshift) or with upper limits on \xray\ luminosities have not
been included in the plotting for panel (b).
The insets show results for the old main-catalog sources.
[{\it see the electronic edition of the Supplement for a color version of this figure.}]
\label{f-lx}}
\end{figure}

Figure~\ref{f-lx} shows  
distributions of observed-frame full-band flux (given in Column~74) 
and absorption-corrected, rest-frame \hbox{0.5--8 keV} luminosity
(given in Column~77) for new sources (main panels) and old sources
(insets), separated by source type.
Based on our source-classification scheme,
it is clear that sources with different types have disparate distributions of
flux and luminosity when either new or old sources are considered,
and that overall galaxies become the numerically
dominant population at full-band fluxes less than $\approx 10^{-16}$ \flux\ or 
\hbox{0.5--8 keV} luminosities less than $\approx 10^{42}$ erg~s$^{-1}$;\footnote{There
may be a selection effect that can potentially contribute to the result that
galaxies numerically dominate over AGNs at $L_{\rm 0.5-8\ keV}\lsim 10^{42}$~erg~s$^{-1}$ since
we used $L_{\rm 0.5-8\ keV}\ge 3\times 10^{42}$~erg~s$^{-1}$ as one of the AGN 
identification criteria.
However, as shown in \S~\ref{sec:maincat} (see the description of Column~78),
$\gsim 92\%$ of the AGNs in the main catalog can be identified by the criteria
other than the luminosity criterion; therefore, such a selection effect should be minimal.}
this trend is more pronounced when only new sources are considered.
It is also clear that
(1) new sources (either AGNs or galaxies)
have similar ranges of flux and luminosity
to those of old sources (either AGNs or galaxies);
(2) new sources (either AGNs or galaxies), as expected, 
typically have smaller fluxes (i.e., have a smaller median flux) than 
old sources (either AGNs or galaxies);
(3) compared to old AGNs,
new AGNs typically have smaller luminosities
(i.e., have a smaller median luminosity); and
(4) compared to old galaxies,
new galaxies have comparable luminosities
(i.e., have about the same median luminosity).

\begin{figure*}
\centerline{\includegraphics[scale=0.9]{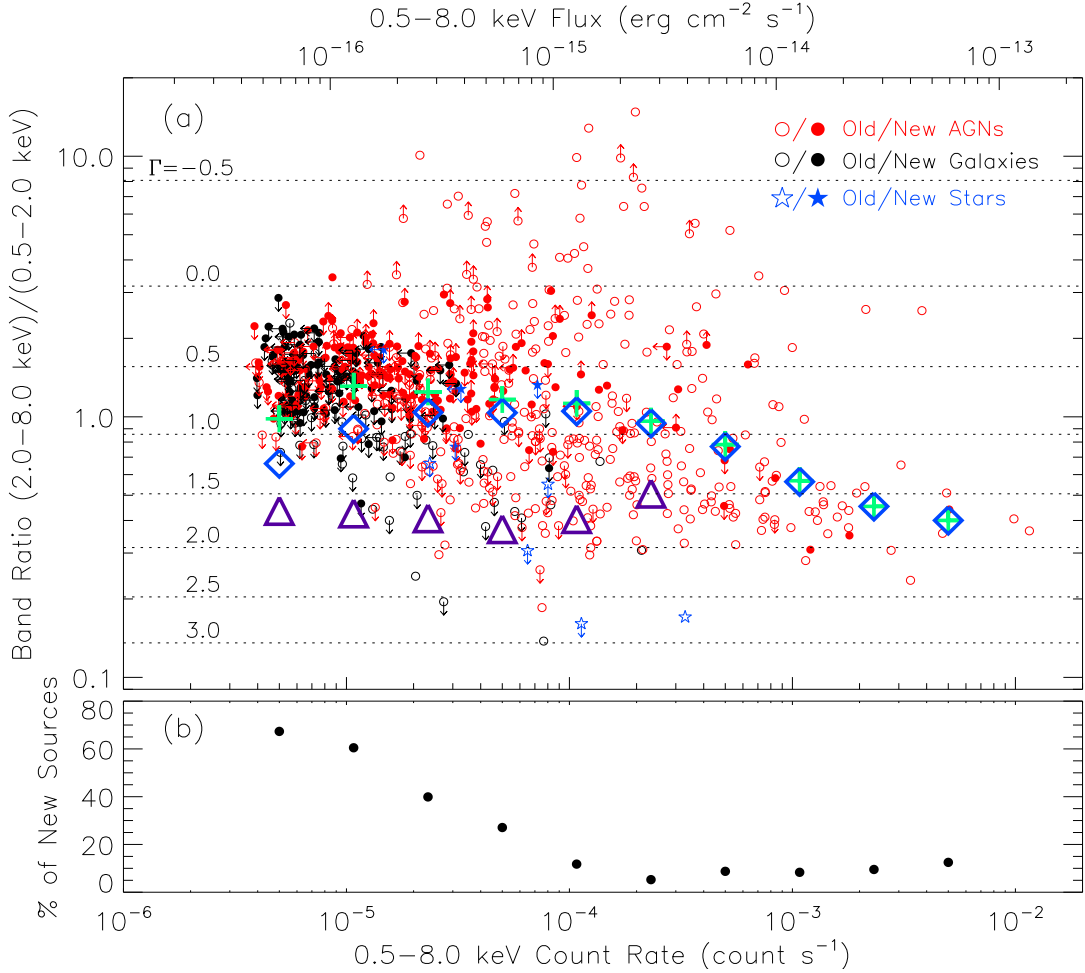}
}
\figcaption{
(a) Band ratio vs. full-band count rate for the main-catalog sources (for reference,
the top $x$-axis shows representative full-band fluxes, which are derived
from full-band count rates assuming a $\Gamma=1.4$ power law). 
Sources that are considered AGNs, galaxies, and stars are colored red, black, and blue, respectively.
Open circles and stars indicate AGNs/galaxies and stars that were previously detected in the L08
main catalog;
filled circles and stars indicate new AGNs/galaxies and stars, respectively.
Arrows indicate upper or lower limits, which mostly lie in the area of low count rates.
Sources detected only in the full band cannot be plotted.
Large crosses, triangles, and diamonds show average band ratios as a function of full-band count rate
derived in bins of $\Delta {\rm log(Count\hspace{0.1cm} Rate)}=0.6$ from stacking 
analyses, for all AGNs, all galaxies, and all sources (including both AGNs and galaxies), respectively.
Horizontal dotted lines show the band ratios
corresponding to given effective photon indexes.
(b) Fraction of new sources as a function of full-band count rate for the main-catalog sources.
The fractions are calculated in bins of $\Delta {\rm log(Count\hspace{0.1cm} Rate)}=0.6$.
[{\it see the electronic edition of the Supplement for a color version of this figure.}]
\label{bratio}}
\end{figure*}

We show in Figure~\ref{bratio}(a) the band ratio as a function of full-band count
rate for new sources (shown as filled symbols) and old sources (shown as open symbols).
The sources are color-coded according to their likely types, 
with red, black, and blue colors indicating AGNs, galaxies, and stars, respectively.
Also shown in Figure~\ref{bratio}(a) are the average band ratios
derived from stacking analyses following the procedure 
described in Luo et~al. (2011),
for all AGNs, all galaxies, and all sources (including both AGNs and galaxies),
shown as large crosses, triangles, and diamonds, respectively.
As expected, the overall average band ratio is dominated by AGNs because
most of the main-catalog sources are AGNs and AGNs typically are more \xray\ 
luminous than galaxies (see Fig.~\ref{f-lx}).
The overall average band ratio rises between 
full-band count rates of $\approx10^{-2}$ and $\approx 10^{-4}$ count~s$^{-1}$,
and it levels off and subsequently decreases
below full-band count rates of $\approx 10^{-4}$ count~s$^{-1}$.
The former increasing trend of the average band ratio
is due to an increase in the number of absorbed AGNs
detected at fainter fluxes and has been reported previously
(e.g., Tozzi et~et. 2001; A03; L05; L08);
the latter decreasing trend of the average band ratio
is partly because the contribution from normal and starburst galaxies 
increases at these lowest count rates (e.g., Bauer et~al. 2004).
Note that, at the lowest count rates studied, 
most of the sources have only band-ratio upper limits;
thus the average band ratio lies below the individual-source upper limits.
We show in Figure~\ref{bratio}(b) the fraction of new sources as a function of
full-band count rate for the main-catalog sources.
Above full-band count rates of $\approx 10^{-4}$ count~s$^{-1}$,
the fraction of new sources is small and roughly constant (\hbox{$\approx 5$--13\%});
below full-band count rates of $\approx 10^{-4}$ count~s$^{-1}$,
the fraction of new sources rises from $\approx 12\%$ to $\approx 67\%$ toward smaller full-band count rates.

\begin{figure}
\centerline{\includegraphics[scale=0.48]{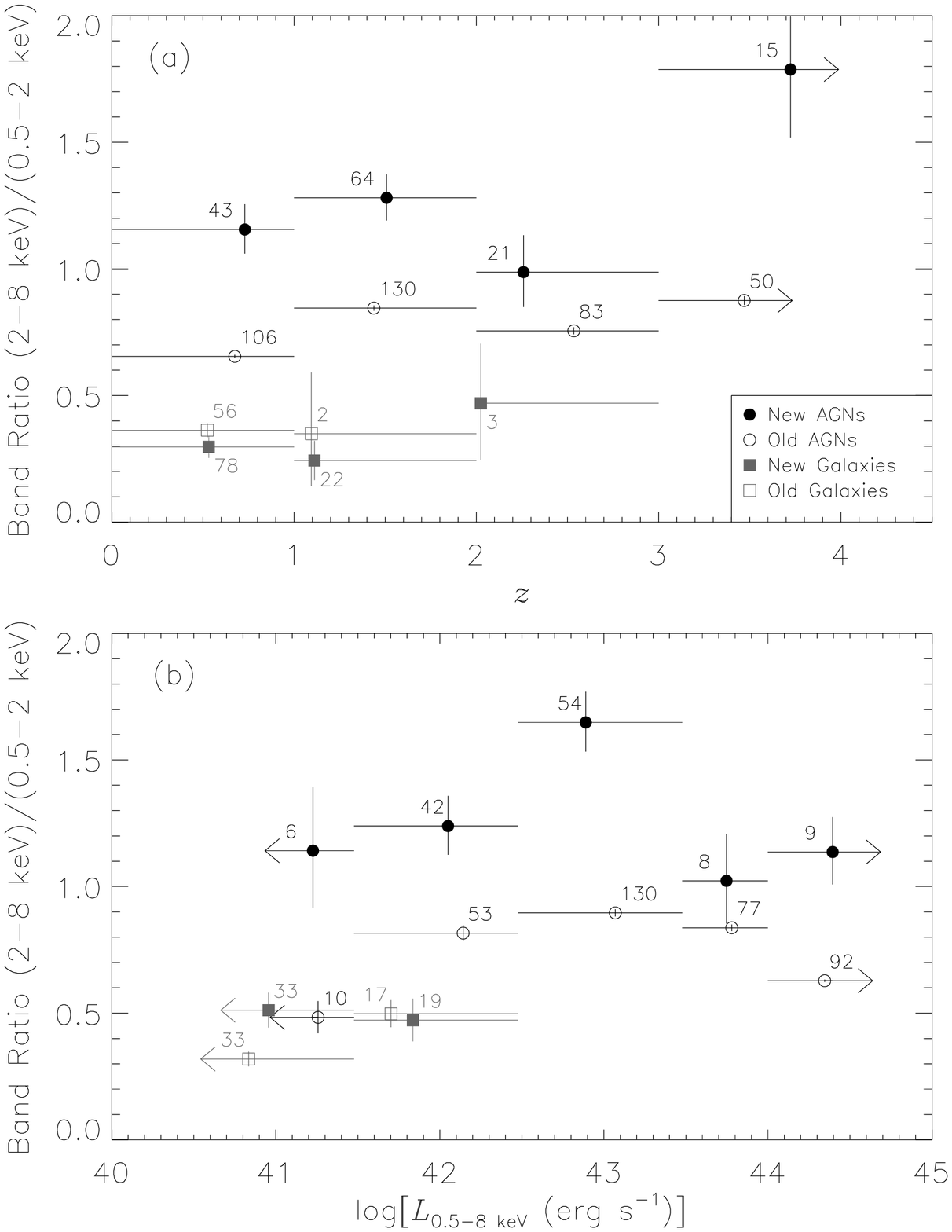}%{/bulk/wildrice1/xuey/cdfsultimate/ms/report/stack/plotstack.eps}
}
\figcaption{Average (i.e., stacked) band ratio in bins of 
(a) redshift ($0<z<1$, $1\le z<2$, $2\le z<3$, and $z\ge 3$) and
(b) \xray\ luminosity [$\log(L_{\rm X})<41.5$,
$41.5\le \log(L_{\rm X})<42.5$,
$42.5\le \log(L_{\rm X})<43.5$,
$43.5\le \log(L_{\rm X})<44.0$, and
$\log(L_{\rm X})\ge 44.0$] for new AGNs (filled circles),
old AGNs (open circles), new galaxies (filled squares), and old galaxies (open squares).
The median redshift or \xray\ luminosity in each bin is used for plotting.
The number of stacked sources in each redshift or luminosity bin is annotated.
\label{fig:stack}}
\end{figure}

To examine further the band-ratio behavior of new and old sources, 
we show in Fig.~\ref{fig:stack} the average (i.e., stacked) band ratio 
in bins of redshift and \xray\ luminosity for new AGNs, old AGNs, new galaxies, and old
galaxies, respectively.
According to Fig.~\ref{fig:stack},
(1) new AGNs have larger band ratios than old AGNs no matter which bin
of redshift or \xray\ luminosity is considered
(presumably due to the fact that
the detection of highly absorbed AGNs with large band ratios requires
deep observations given the small ACIS-I effective area
at high energies);
(2) in the two lower redshift bins ($0<z<1$ and $1\le z<2$),
the band ratios of new and old galaxies
appear roughly consistent and constant within errors
(hinting at no evolution in the \xray\ spectral shape of 
the accreting binary populations that dominate the \xray\ emission of normal and starburst galaxies);
(3) new AGNs and old AGNs have similar patterns of band ratio versus \xray\ luminosity,
both peaking at the bin of $42.5\le \log(L_{\rm X})<43.5$; and
(4) in the lowest luminosity bin [$\log(L_{\rm X})<41.5$],
new galaxies have a larger average band ratio than old galaxies, while in
a higher luminosity bin [$41.5\le \log(L_{\rm X})<42.5$],
new and old galaxies have consistent band ratios.

\begin{figure*}
\centerline{\includegraphics[scale=0.77]{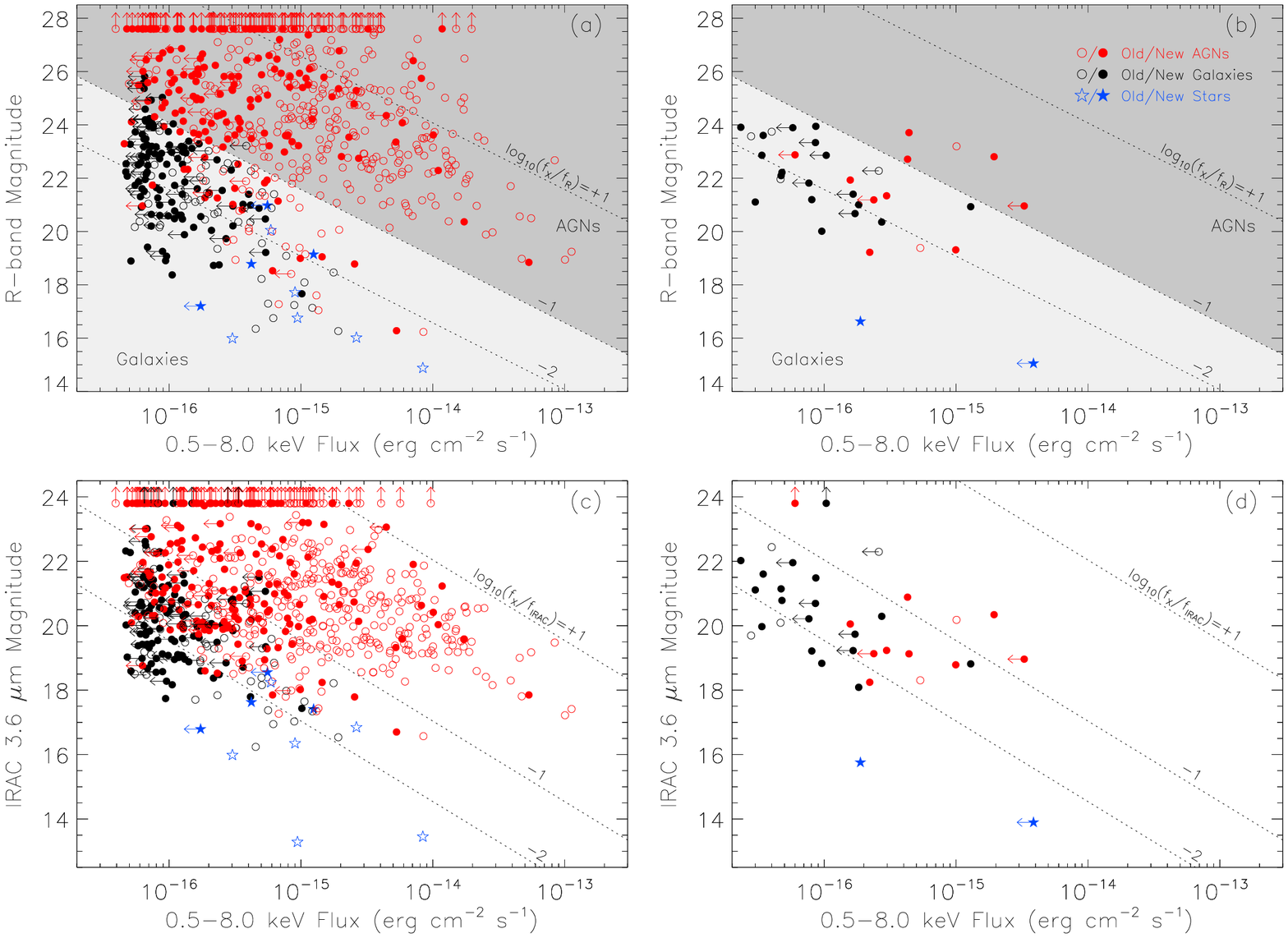}
}
\figcaption{
(Top) WFI $R$-band magnitude vs. full-band flux for sources in 
(a) the main catalog and (b) the supplementary optically bright catalog
[note that the legend in Panel (b) applies for all the panels in this figure].
Sources that are considered AGNs, galaxies, and stars are colored red, black, and blue, respectively.
Open circles indicate AGNs/galaxies
that were previously detected in (a) the L08 main catalog or 
(b) the L08 main or supplementary optically bright catalog;
open stars in (a) indicate stars 
that were previously detected in the L08 main catalog;
filled circles and stars indicate new AGNs/galaxies and stars, respectively.
Arrows indicate limits.
Diagonal lines indicate constant flux ratios between the WFI $R$-band and the full band, with the shaded areas showing the
approximate flux ratios for AGNs (dark gray) and galaxies (light gray).
(Bottom) IRAC 3.6~$\mu$m magnitude vs. full-band flux for sources in 
(c) the main catalog and (d) the supplementary optically bright catalog.
All the symbols are the same as those in Panels (a) and (b).
The diagonal lines indicate constant flux ratios between the IRAC 3.6~$\mu$m band and the
full band.
Note that several galaxies that have $R$-band detections were not detected
in the IRAC 3.6~$\mu$m band, probably due to source blending in the IRAC 3.6~$\mu$m band
and/or these galaxies being very blue systems. 
[{\it see the electronic edition of the Supplement for a color version of this figure.}]
\label{fox}}
\end{figure*}

We show in Figure~\ref{fox}(a) the WFI $R$-band magnitude versus
the full-band flux for new sources (filled symbols) and old sources (open symbols), as well as
the approximate flux ratios for AGNs and galaxies
(e.g., Maccacaro et~al. 1988; Stocke et~al. 1991; Hornschemeier et~al. 2001; 
Bauer et~al. 2004; also see the description of Column~78 for AGN identification).
The sources are color-coded according to their likely types, with red, black, and blue colors indicating AGNs, galaxies, and stars, respectively.
For comparison, we also show in Fig.~\ref{fox}(c) the IRAC 3.6~$\mu$m magnitude versus the full-band flux
for new sources (filled symbols) and old sources (open symbols), since a higher fraction of the main-catalog sources have counterparts in the
IRAC 3.6~$\mu$m band than in the WFI $R$-band (i.e., $\approx 88\%$ vs. $\approx 75\%$; 
see the description of \hbox{Columns~23--43}).
Overall, a total of 568 (76.8\%) of the main-catalog sources are likely AGNs,
and the majority of them lie in the region expected for 
relatively luminous AGNs [i.e., $\log (f_{\rm X}/f_{\rm R})>-1$; dark gray areas in Fig.~\ref{fox}(a)];
of these 568 AGNs, 192 (33.8\%) are new.
A total of 162 (21.9\%) of the main-catalog sources are likely galaxies,
and the majority of them lie in the region expected for 
normal galaxies, starburst galaxies, and low-luminosity AGNs
[i.e., $\log (f_{\rm X}/f_{\rm R})\le -1$; light gray areas in Fig.~\ref{fox}(a)];
of these 162 sources, 104 (64.2\%) are new.
Only 10 (1.3\%) of the main-catalog sources are likely stars
with low \hbox{X-ray}-to-optical flux ratios;
of these 10 stars, 4 are new.
Among new sources, normal and starburst galaxies
account for a fraction of $\approx 35\%$,
in contrast to $\approx 13\%$ if old sources are considered.
The above source-classification analysis indicates that, as expected, 
the \hbox{4~Ms} \mbox{CDF-S} survey is detecting sources in or close to a regime
where the galaxy number counts approach the AGN number counts,
due to the steeper number-count slope expected for galaxies (e.g., Bauer et~al. 2004).

\begin{figure}
\centerline{\includegraphics[scale=0.48]{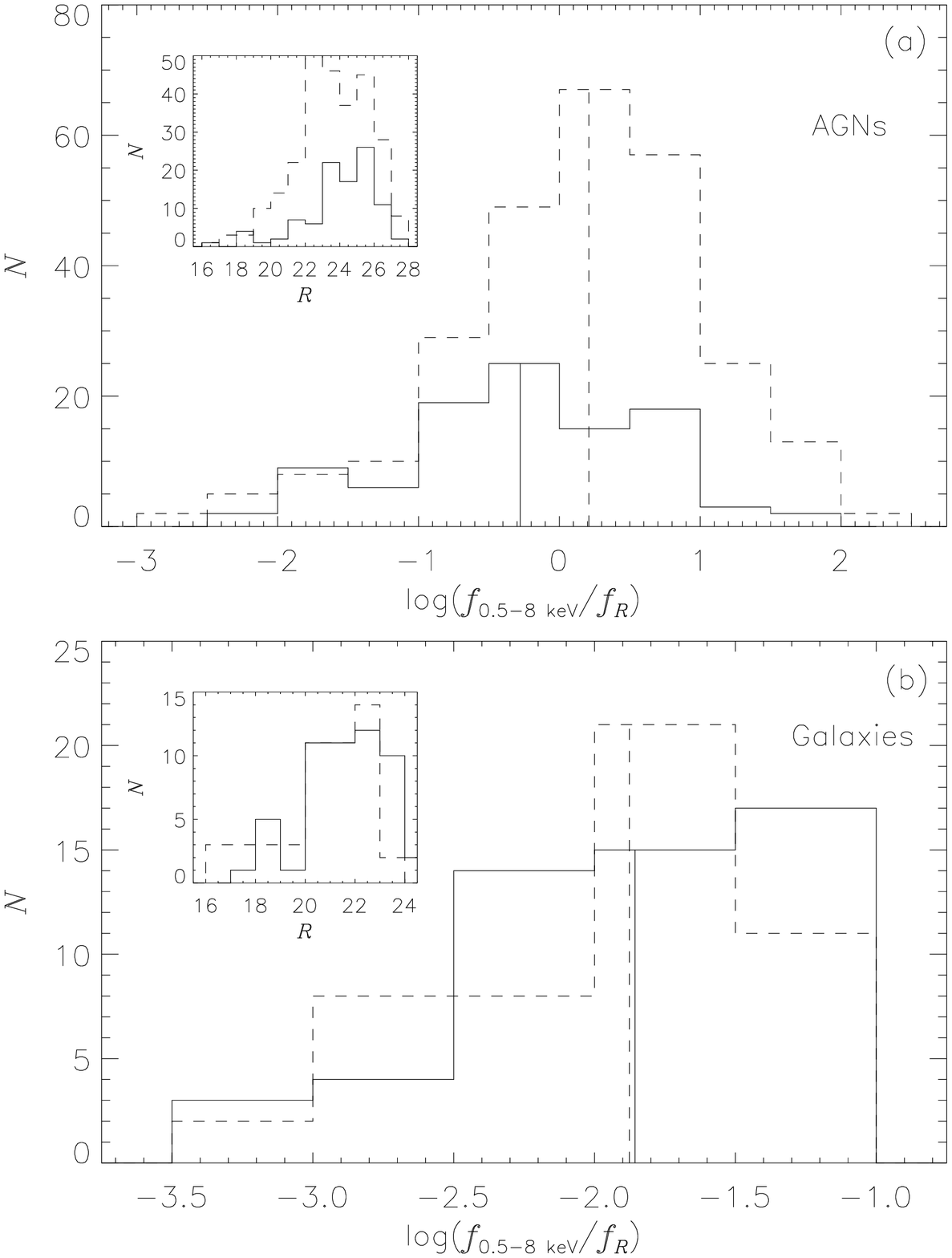}%{figs/x-to-R.eps}
}
\figcaption{Distributions of \xray-to-optical flux ratio for (a) new AGNs (solid histogram)
and old AGNs (dashed histogram) and (b) new galaxies (solid histogram) and
old galaxies (dashed histogram) with median flux ratios indicated by vertical lines.
Shown in the insets are the distributions of $R$-band magnitude for new AGNs/galaxies
(solid histograms) and old AGNs/galaxies (dashed histograms).
Only sources with both full-band and $R$-band detections are shown.
\label{xtor}}
\end{figure}

Figure~\ref{xtor} shows the 
distributions of \xray-to-optical flux ratio for new AGNs, old AGNs, new galaxies,
and old galaxies, respectively.
It is clear that 
(1) new AGNs generally have smaller \xray-to-optical flux ratios than old AGNs
and (2) new and old galaxies have similar distributions of \xray-to-optical flux ratio.

\section{Supplementary Optically Bright Chandra Source Catalog}\label{sec:supp2}

\subsection{Supplementary Catalog Production}

Of the 152 (i.e., $892-740=152$) candidate-list \hbox{X-ray} sources that
were not included in the main \chandra\ catalog (see \S~\ref{sec:srcselect}),
119 were of moderate significance ($0.004<P<0.1$).
To recover genuine \xray\ sources from this sample,
we constructed a supplementary \chandra\ source catalog consisting of
the subset of these sources 
that have bright optical counterparts.
Given that the density of optically 
bright sources on the sky is comparatively low,
it is likely that the \chandra\ sources with such counterparts are real.
We selected bright optical sources from the WFI $R$-band source
catalog described in \S~\ref{sec:img}.
We searched for bright optical counterparts (having $R\le 24.0$) to the 119
\hbox{X-ray} sources of interest using a matching radius of $1.2\arcsec$.
The choices of $0.004<P<0.1$, the $R$-band cutoff magnitude, and the matching radius
were made to ensure a good
balance between the number of detected sources and the 
expected number of false sources.
We find a total of 36 optically bright \hbox{X-ray} sources,
of which 3 are L08 main-catalog sources that were not included in our main catalog
and 3 are L08 supplementary optically bright sources (i.e., 30 new sources
in the 4~Ms supplementary catalog).
We note that the majority (22 out of 30) of the L08 supplementary optically 
bright sources are included in our main catalog (see the description of 
Column~59); this explains the small number of L08 supplementary sources 
included in our supplementary catalog.
We estimated the expected number of false matches to be $\approx 2.2$ 
(i.e., $\approx6.1\%$) by
manually shifting the \hbox{X-ray} source positions in right
ascension and declination 
and recorrelating with the optical sources.

We present these 36 \hbox{X-ray} sources in Table~\ref{tbl-sp2}
as a supplementary optically bright \chandra\ source catalog.
The format of Table~\ref{tbl-sp2} is identical to that of Table~\ref{tab:main}
(see \S~\ref{sec:maincat} for the details of each column).
We note that the source-detection criterion is $P<0.1$ for the sources in this
supplementary catalog, as opposed to $P<0.004$ for the main-catalog sources.
Additionally, we set the multiwavelength identification-related columns 
(i.e., Columns~18--22) to the WFI $R$-band matching results.

\begin{deluxetable*}{lllcccccccccc}
\tabletypesize{\scriptsize}
\tablewidth{0pt}
\tablecaption{Supplementary Optically Bright \chandra\ Source Catalog}

\tablehead{
\colhead{} &
\multicolumn{2}{c}{X-ray Coordinates} &
\multicolumn{2}{c}{Detection Probability} &
\colhead{}                   &
\colhead{}                   &
\multicolumn{6}{c}{Counts}      \\
\\ \cline{2-3} \cline{4-5} \cline{8-13} \\
\colhead{No.}                    &
\colhead{$\alpha_{2000}$}       &
\colhead{$\delta_{2000}$}       &
\colhead{$\log P$} &
\colhead{{\sc wavdetect}} &
\colhead{Pos Err}       &
\colhead{Off-Axis}       &
\colhead{FB}          &
\colhead{FB Upp Err}          &
\colhead{FB Low Err}          &
\colhead{SB}          &
\colhead{SB Upp Err}          &
\colhead{SB Low Err}          \\
\colhead{(1)}         &
\colhead{(2)}         &
\colhead{(3)}         &
\colhead{(4)}         &
\colhead{(5)}         &
\colhead{(6)}         &
\colhead{(7)}         &
\colhead{(8)}         &
\colhead{(9)}        &
\colhead{(10)}        &
\colhead{(11)}        &
\colhead{(12)}        &
\colhead{(13)}
}

\startdata
1 \dotfill \ldots & 03 31 44.64 &$-$27 45 19.4 &  $-1.9$ &  $-$5 &  0.8 &  10.10 &    35.4 &  19.2 &  18.0 &    25.8 &  $-$1.0 &  $-$1.0 \\
2 \dotfill \ldots & 03 31 55.98 &$-$27 39 42.8 &  $-2.0$ &  $-$5 &  1.1 &  11.25 &    24.2 &  14.1 &  12.9 &    14.6 &   \phantom{0}8.1 &   \phantom{0}6.9 \\
3 \dotfill \ldots & 03 31 56.42 &$-$27 44 11.4 &  $-1.9$ &  $-$5 &  0.6 &   \phantom{0}8.19 &    39.5 &  21.5 &  18.9 &    16.6 &  12.4 &   \phantom{0}9.8 \\
4 \dotfill \ldots & 03 31 57.24 &$-$27 45 37.2 &  $-1.3$ &  $-$5 &  0.8 &   \phantom{0}7.38 &    64.8 &  $-$1.0 &  $-$1.0 &    16.7 &  11.5 &  10.3 \\
5 \dotfill \ldots & 03 32  07.63 &$-$27 49 27.2 &  $-2.3$ &  $-$8 &  0.5 &   \phantom{0}4.63 &    23.4 &  11.2 &  10.0 &    12.2 &   \phantom{0}6.7 &   \phantom{0}5.5 \\
\enddata
\tablecomments{
Units of right
ascension are hours, minutes, and seconds, and units of declination are
degrees, arcminutes, and arcseconds.
Table~\ref{tbl-sp2} is presented in its entirety in the electronic edition.
A portion is shown here for guidance regarding
its form and content. The full table contains 79 columns of
information for the 36 \hbox{X-ray} sources.}
\label{tbl-sp2}
\end{deluxetable*}

\subsection{Properties of Supplementary-Catalog Sources}

We show in Fig.~\ref{pos}(b) the positions of the 36 sources in the
supplementary optically bright \chandra\ catalog, with the 30 new 
sources shown as filled circles.
These 36 supplementary sources have $R$-band AB magnitudes ranging from
15.1 to 23.9.
We show in Fig.~\ref{fox}(b) the $R$-band magnitude versus the full-band flux
for these 36 sources, with the sources being color-coded based on their likely types.
For comparison, Fig.~\ref{fox}(d) shows the IRAC
3.6~$\mu$m magnitude versus the full-band flux for these 36 sources.
A total of 12 (33.3\%) of these 36 sources are likely AGNs;
22 (61.1\%) of these 36 sources are likely galaxies and they all
lie in the region expected for normal galaxies, starburst galaxies, and 
low-luminosity AGNs;
2 (5.6\%) of these 36 sources are likely stars.
The majority of these 36 supplementary sources appear to be 
optically bright, \mbox{X-ray} faint non-AGNs (e.g., A03; Hornschemeier et~al. 2003)
as a result of our selection criteria,
and thus they are not representative of the faintest \hbox{X-ray} sources as a whole.
A total of 31 (86.1\%) of these 36 sources have either spectroscopic or 
photometric redshifts.
Of the 5 sources that have no redshift estimate,
2 are bright stars with their redshifts set to $-1.000$; the other 
3 have their photometry severely affected by a nearby bright source, 
thus no redshift estimates were available.

\section{Completeness and Reliability Analysis}\label{sec:comp}

We performed simulations to assess the completeness and reliability of our
main catalog; such practice has been common among \xray\ surveys 
(e.g., Cappelluti et~al. 2007, 2009; Puccetti et~al. 2009).

\subsection{Generation of Simulated Data}\label{sec:simdata}

First, we produced a mock catalog that covers the entire \cdfs\ and extends well 
below the detection limit of the 4~Ms exposure [i.e., mock \hbox{0.5--2 keV} 
flux limits of \hbox{(2--3)$\times 10^{-18}$} \flux].
Source coordinates were assigned using a recipe by Miyaji et~al. (2007) to
include realistic source clustering.
In this mock catalog, each simulated AGN was assigned a soft-band flux that was
drawn randomly from the soft-band \hbox{log $N$ -- log $S$} relation in
the AGN population synthesis model by Gilli, Comastri, \& Hasinger (2007).
Each simulated galaxy has a soft-band flux drawn randomly from 
the soft-band galaxy \hbox{log $N$ -- log $S$} relation of
the ``peak-M'' model by Ranalli, Comastri, \& Setti (2005).
The AGN and galaxy integrated fluxes match the cosmic \xray\ background fluxes.
The minimum soft-band fluxes simulated
($\approx 3\times 10^{-18}$ \flux\ for AGNs and $\approx 2\times 10^{-18}$ \flux\
for galaxies) are a factor of \hbox{$\approx 3$--5} below the detection limit of the central
4~Ms \cdfs\ (see \S~\ref{sec:smap}); inclusion of these undetectable sources 
simulates the spatially non-uniform background component due to undetected sources.
The soft-band fluxes of the simulated AGNs and galaxies were converted into
full-band fluxes assuming power-law spectra with $\Gamma=1.4$ and $\Gamma=2.0$, respectively.

Second, we constructed event lists from 54 simulated \hbox{ACIS-I} observations of the mock catalog,
each configured to have the same aim point, roll angle, exposure time, and aspect solution 
file as one of the \cdfs\ observations
(see Table~\ref{tbl-obs}).
The MARX simulator was used to convert source fluxes to a Poisson stream of dithered photons, 
and to simulate their detection by ACIS. 
These event lists represent only events arising from the mock point sources.

Third, we extracted the corresponding background event files that are appropriate to the
simulated source event files from the real 4~Ms \cdfs\ event files.
For each real event file, we masked all the events relevant to the main-catalog and
supplementary-catalog sources and
then filled the masked regions with events that obey the local probability distribution
of background events.
The resulting background event files include the contribution ($\approx 0.5$\%) of unresolved faint
sources that was also present in the MARX-simulated source event files.
To avoid counting the contribution of unresolved faint sources twice,
we removed 0.5\% of the events at random in each background event file and then
combined it with the corresponding source event file.
Thus we produced a set of 54 simulated ACIS-I observations that closely
mirror the 54 real \cdfs\ observations. 

Finally, we obtained a simulated merged event file (i.e., sum of source and background events) 
following \S~\ref{sec:img}, 
constructed images from this simulated merged event file for the three standard bands following \S~\ref{sec:img},
ran {\sc wavdetect} on each simulated combined raw image
at a false-positive probability threshold of $10^{-5}$ to produce a candidate-list catalog following \S~\ref{sec:list},
and utilized AE to perform photometry (and thus compute $P$ values) for the sources in this candidate-list catalog
following \S~\ref{sec:list}.

\subsection{Completeness and Reliability}\label{sec:comprel}

\begin{figure*}
\centerline{
\includegraphics[scale=0.64]{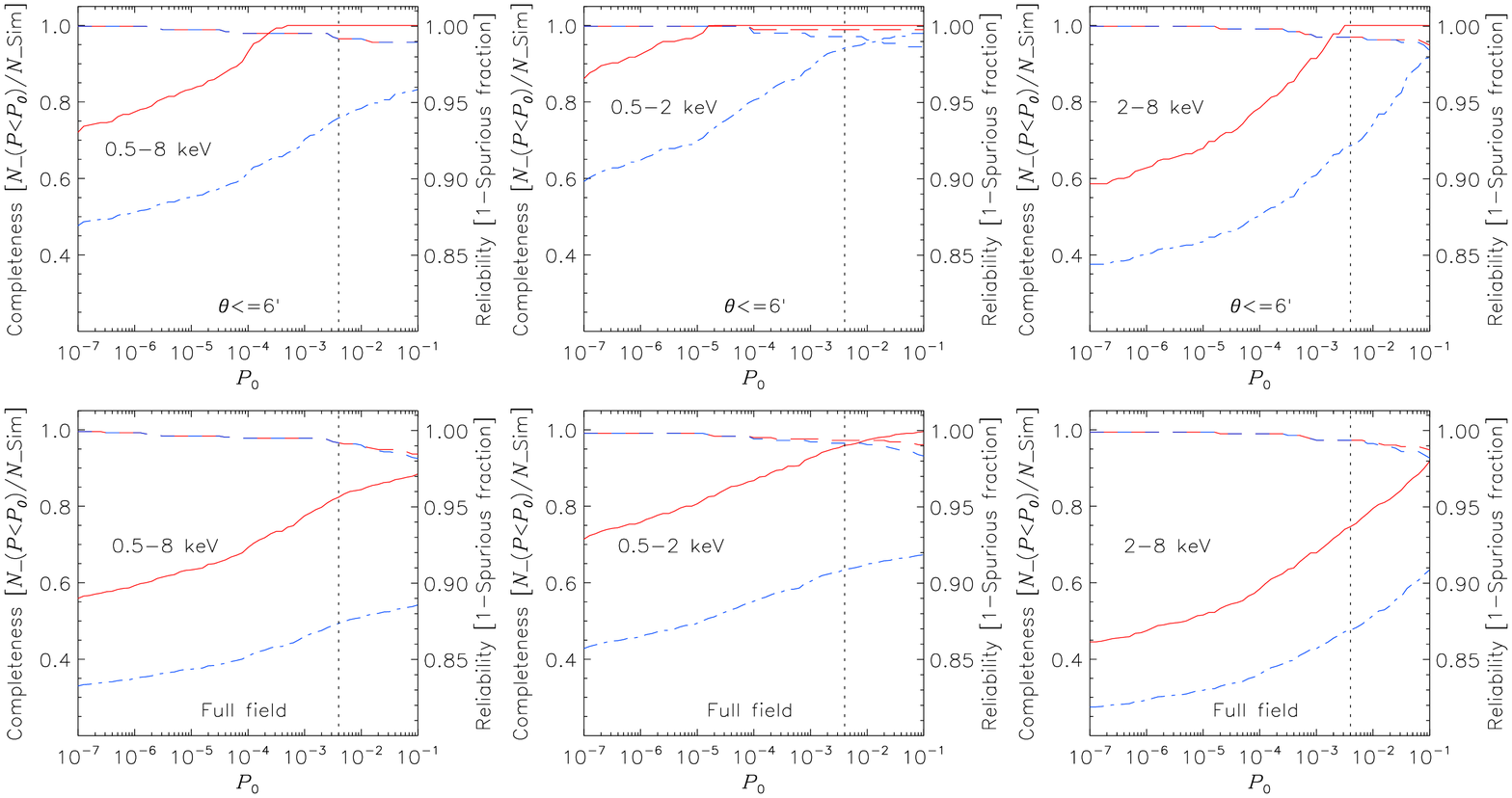}%{/bulk/wildrice1/xuey/cdfsultimate/ms/report/simulation/comp-rel.eps}
}
\figcaption{(Top) Case of $\theta\le 6\arcmin$: completeness (solid and dashed-dot curves; left $y$-axis) and reliability 
(long dashed and short dashed curves; right $y$-axis) as a function of $P_0$, 
the AE-computed binomial no-source probability threshold chosen for detection,
for the simulations in the full, soft, and hard bands, for sources with at least 
15 counts (red solid and long dashed curves)
and at least 8 counts (blue dashed-dot and short dashed curves), respectively.
Note that the short dashed curves overlap almost exactly along the long dashed curves
in some cases (e.g., top-left and top-right panels).
The vertical dotted lines indicate the chosen main-catalog source-detection 
threshold of $P_0=0.004$.
(Bottom) Same as top panels, but for the entire \cdfs\ field.\label{fig:comp1}
[{\it see the electronic edition of the Supplement for a color version of this figure.}]
}
\end{figure*}

Our simulations allow us to assess the completeness and reliability of our main catalog.
Completeness is defined as
the ratio between the number of detected sources (given
a specific detection criterion \hbox{$P<P_0$}) and the number of input simulated sources,
above a specific source-count limit (this source-count limit applies to both
the detected sources and the input simulated sources).
Reliability is defined as 
1 minus the ratio between the number of spurious sources
and the number of input simulated sources, above a specific source-count limit 
(again, this source-count limit applies to both
the spurious sources and the input simulated sources).
The top panels of Fig.~\ref{fig:comp1} 
show the completeness and reliability as a function of the AE-computed
binomial no-source probability $P$ within the central $\theta\le 6\arcmin$ area
for the simulations in the full, soft, and hard bands,
for sources with at least 15 counts and 8 counts.
The bottom panels of Fig.~\ref{fig:comp1} correspond to the case for the entire \cdfs\ field.
8 counts is close to our source-detection limit in the soft band.
In each energy band, the completeness level for the case of 8 counts is, as expected, 
lower than that for the case of 15 counts, for both the central 
$\theta\le 6\arcmin$ area and the entire \cdfs\ field;
and the completeness level for the case of either 8 counts or 15 counts within
the central $\theta\le 6\arcmin$ area is higher than the corresponding 
completeness level in the entire \cdfs\ field.
At the chosen main-catalog $P$ threshold of 0.004,
the completeness levels within the central $\theta\le 6\arcmin$ area are 
100.0\% and 75.8\% (full band), 100.0\% and 94.1\% (soft band),
and 100.0\% and 68.6\% (hard band) for sources with at least 15 and 8 counts, respectively.
The completeness levels for the entire \cdfs\ field are
82.4\% and 49.3\% (full band), 95.9\% and 63.5\% (soft band),
and 74.7\% and 47.6\% (hard band) for sources with at least 15 and 8 counts, respectively.
The reliability level ranges from 99.2\% to 99.8\% for each energy band and each source-count limit,
which implies that, in the main catalog (i.e., the entire \cdfs\ field), there are
about 4, 4, and 3 spurious detections with $\ge 15$ counts in the full,
soft, and hard bands, and 
about 4, 5, and 3 spurious detections with $\ge 8$ counts in the full,
soft, and hard bands, respectively.

\begin{figure}
\centerline{
\includegraphics[scale=0.55]{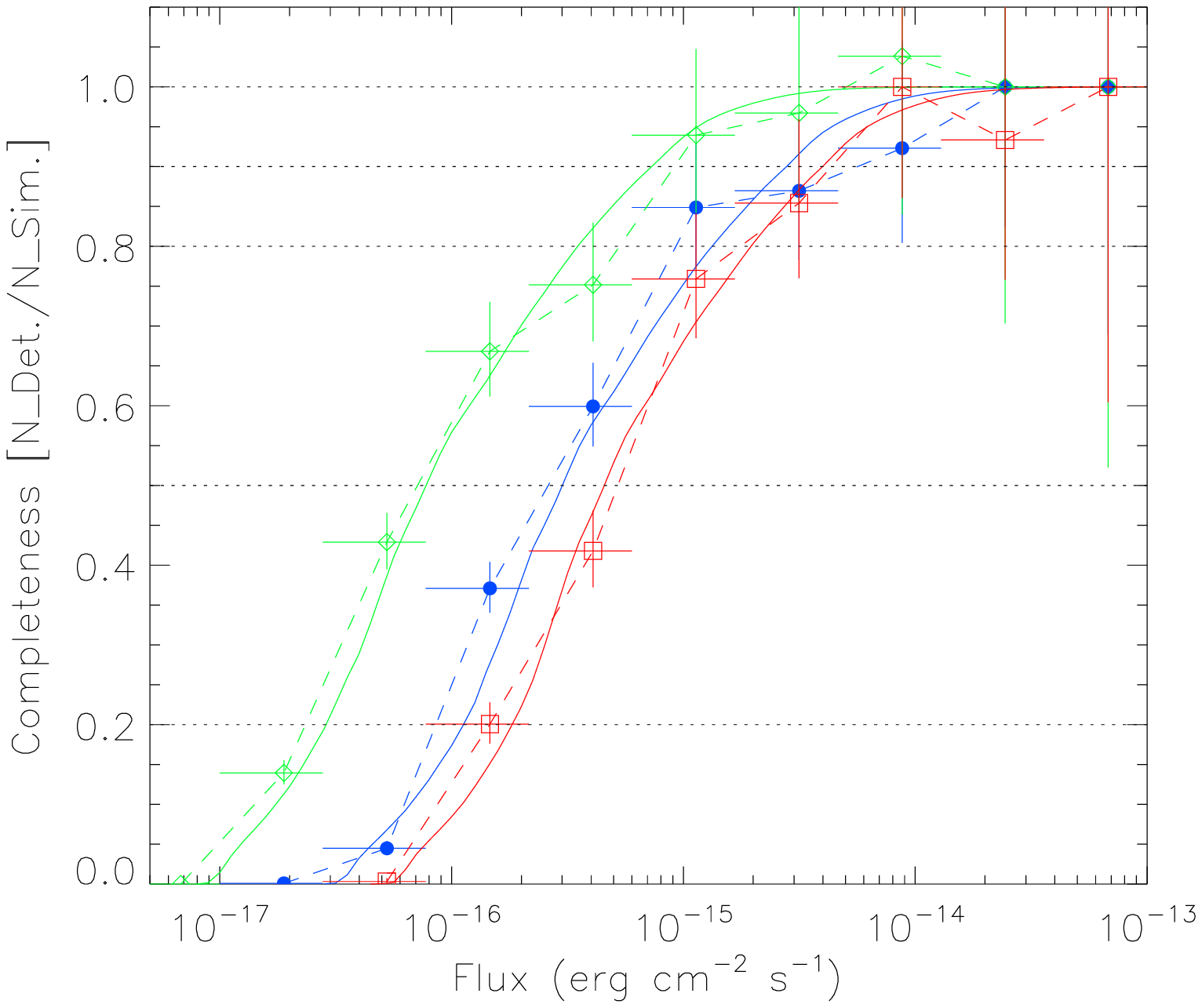}%{/bulk/wildrice1/xuey/cdfsultimate/ms/report/simulation/completeness-flux.eps}
}
\figcaption{Completeness as a function of flux under the main-catalog $P<0.004$ criterion
for the simulations in the full (blue filled circles), soft (green open diamonds), and hard (red open squares) bands, overlaid with the corresponding sky coverage curves
(solid curves) 
calculated in \S~\ref{sec:longexp} and normalized to the maximum sky coverage (see the solid
curves in Fig.~\ref{senhist}).
The dashed lines connect the corresponding cross points.
The horizontal dotted lines indicate five completeness levels.
[{\it see the electronic edition of the Supplement for a color version of this figure.}]
\label{fig:comp2}}
\end{figure}

We show in Fig.~\ref{fig:comp2} the completeness as a function of flux
under the main-catalog $P<0.004$ criterion
for the simulations in the full, soft, and hard bands.
These curves of completeness versus flux derived from the simulations approximately track
the normalized sky coverage curves (i.e., the curves of survey solid angle versus flux limit; 
shown as solid curves in Fig.~\ref{fig:comp2})
derived from the real \cdfs\ data (see \S~\ref{sec:smap}).
Table~\ref{tab:comp} gives the flux limits corresponding to four completeness
levels in the full, soft, and hard bands, as shown as horizontal dotted lines in Fig.~\ref{fig:comp2}.

\begin{deluxetable}{cccc}
\tabletypesize{\small}
\tablewidth{0pt}
\tablecaption{Flux Limit and Completeness \label{tab:comp}}
\tablehead{
\colhead{Completeness} &
\colhead{$f_{\rm 0.5-8\ keV}$} &
\colhead{$f_{\rm 0.5-2\ keV}$} &
\colhead{$f_{\rm 2-8\ keV}$} \\
\colhead{(\%)} &
\colhead{(\flux)} &
\colhead{(\flux)} &
\colhead{(\flux)} 
}
\startdata
90 & $2.8\times 10^{-15}$ & $7.3\times 10^{-16}$ & $4.0\times 10^{-15}$ \\
80 & $1.3\times 10^{-15}$ & $3.5\times 10^{-16}$ & $2.0\times 10^{-15}$ \\
50 & $3.0\times 10^{-16}$ & $7.8\times 10^{-17}$ & $4.6\times 10^{-16}$ \\
20 & $1.1\times 10^{-16}$ & $2.9\times 10^{-17}$ & $1.8\times 10^{-16}$ \\
\enddata
\end{deluxetable}

\section{BACKGROUND AND SENSITIVITY ANALYSIS}\label{sec:bkg}

\subsection{Background Map Creation}\label{sec:bmap}

We created background maps for the three standard-band images as follows.
We first masked the 740 main-catalog sources and the 36 supplementary catalog
sources using circular apertures
with radii of 1.5 (2.0) times the $\approx 99\%$ PSF EEF radii
for sources with full-band counts below (above) 10,000.
Larger masking radii were used for the brightest sources (there are 3 main-catalog
sources with full-band counts above 10,000)
to ensure their source photons were fully removed.
Approximately 18.3\% of the pixels were masked.
By design, the background maps include minimal or no
contributions from the sources in the main and supplementary catalogs; 
however, the background in the regions of a few extended sources 
(e.g., Bauer et~al. 2002; L05; A. Finoguenov et~al. in preparation)
will be slightly elevated.
We then filled in the masked regions for each source 
with background counts that obey the local probability distribution of counts
within an annulus with an inner radius being the aforementioned 
masking radius and an outer radius of 2.5 (3.0) times the \hbox{$\approx 99\%$}
PSF EEF radius for sources with full-band counts below 
(above) 10,000.
We summarize in Table~\ref{tbl-bkg} the background properties.
We find our mean background count rates to be in agreement with those presented in L08.
Our background is the sum of contributions
from the unresolved cosmic background, particle background, and instrumental
background (e.g., Markevitch 2001; Markevitch et~al. 2003).
We do not distinguish between these different background contributions
because we are here only interested in the total background.
Even with a 4~Ms exposure, 
the majority of the pixels have no background counts; i.e.,
in the full, soft, and hard bands, $\approx 65\%$, 89\%, and 72\% of the pixels are zero,
respectively.

\begin{deluxetable*}{lcccc}
%\tabletypesize{\small}

\tabletypesize{\footnotesize}
\tablecaption{Background Parameters}
\tablehead{
\colhead{}                                 &
\multicolumn{2}{c}{Mean Background}                 &
\colhead{Total Background$^{\rm c}$}                                 &
\colhead{Count Ratio$^{\rm d}$}                                \\
\cline{2-3}
\colhead{Band (keV)}                                 &
\colhead{(count pixel$^{-1}$)$^{\rm a}$}                         &
\colhead{(count Ms$^{-1}$ pixel$^{-1}$)$^{\rm b}$}               &
\colhead{(10$^5$ counts)}                &
\colhead{(Background/Source)}                
}
\tablewidth{0pt}
\startdata
Full (0.5--8.0)  & 0.482 & 0.252  & 33.3 & 10.6  \\
Soft (0.5--2.0)   & 0.119 & 0.063  & \phantom{0}8.2 & \phantom{0}4.3  \\
Hard (2--8)   & 0.363 & 0.178  & 25.1 & 20.5  \\
\enddata
\par 
\tablenotetext{a}{The mean numbers of background counts per pixel
measured from the background maps (see \S~\ref{sec:bmap}; note that a pixel has a size of $0.492\arcsec$), which were not corrected for vignetting or exposure-time variations.}
\tablenotetext{b}{The mean numbers of counts per pixel 
divided by the mean effective exposures
(i.e., 1.909~Ms, 1.877 Ms, and 2.040 Ms for the full band, soft band, and hard band, respectively)
that are measured from the background maps (see \S~\ref{sec:bmap}) and exposure maps (see \S~\ref{sec:img}), respectively; these calculations take into account the effects of vignetting and exposure-time variations.}
\tablenotetext{c}{Total numbers of background counts in the background maps.}
\tablenotetext{d}{Ratio between the total number of background 
counts and the total number of detected source counts.}
\label{tbl-bkg}
\end{deluxetable*}

\subsection{Sensitivity Map Creation}\label{sec:smap}

According to Table~\ref{tbldet},
the minimum detected source counts are $\approx 11.4$, 6.0, and 10.7 in the
full, soft, and hard bands for the main-catalog sources, which
correspond to full, soft, and hard-band fluxes of
\hbox{$\approx 3.5\times 10^{-17}$}, $8.8\times 10^{-18}$, and $6.4\times 10^{-17}$ \flux,
respectively,
assuming that sources having these minimum counts are located
at the average aim point and have a $\Gamma=1.4$ power law spectrum with Galactic absorption.
This calculation provides a measure of the ultimate sensitivity of the main catalog, 
which, however, is only relevant for a small central region 
near the average aim point.
We created sensitivity maps in the three standard bands for the main catalog in order to
determine the sensitivity as a function of position across the field.

In the binomial no-source probability equation [i.e., eq.~(\ref{equ:bi}) in \S~\ref{sec:srcselect}], we need to measure $B_{\rm src}$ and $B_{\rm ext}$ to obtain the minimum number of counts required
for a detection ($S$),
given the criterion of $P_{\rm threshold}=0.004$.
We determined $B_{\rm src}$ in the background maps for the main catalog using circular apertures with $\approx 90\%$ PSF EEF radii.
Due to the PSF broadening with off-axis angle,
the value of $B_{\rm ext}$ has an off-axis angle dependency, i.e.,
the larger the off-axis angle, the larger the value of $B_{\rm ext}$.
To follow the behavior of AE when extracting background counts of the main-catalog sources,
we derived the value of $B_{\rm ext}$ as follows:
for a given pixel in the background map, we computed its off-axis angle $\theta_p$ and
set the value of $B_{\rm ext}$ to the maximum $B_{\rm ext}$ value of the main-catalog sources
that are located in an annulus with the inner/outer radius being 
$\theta_p-0.25\arcmin$/$\theta_p+0.25\arcmin$ (note that the adopted maximum $B_{\rm ext}$ value 
corresponds to the highest sensitivity).  
Given the computed $B_{\rm src}$ and $B_{\rm ext}$, we numerically solved eq.~(\ref{equ:bi}) to obtain
the minimum counts $S$ (in the source-extraction region) required for detections under the
main-catalog source-detection criterion $P<0.004$.
We then created sensitivity maps for the main catalog using the
exposure maps, assuming a $\Gamma=1.4$ power-law model with Galactic
absorption.
The above procedure takes into account effects such as the PSF broadening with 
off-axis angle,
the varying effective exposure (due to, e.g., vignetting and CCD gaps; see Fig.~\ref{fbemap}), 
and the varying background rate across the field. 
There are 11 main-catalog sources lying \hbox{$\approx 1$--9\%} below the derived
sensitivity limits, i.e., 5 sources in the full band,
6 sources in the soft band, and  
none in the hard band,
probably due to background fluctuations and/or their real $\Gamma$ values
deviating significantly from the assumed value.

\begin{figure}
\centerline{\includegraphics[scale=0.4]{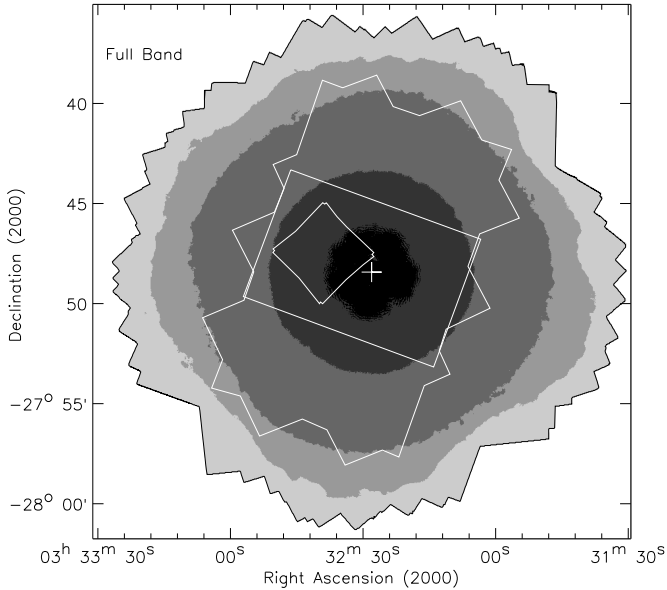}}
\figcaption{
Full-band sensitivity map for the main catalog, created 
following \S~\ref{sec:smap}.
The gray-scale levels, from black to light gray, represent areas with flux limits
of $<4.0\times 10^{-17}$, \hbox{$4.0\times 10^{-17}$} to $10^{-16}$, $10^{-16}$ to 
\hbox{$3.3\times10^{-16}$}, \hbox{$3.3\times 10^{-16}$} to $10^{-15}$, and $>10^{-15}$ \flux, respectively.
The regions and the plus sign are the same as those in Fig.~\ref{fbimg}.
\label{senmap}}
\end{figure}

We show in Figure~\ref{senmap} the full-band sensitivity map for the main catalog.
It is apparent that higher sensitivities are achieved at smaller off-axis angles.
The $\approx$1~arcmin$^2$ region at the average
aim point has mean sensitivity limits of 
$\approx 3.2\times 10^{-17}$, $9.1\times 10^{-18}$, and $5.5\times 10^{-17}$ \flux\ 
for the full, soft, and hard bands, respectively.

\subsection{Sensitivities of and Prospects for Longer \chandra\ Exposures}\label{sec:longexp}

To investigate the improvement in sensitivity due to additional exposure,
we also created exposure maps, background maps, and sensitivity maps for the 1~Ms and 2~Ms
\mbox{CDF-S} and simulated exposure maps, background maps, and sensitivity maps for the
8~Ms \mbox{CDF-S}.
We followed the procedure detailed in \S~\ref{sec:img} to create exposure maps for
the 1~Ms and 2~Ms \mbox{CDF-S}.
We simulated the 8~Ms \mbox{CDF-S} exposure maps by rotating the 4~Ms \mbox{CDF-S} exposure maps
90 degrees clockwise about the average aim point 
(this rotation approach simulates the variations of roll angles between observations)
and then adding the rotated 4~Ms exposure maps 
to the real 4~Ms exposure maps.
We followed \S~\ref{sec:bmap} to create background maps for
the 1~Ms and 2~Ms \mbox{CDF-S}.
To obtain the 8~Ms \mbox{CDF-S} background maps,
we first simulated a set of 4~Ms \mbox{CDF-S} background maps
by filling in each pixel in a simulated background map with background counts
that obey the local probability distribution of counts derived from 
the corresponding real 4~Ms \mbox{CDF-S} background map;
we then rotated the simulated 4~Ms background maps 90 degrees clockwise 
about the average aim point and added the rotated 4~Ms background maps
to the real 4~Ms background maps.
We followed the procedure detailed in \S~\ref{sec:smap} to create sensitivity maps
for the 1~Ms, 2~Ms, and 8~Ms \mbox{CDF-S} under the source-detection
criterion $P<0.004$,
where we appropriately scaled the value of $B_{\rm ext}$ that was derived
when creating the 4~Ms sensitivity maps (i.e., scaling factors of 0.25, 0.50, and 2.0
were adopted for the 1~Ms, 2~Ms, and 8~Ms \mbox{CDF-S}, respectively).

\begin{figure}
\centerline{\includegraphics[scale=0.5]{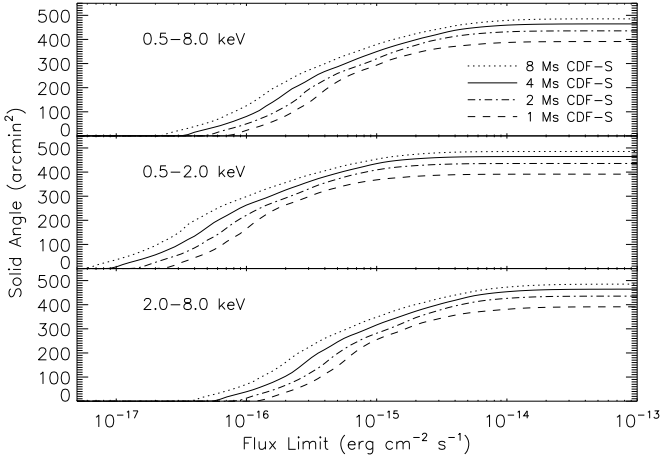}}
\figcaption{
Survey solid angle as a function of flux limit in the three standard bands
for the main catalog (shown as solid curves),
determined following \S~\ref{sec:smap}.
For comparison, the curves for 
the 1~Ms \hbox{CDF-S} (shown as dashed curves),
the 2~Ms \hbox{CDF-S} (shown as dash-dot curves), and
the simulated 8~Ms \hbox{CDF-S} (shown as dotted curves)
are also plotted; these curves are calculated/simulated consistently
(see \S~\ref{sec:longexp}).
\label{senhist}}
\end{figure}

\begin{figure}
\centerline{\includegraphics[scale=0.5]{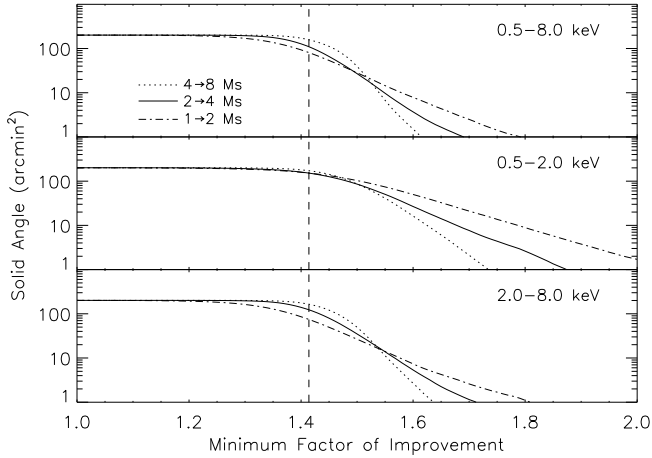}}
\figcaption{Survey solid angle within the central $\theta=8\arcmin$ area as a function of   
minimum factor of improvement in sensitivity 
in the three standard bands under the source-detection criterion $P<0.004$.
The improvement curves for the cases of 1 to 2~Ms, 2 to 4~Ms, and 4 to 8~Ms
are shown as dash-dot, solid, and dotted curves, respectively;
the curves for the first two cases are calculated using real data, while
the curves for the third case are simulated (see \S~\ref{sec:longexp}).
The vertical dashed lines indicate a factor of $\sqrt{2}$ improvement in sensitivity
that is expected for each doubling of exposure time in a background-limited case under
the assumption, here inapplicable, of Gaussian statistics.
\label{impro}}
\end{figure}

We show in Figure~\ref{senhist} plots of solid angle versus flux limit
in the three standard bands for the 1--8~Ms \mbox{CDF-S} under the 
source-detection criterion $P<0.004$.
It is clear that, for each of the three standard bands,
the quantitative increases in sensitivity are comparable 
between the cases of 1~to~2~Ms, 2~to~4~Ms, and 4~to~8~Ms.
To examine the improvement in sensitivity more clearly,
we created sensitivity improvement maps by dividing 
the 1~Ms, 2~Ms, and 4~Ms sensitivity maps by the 2~Ms, 4~Ms, and 8~Ms sensitivity maps, respectively.
We show in Figure~\ref{impro} plots of solid angle versus
minimum factor of improvement in sensitivity in the three standard bands
between the 1~Ms, 2~Ms, and 4~Ms \mbox{CDF-S} and the 2~Ms, 4~Ms, and 8~Ms \mbox{CDF-S}, respectively.
Figure~\ref{impro} only considers the central $\theta=8\arcmin$ area, since such an area
will be covered by any individual \mbox{CDF-S} observation.
It is clearly shown in Fig.~\ref{impro} that
(1) for the three standard bands,
the majority of the central \mbox{CDF-S} area generally 
has a factor of $> \sqrt{2}$ improvement in sensitivity 
for each doubling of exposure time (note that $\sqrt{2}=1.414$ corresponds to the
background-limited case under the assumption, here inapplicable, of Gaussian statistics); and
(2) among the three standard bands, 
the improvement in sensitivity is most pronounced in the soft band 
for each doubling of exposure time,
due to the fact that the soft band has the lowest background level (see, e.g., Table~\ref{tbl-bkg}).
We note that, for each of the three standard bands during each doubling of exposure time,
the improvement in sensitivity greater than a factor of \hbox{1.5--1.6}
generally occurs in the ACIS-I CCD gap areas (see Fig.~\ref{fbemap}) where the improvement 
in exposure time is often greater than a factor of two.
For the central $\approx 100$~arcmin$^2$ area,
the average improvement in sensitivity is typically a factor of \hbox{1.4--1.6}
for each of the three standard bands,
no matter which case of 1 to 2~Ms, 2 to 4~Ms, or 4 to 8~Ms is considered.
Based on the above analyses, 
we conclude that additional
exposure over the \hbox{CDF-S} region, e.g., doubling the current 4~Ms exposure, 
will still yield higher sensitivities in the central
area of the field by a comparable amount to any previous doubling 
of exposure time (i.e., 1 to 2~Ms, or 2 to 4~Ms).
The faintest sources detected in an 8~Ms \cdfs\ should have full, soft, and hard-band 
fluxes of $\approx 2.1\times 10^{-17}$, $6.0\times 10^{-18}$, and 
$3.7\times 10^{-17}$ \flux, respectively.
Based upon the derived sensitivity maps and CXRB synthesis models 
(e.g., Gilli, Comastri, \& Hasinger 2007; Treister, Urry, \& Virani 2009),
a total of $\approx 1000$ sources, including \hbox{$\approx 120$--130} new AGNs
and \hbox{$\approx 90$--100} new galaxies,
are expected to be detected in an 8~Ms \cdfs.

In addition to the improvements in sensitivity described above
that would probe unexplored discovery space, significant additional 
\hbox{CDF-S} exposure could greatly improve the \xray\ spectra, 
light curves, and positions for the nearly 800 known \xray\ sources 
in our main and supplementary catalogs. This would provide improved 
physical understanding of these sources; e.g., AGN content and 
luminosity, level and nature of AGN obscuration, shape of the 
X-ray continuum, and level of X-ray emission from X-ray binaries
and supernova remnants.

\section{SUMMARY}\label{sec:summary}

We have presented catalogs and basic analyses of \hbox{X-ray} sources
detected in the deepest \chandra\ survey: the 4~Ms \hbox{CDF-S}.
We summarize the most-important results as follows.

\begin{enumerate}
\item
The entire \hbox{CDF-S} consists of 54 individual observations, with
a summed exposure of 3.872~Ms and a total solid angle coverage of 464.5 arcmin$^{2}$.

\item
The main \chandra\ source catalog contains 740 sources that
were detected with {\sc wavdetect} at a false-positive probability
threshold of $10^{-5}$ and satisfy our binomial-probability source-selection 
criterion of \mbox{$P<0.004$};
this approach is designed to maximize the number of reliable sources detected.
These 740 sources were detected in up to three standard \mbox{X-ray} 
bands: 0.5--8.0~keV (full band), 0.5--2.0~keV (soft band),
and 2--8~keV (hard band).
716 (96.8\%) of these 740 sources have multiwavelength counterparts,
with 673 (94.0\% of 716) having either spectroscopic or photometric redshifts.

\item
The supplementary \chandra\ source catalog consists of 36 sources that
were detected with {\sc wavdetect} at a false-positive probability
threshold of $10^{-5}$ and satisfy the conditions of having
$0.004<P<0.1$ and having bright optical counterparts ($R<24.0$).

\item
\mbox{X-ray} source positions for the main
and supplementary \chandra\ source catalogs
have been determined using centroid and matched-filter techniques.
The absolute astrometry of the combined
\xray\ images and \xray\ source positions has been established using a
VLA 1.4~GHz radio catalog.
The median positional uncertainty at the $\approx 68\%$ confidence level 
is $0.42\arcsec$/$0.72\arcsec$
for the main/supplementary \chandra\ source catalog.

\item
Basic analyses of the X-ray and optical properties of the sources indicate
that they represent a variety of source types. 
More than 75\% of the sources in the
main \chandra\ catalog are likely AGNs.
Near the center of the 4~Ms \hbox{CDF-S} (i.e., within an off-axis angle of $3\arcmin$),
the observed AGN and galaxy source densities have reached
$9800_{-1100}^{+1300}$~deg$^{-2}$ and 
$6900_{-900}^{+1100}$~deg$^{-2}$, respectively.
The majority of the sources in the
supplementary optically bright catalog are likely normal
and starburst galaxies.

\item A total of 300 main-catalog sources are new,
compared to the 2~Ms main-catalog sources. 
Of the 300 new main-catalog sources,
$\approx 64\%$ are likely AGNs while $\approx 35\%$ are likely normal
and starburst galaxies (the remaining $\approx 1\%$ are likely stars), reflecting the rise of normal and starburst
galaxies at these very faint fluxes.
Indeed, based on our source-classification scheme, galaxies become the numerically dominant population of sources appearing
at \hbox{0.5--8 keV} fluxes less than $\approx 10^{-16}$ \flux\ or
luminosities less than $\approx 10^{42}$ erg~s$^{-1}$.

\item Simulations show that our main catalog is highly reliable 
(e.g., \hbox{$\lsim 5$} spurious detections are expected in the soft band) 
and is reasonably complete
(e.g., the completeness level for the soft band is $>94\%$ for sources with $\ge 8$~counts in the central $\theta\le 6\arcmin$ area).

\item
The mean background (corrected for vignetting and exposure-time variations) 
is 0.252, 0.063, and 0.178
count~Ms$^{-1}$~pixel$^{-1}$ for the full, soft, and hard bands, respectively;
the majority of the pixels have zero background counts.

\item
The 4~Ms \mbox{CDF-S} reaches  
on-axis flux limits of  
\hbox{$\approx 3.2\times 10^{-17}$}, $9.1\times 10^{-18}$, and $5.5\times 10^{-17}$ \flux\
for the full, soft, and hard bands, respectively,
a factor of 1.5--1.6 improvement over the 2~Ms \mbox{CDF-S}.
Another doubling of the \hbox{CDF-S} exposure time would still 
yield higher sensitivities in the central area of the field
by a comparable amount to any previous doubling of exposure time, 
thus providing a significant number of new \xray\ sources that
probe the key unexplored discovery space.

\end{enumerate}

The \hbox{CDF-S} source catalogs and data products provided by this paper will be
beneficial to many ongoing and future studies; 
e.g., a search for a population of heavily obscured AGNs at intermediate redshifts
(Luo et~al. 2011),
\mbox{X-ray} spectral constraints on heavily obscured and Compton-thick AGNs 
at high redshifts (Alexander et~al. 2011; Gilli et~al. 2011),
derivation of \hbox{X-ray} number counts for different source types
and the evolution of normal-galaxy luminosity functions
(B.~D.~Lehmer et~al., in preparation), and
a study of extended sources (A.~Finoguenov et~al., in preparation).
The \hbox{CDF-S} will continue to be a premiere deep-survey field over the coming decades;
the \hbox{CDF-S} imaging and spectroscopic coverage are superb and continue to improve.
For example, 
the Cosmic Assembly Near-IR Deep Extragalactic Legacy Survey (CANDELS)\footnote{See http://candels.ucolick.org/index.html for details on the CANDELS.} 
will utilize {\it HST}/WFC3 to image the GOODS-S; 
the highest sensitivity will be achieved with 5-orbit observations 
in the central region of this field.
Together, deeper \chandra\ and multiwavelength data will be critical to allow
comprehensive understanding of faint \xray\ sources.

\vspace{0.5cm}

We thank the referee for helpful feedback that improved this work.
We thank the \chandra\ Director's Office for allocating the time for these observations.
We also thank L.~K. Townsley for helpful discussions on data reduction and N.~A.~Miller for kindly providing the $5\sigma$ VLA 1.4~GHz radio catalog.
Support for this work was provided by NASA through \chandra\ Award SP1-12007A (YQX, BL, WNB, FEB) issued by the \chandra\ X-ray Observatory Center, which is operated by the Smithsonian Astrophysical Observatory, and by NASA ADP grant NNX10AC996 (YQX, BL, WNB).
We also acknowledge the financial support of
the Chile CONICYT under grants FONDECYT 1101024 and FONDAP (CATA) 15010003 (FEB),
the Royal Society (DMA),
the Philip Leverhulme Prize (DMA),
the Science and Technology Facilities Council (DMA, IRS),
the Italian Space Agency (ASI) under the ASI-INAF contracts I/009/10/0 and I/088/06/0 (AC, RG, PT, CV),
and the German Deutsche Forschungsgemeinschaft Leibniz Prize FKZ HA 1850/28-1 (GH).

\end{document}